\documentclass[reprint,superscriptaddress,amsmath,amssymb,floatfix,aps,prd]{revtex4-2}

\usepackage{graphicx}
\usepackage{dcolumn}
\usepackage{bm}
\usepackage{array}
\usepackage{xfrac}
\usepackage{caption}
\usepackage{subcaption}
\usepackage{gensymb}
\DeclareMathOperator{\sinc}{sinc}

\begin{document}

\title{Detection of gravitational waves in circular particle accelerators \\ II. Response analysis and parameter estimation using synthetic data}

\author{Suvrat Rao}
 \email{suvrat.rao@uni-hamburg.de}
 \affiliation{Hamburger Sternwarte, Universit\"at Hamburg, Gojenbergsweg 112, 21029 Hamburg, Germany}
\author{Julia Baumgarten}
 \affiliation{Physics Department, Jacobs University Bremen, Campus Ring 1, 28759 Bremen, Germany}
\author{Jochen Liske} 
 \affiliation{Hamburger Sternwarte, Universit\"at Hamburg, Gojenbergsweg 112, 21029 Hamburg, Germany}
\author{Marcus Br\"uggen}
 \affiliation{Hamburger Sternwarte, Universit\"at Hamburg, Gojenbergsweg 112, 21029 Hamburg, Germany}

\date{\today}

\begin{abstract}
 We simulate the response of a Storage Ring Gravitational Wave Observatory (SRGO) to astrophysical millihertz (mHz) gravitational waves (GWs), numerically obtaining its sensitivity curve and optimal choices for some controllable experiment parameters. We also generate synthetic noisy GW data and use Markov Chain Monte Carlo (MCMC) methods to perform parameter estimation of the source properties and their degeneracies. We show that a single SRGO could potentially localize the GW source in the sky using Earth’s rotation. We also study the source sky localization area, mass and distance estimation errors as functions of noise, data sampling rate, and observing time. Finally, we discuss, along with its implications, the capacity of an SRGO to detect and constrain the parameters of mHz GW events. 
\end{abstract}

\maketitle

\section{Introduction}
\label{intro}

Theoretical studies of gravitational waves (GWs) interacting with storage rings (circular particle accelerators where beams of charged particles circulate without collisions for long periods of time), intending to explore the possibility of using storage rings as GW detectors, have been conducted independently by several authors over the past decades \cite{Zer-Zion:355151, gr-qc/9906117, gr-qc/0210091, Dong:2002sr, Faber2018}.

However, these studies only considered the scenario of GWs propagating perpendicularly to the plane of the storage ring (i.e., a ``face-on" orientation). This particular orientation maximizes the GW-induced oscillations of the charged test mass particles circulating in the ring (which we will refer to as ``ions" henceforth) along the ring's radial direction. Thus, one can hope to exploit resonances with the beam's betatron oscillations and detect the presence of GWs using beam position monitors. However, as storage ring betatron frequencies generally fall in a range where no significant astrophysical GW sources exist, and since these radial oscillations are expected to be extremely small, this idea did not seem very promising. 

In our previous paper \cite{PhysRevD.102.122006}, henceforth referred to as paper~I, we instead discussed the possibility of detecting GWs by monitoring the circulation time of the ions instead of their orbit shape. We expect the circulation time of the ions to be affected by a passing GW not just because of the above-mentioned orbit shape distortion but also because of a GW-induced change in the velocity of the ions. While the former effect is, in general, proportional to $h^2$ (where $h$ is the dimensionless GW strain, see Appendix \ref{app_A}), the key insight from paper~I was that the latter is, in general, proportional to $h$. It is only in the special case of a face-on orientation of the storage ring, where a GW can only cause a beam shape distortion, that the latter effect vanishes, leaving only the former, much smaller effect. Since previous studies had only considered this special case, it was prematurely concluded that a detection of GWs by monitoring the circulation time of the ions was unfeasible. The realisation from paper~I that the effect of the GW-induced change in the velocity of the ions on their circulation time is in general proportional to $h$ clearly changes this assessment.

Moreover, in paper~I we also showed that the circulation time deviation induced by a GW has a periodicity equal to that of the GW, suggesting that it builds up over time during the first half of a GW period and then wanes during its second half. 
In addition, the peak value of the signal is proportional to the GW period, making a storage ring GW detector more sensitive to lower frequency GWs. Importantly, we showed in paper~I that, due to an overlap of several conditions, a Storage Ring Gravitational Wave Observatory (SRGO) would be most sensitive to the yet undetected millihertz (mHz) GWs from astrophysical sources that are also targeted by future space-based GW detectors such as the Laser Interferometer Space Antenna (LISA) \cite{2017arXiv170200786A, 2019arXiv190706482B}. 

The quantity measured by an SRGO would be analogous to the ``timing residuals" measured by pulsar timing arrays (PTAs), which are the GW-induced arrival-time deviations of radio pulses produced by millisecond pulsars from their expected, highly regular arrival times in the absence of GWs and noise sources \cite{Dahal2020}. Moreover, using ions as moving test masses in an SRGO to detect GWs is similar to the idea of GW detection by atom interferometry, where ballistic atoms are used as test masses \cite{Geiger2017, Badurina_2020, Canuel2018, Canuel_2020, Abe2021}.

Recently, D'Agnolo et al.\ independently derived the expected GW-induced circulation time deviation of ions in an SRGO which is in agreement with our results from paper~I, under the condition that no radio frequency (RF) system is present in the storage ring \cite{2105.00992}. Therefore, the general relativistic calculations for an SRGO using two different approaches (metric formalism in our paper~I, and Riemann tensor formalism by D'Agnolo et al.), give, effectively, the same result.

In paper~I, we considered the Large Hadron Collider (LHC) at CERN as an existing facility that could potentially be turned into an SRGO. However, the LHC is not the ideal facility to realize the SRGO detector because the presence of an RF system in the storage ring would dampen the GW signal that we hope to detect \cite{2105.00992}. Instead, rings capable of storing coasting (i.e., without interference by an RF system) ``crystalline ion beams" \cite{Habs1995, Schtz2001, PhysRevLett.87.184801} or rings that could potentially store a single circulating ion \cite{Nagaitsev:IPAC2019-MOPRB089} may be better laboratories for detecting GWs. Moreover, there are better options for the ion time-tagging detector technology than that proposed in paper~I, such as ``beam arrival time monitors" \cite{Frisch:2016hju}, which are electro-optic charge centroid monitors providing femtosecond timing precision. Also, improving the vacuum quality inside storage rings would enable sustaining stable, coasting ion beams or single ions for longer periods, allowing for longer SRGO observation runs.

In this paper we continue our investigation from paper~I, improving on some previous results, and exploring new areas with an updated theoretical formalism and numerical simulations. The rest of this paper is organized as follows: in Section~\ref{review}, we provide an updated theoretical framework for an SRGO. In Section~\ref{models}, we provide the mathematical models and numerical procedures used in this investigation. Section~\ref{response_analysis} presents the SRGO response function and its variation with various parameters. In Section~\ref{sect_param_estim}, we study the estimation of GW source parameters with an SRGO using artificial noisy data. Finally, in Section~\ref{discuss}, we discuss various aspects of our study, and we summarize our main results in Section~\ref{summary}.

\section{Review and revision of SRGO basics}
\label{review}

Using the metric formalism of general relativity, in paper~I, we derived  the circulation time deviation of test masses in a storage ring due to GWs relative to the circulation time in the absence of any GWs.
Here, we revise some of these results and display them in a more concise form. We recall from Eq.~(10) of paper~I that the circulation time deviation (relative to the case of no GWs) is given by the expression
\begin{eqnarray}
\lefteqn{\Delta T_{\rm GW}(T) = } \nonumber \\
& & \Bigl(1-\frac{v_{0}^2}{2c^2}\Bigr)\int_{t_{0}}^{t_{0}+T}\bigl(h_{\theta\phi\psi}(t, \alpha(t))-h_{\theta\phi\psi}(t_{0}, \alpha_0)\bigr)dt ,
   \label{dtcircular}
\end{eqnarray}
where $v_{0}$ is the constant speed of the ions in the absence of any GWs, $c$ is the speed of light, $t_0$ is the time of the start of the monitoring campaign, and $T$ is the duration of the monitoring campaign. $h_{\theta \phi \psi}$ is a function of the GW strain amplitudes, $h_{+,\times}(t)$, the orientation of the GW with respect to the ring, parameterized by the three angles $\theta(t), \phi(t)$, and $\psi(t)$, which vary with time due to the Earth's rotation (see below), and the angular trajectory of the ions in the ring, $\alpha(t)$.

As it is not possible to know a priori whether GWs are present at any given time, $v_0$ in Eq.~(\ref{dtcircular}) cannot be measured in practice. We therefore reformulate Eq.~(\ref{dtcircular}) in terms of $v(t_0) = v_i$, the instantaneous initial speed of the ions, which {\em can} be measured. Note that $v_i$ denotes the {\em instantaneous} speed of the ions, as opposed to the speed averaged over one or more revolutions. We start from Eq.~(6) of paper~I, and keeping the term $v(t_0)$ as it is, we follow through with the remaining derivation as done in paper~I, to get a reformulated expression for the circulation time deviation:
\begin{multline}
   \Delta T_{\rm GW} \left( T \right) = \int_{t_{0}}^{t_{0}+T}\biggl[\Bigl(1-\frac{v_{i}^2}{2c^2}\Bigr) h_{\theta\phi\psi}(t, \alpha(t)) \\ - \frac{1}{2}\Bigl(1-\frac{v_{i}^2}{c^2}\Bigr) h_{\theta\phi\psi}(t_{0}, \alpha_0)\biggr]dt.
   \label{dtcircularnew}
\end{multline}
Thus, $\Delta T_{\rm GW} \left( T \right)$ is now the timing deviation relative to the expected ion arrival times calculated using $v_i$ instead of $v_0$. 

The first term in Eq.~(\ref{dtcircularnew}) is an oscillatory function containing three different frequencies, viz.\ the GW frequency, Earth's rotational frequency and the revolution frequency of the storage ring ions. The latter frequency is always much greater than the former two in the case of mHz GWs. This allows us to analytically integrate out the rapidly oscillating terms corresponding to the ion revolution frequency. From paper~I, we recall that
\begin{equation}
\begin{aligned}
   h_{\theta\phi\psi}(t, \alpha) = {} & h_{+}(t)\Bigl(f_{s}^{+}\sin^2{\alpha}+f_{c}^{+}\cos^2{\alpha}+f_{sc}^{+}\sin{2\alpha}\Bigr) +\\
   &h_{\times}(t)\Bigl(f_{s}^{\times}\sin^2{\alpha}+f_{c}^{\times}\cos^2{\alpha}+f_{sc}^{\times}\sin{2\alpha}\Bigr),
\end{aligned}
\label{strain}
\end{equation}
where $h_{+}(t)$ and $h_{\times}(t)$ are the plus and cross polarization GW strain components, $\alpha(t)=\alpha_0 + \frac{v_i}{R}(t-t_0)$, and $R$ is the radius of the storage ring. The terms $\sin^2{\alpha}$, $\cos^2{\alpha}$ and $\sin{2\alpha}$ integrate out to yield the constant factors $\frac{1}{2}$, $\frac{1}{2}$ and $0$ respectively. The `$f$'-terms inside the parentheses are functions of cosines and sines of the angles $\theta, \phi, \psi$ (see paper~I, Section~2). 

We now make the assumption that $v_{i}$ is large enough such that the second term in Eq.~(\ref{dtcircularnew}), which is linear in $T$, is negligible compared to the amplitude of the oscillation of the first term, even for integration times of $T \approx$ a few days. Although the best ion energy for a successful GW detection is still under discussion, the above is a reasonable assumption since relativistic ions may indeed be advantageous (Schmirander et al., in preparation).

Therefore, in the end, we obtain the reformulation,
\begin{equation}
   \Delta T_{\rm GW} \left( T \right) = -\frac{1}{2}\Bigl(1-\frac{v_{i}^2}{2c^2}\Bigr)\int_{t_{0}}^{t_{0}+T}\bigl(F_{+}h_{+} + F_{\times}h_{\times}\bigr)dt ,
   \label{dtcircular2}
\end{equation}
where $F_{+}$ and $F_{\times}$ are the plus and cross polarization antenna pattern functions of an SRGO, with
$F_{+}=\sin^2{\theta}\cos{2\psi}$ and $F_{\times}=\sin^2{\theta}\sin{2\psi}$. 

Note that in our derivation of $\Delta T_{\rm GW} \left( T \right)$, we only considered terms that are linear in $h$ (see paper~I, Section~2) since all contributions by higher order terms are usually negligible. However, in the specific case of a face-on orientation of the storage ring relative to the GW source (corresponding to $\theta = 0$), we see that $\Delta T_{\rm GW} \left( T \right) =0$. In this specific case, the signal is therefore dominated by higher order $h$-terms, as discussed above. We will nevertheless continue to disregard these terms even in the face-on case as these much smaller higher order terms could in any case never be detected by an SRGO that is just sensitive enough to detect the first order $h$-term in other orientations.

Note that the integrand in Eq.~(\ref{dtcircular2}) now has the same form as the GW response of the Laser Interferometer Gravitational-wave Observatory (LIGO) \cite{Abbott2009}, but the response signal in our case is a time-integral of this integrand. Also, while the antenna pattern functions of an SRGO are very different compared to those of LIGO, interestingly, they happen to be exactly the same as those of a bar detector \cite{PhysRevLett.17.1228, PhysRevLett.18.498, PhysRevD.87.082002, PhysRevD.54.1264}, the longitudinal axis of which is aligned with the axis of the storage ring (see Section~4.2.1 of \cite{Sathyaprakash2009}). Further, the SRGO antenna pattern shown in Fig.~2 of paper~I (averaged over all polarization angles), which was derived after making several approximations, happens to be exactly valid even for the general case derived here.

For GWs in the mHz regime, the effect of Earth's rotation must be taken into account, as it causes the angles $\phi$, $\theta$ and $\psi$ to be periodic functions of time with a period equal to a sidereal Earth day. We recall from paper~I that $\phi, \theta$ and $\psi$ are the Euler angles which transform the observer's coordinate system with the $z$-axis coinciding with the storage ring axis and the $x$-axis pointing towards the timing detector to a coordinate system with the new $z$-axis coinciding with the GW propagation direction, and the $xy$ axes being aligned with the GW polarization axes. Specifically, $\phi, \theta$ and $\psi$ are the rotations around the $z$, $y'$ and $z''$ axes, respectively. Both coordinate systems are right-handed, and, strictly speaking, their origin is at the center of the storage ring. However, given the large distances to any GW sources, we may consider the storage ring to be located at the Earth's centre, in which case the ring's geographical coordinates only specify the ring's orientation in space. $\phi, \theta$ and $\psi$ are thus given by:
\begin{center}
\begin{widetext}
\begin{multline}
\begin{pmatrix}
\cos{\psi} & -\sin{\psi} & 0\\
\sin{\psi} & \cos{\psi} & 0\\
0 & 0 & 1
\end{pmatrix} \cdot \begin{pmatrix}
\cos{\theta} & 0 & \sin{\theta} \\
0 & 1 & 0\\
-\sin{\theta} & 0 & \cos{\theta}
\end{pmatrix} \cdot \begin{pmatrix}
\cos{\phi} & -\sin{\phi} & 0\\
\sin{\phi} & \cos{\phi} & 0\\
0 & 0 & 1
\end{pmatrix} = 
\begin{pmatrix}
\cos{\psi_{\rm eq}} & -\sin{\psi_{\rm eq}} & 0\\
\sin{\psi_{\rm eq}} & \cos{\psi_{\rm eq}} & 0\\
0 & 0 & 1
\end{pmatrix} \cdot \begin{pmatrix}
-\sin{\delta_{\rm src}} & 0 & \cos{\delta_{\rm src}} \\
0 & 1 & 0\\
-\cos{\delta_{\rm src}} & 0 & -\sin{\delta_{\rm src}}
\end{pmatrix} \\ \cdot \begin{pmatrix}
-\cos{\omega_\oplus\bigl(\alpha_{\rm src}-l_0-(t-t_0)\bigr)} & \sin{\omega_\oplus\bigl(\alpha_{\rm src}-l_0-(t-t_0)\bigr)} & 0\\
-\sin{\omega_\oplus\bigl(\alpha_{\rm src}-l_0-(t-t_0)\bigr)} & -\cos{\omega_\oplus\bigl(\alpha_{\rm src}-l_0-(t-t_0)\bigr)} & 0\\
0 & 0 & 1
\end{pmatrix} \cdot \begin{pmatrix}
\sin{\theta_{\rm SRGO}} & 0 & -\cos{\theta_{\rm SRGO}} \\
0 & 1 & 0\\
\cos{\theta_{\rm SRGO}} & 0 & \sin{\theta_{\rm SRGO}}
\end{pmatrix} \cdot \begin{pmatrix}
\cos{\phi_0} & -\sin{\phi_0} & 0\\
\sin{\phi_0} & \cos{\phi_0} & 0\\
0 & 0 & 1
\label{angles}
\end{pmatrix} 
\end{multline}
\end{widetext}
\end{center}
$\omega_\oplus$ is the angular speed of Earth's rotation. $\alpha_{\rm src}$ and $\delta_{\rm src}$ are the right ascension (in units of time) and declination of the GW source, and $\psi_{\rm eq}$ is the GW's polarization angle in the equatorial coordinate system.
$\phi_0$ 
is the angle between the line joining the center of the storage ring to the timing detector, and the meridian passing through the center of the storage ring, measured using the right-hand curl rule starting from the detector position. $l_0$ and $\theta_{\rm SRGO}$ are, respectively, the local sidereal time at the start of the observations, $t_0$, and the geographical latitude of the storage ring.

Note that, without loss of generality, we may assume that the timing detector lies due south (turning the last factor in Eq.~\ref{angles} above into the identity matrix), such that the SRGO coordinate system is similar to the horizon celestial coordinate system, where $\phi = 360 \degree - {\rm azimuth}$ and $\theta = 90 \degree + {\rm altitude}$. The difference is due to the fact that the handedness of the horizon system is opposite to the right-handed convention used above. Also, $\phi$ and $\theta$ are defined to point along the propagation direction of the GW, whereas the azimuth and altitude point towards the source.


Furthermore, the polarization angle in a coordinate system is defined as the angle between a reference line made by projecting the z-axis onto the GW wavefront plane, and a chosen GW polarization axis, following a right-handed rotation convention. In the equatorial celestial coordinate system, the z-axis points towards the north celestial pole. In the SRGO coordinate system, the z-axis points perpendicular to the plane of the storage ring. Therefore, $\psi$ and $\psi_{\rm eq}$ are the same when the SRGO is located at one of the Earth's poles.

Denoting the result of the matrix product on the right-hand side of Eq.~(\ref{angles}) as $R$, we find:
\begin{equation}
\begin{matrix}
\phi(t) = \arctan{\left(\frac{R_{21}}{-R_{20}}\right)},\\
\theta(t) = \arccos {(R_{22})},\\
\psi(t) = \arctan {\left(\frac{R_{12}}{-R_{02}}\right)}.\\
\end{matrix}
\end{equation}
Note that in the above relations, we must take care to use the appropriate functions to obtain the angles in their correct quadrants. 

\begin{figure}[!ht]
\includegraphics[width=\linewidth]{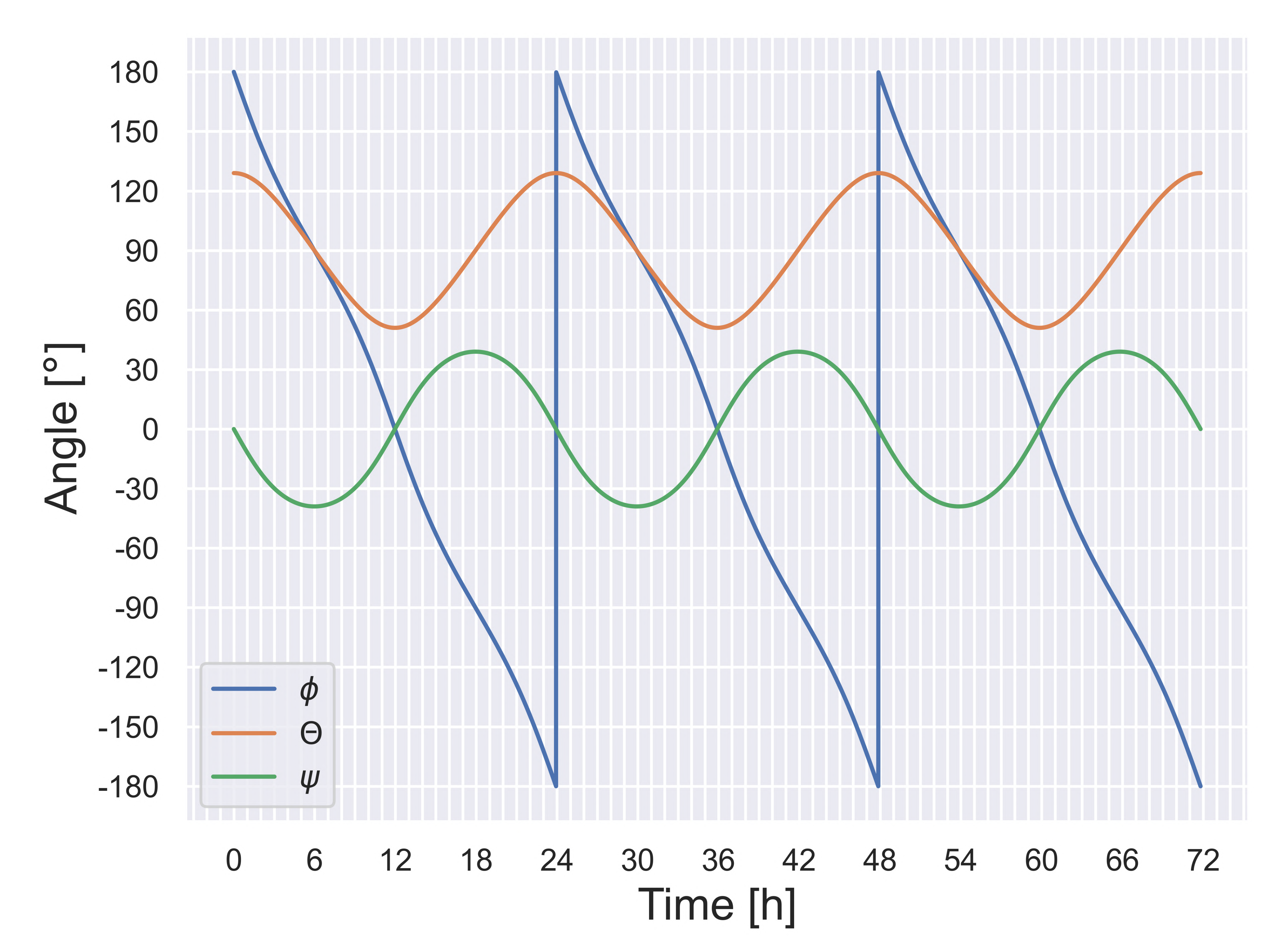}
\caption{The time evolution of the Euler angles, $\phi, \theta, \psi$ for the case of $\theta_{\rm SRGO} = 51\degree$ and all other parameters set to $0$ (cf.\ also Fig.~\ref{fig_response}).}
\label{fig_angles}
\end{figure}

In Fig.~\ref{fig_angles} we show the time evolution of $\phi$, $\theta$ and $\psi$ over the course of three days. The $\theta$ and $\psi$ curves determine the track of the GW source across the SRGO antenna pattern due to Earth's rotation (cf.\ Eq.~\ref{dtcircular2}).

\section{Models and numerical procedures}
\label{models}

We obtain all the numerical results in this work from a computer code written in Python and freely available on GitHub \cite{Rao_Storage_Ring_Gravitational-wave_2022}. Below, we detail the mathematical models and numerical procedures used by the code to obtain our results.

\subsection{GW source model}
\label{subsect_A}

We consider the simplest realistic models for mHz GWs from astrophysical sources viz.\ the dominant harmonic of GWs from the quasi-circular inspiral phase of non-spinning compact object binaries \cite{Blanchet_1996}, accounting for the redshift correction to the GW frequency and chirp mass.   

The inspiral phase of a non-spinning binary system can be modeled using post-Newtonian analysis \cite{Blanchet2014}, which provides relatively simple analytical expressions for the time-varying GW strain amplitudes corresponding to the plus and cross polarizations. The simplest of these corresponds to the quadrupole formula:

\begin{equation}
\label{hplus}
h_{+}= \frac{4}{d_L} \left(\frac{G\mathcal{M}}{c^2}\right)^\frac{5}{3} \left(\frac{\pi f}{c}\right)^\frac{2}{3} \frac{1+\cos^2{(i)}}{2} \cos{\left(2 \pi f t + \delta_0\right)},
\end{equation}

\begin{equation}
\label{hcross}
h_{\times}= \frac{4}{d_L} \left(\frac{G\mathcal{M}}{c^2}\right)^\frac{5}{3} \left(\frac{\pi f}{c}\right)^\frac{2}{3} \cos{(i)} \sin{\left(2 \pi f t + \delta_0\right)}.
\end{equation}

If $m_1$ and $m_2$ are the masses of the objects in the binary, then $\mathcal{M} =  \frac{(1+z)(m_1 m_2)^\frac{3}{5}}{(m_1 + m_2)^\frac{1}{5}}$ is the redshift-corrected chirp mass. $d_L$ is the luminosity distance of the GW source and $z$ is its redshift. For a higher computational speed, we use the approximate analytical relation between $d_{L}$ and $z$ for $\Lambda$CDM cosmology from \cite{Pen_1999}. $i$ is the inclination angle between the observer's line of sight to the GW source and the angular momentum vector of the GW source. $\delta_0$ is the initial phase of the GW at the start of the observing time, $t_0$. The redshift-corrected, time-varying GW frequency is $f = \left(1+z\right)^{-1}\left(f_{0}^{-\frac{8}{3}} - \frac{8}{3}k(t-t_0) \right)^{-\frac{3}{8}} = \left(1+z\right)^{-1}\sfrac{\sqrt{G(m_1+m_2)}}{\pi r^{\frac{3}{2}}}$, where $r$ is the separation between the objects in the binary; $f_0$ is the GW frequency at $t=t_0$, corresponding to an initial separation of $r=r_0$; and $k = \frac{96}{5} \pi ^\frac{8}{3} \left(\sfrac{G\mathcal{M}}{c^3}\right)^\frac{5}{3}$. $G$ is the gravitational constant.  

We use an approximation of the innermost stable circular orbit, $r_{\rm isco} = \frac{6G\max{(m_1,}{ m_2)}}{c^2}$, to mark the end of the inspiral phase. Upon reaching this point, or at the end of our chosen observation time (whichever comes earlier), our computations are halted.

In total, our GW source model thus has $9$ parameters.

\subsection{Storage ring model}
\label{ring_model}

We consider a hypothetical circular storage ring having a circumference of $100$~m and containing a single, rigid, ultrarelativistic particle (not necessarily ionic or subatomic) that is coasting stably at close to the speed of light (making the pre-factor in Eq.~(\ref{dtcircular2}) equal to $-1/4$), with no RF system and a single timing detector present within the ring. The timing detector is placed to the south of the storage ring's center, such that it lies on the meridian of the centre of the storage ring, i.e.\ $\phi_0 = 0\degree$. 

We assume that this storage ring has a sufficiently high vacuum to allow the particle to circulate for the entire duration of the observation run without any collisions with air molecules. We also assume that the particle has a charge-to-mass ratio which minimises the noise due to synchrotron radiation, such that it can circulate stably during the observation run without the RF system. 

\begin{figure*}[!ht]
\includegraphics[width=\linewidth, height=0.53854167\linewidth]{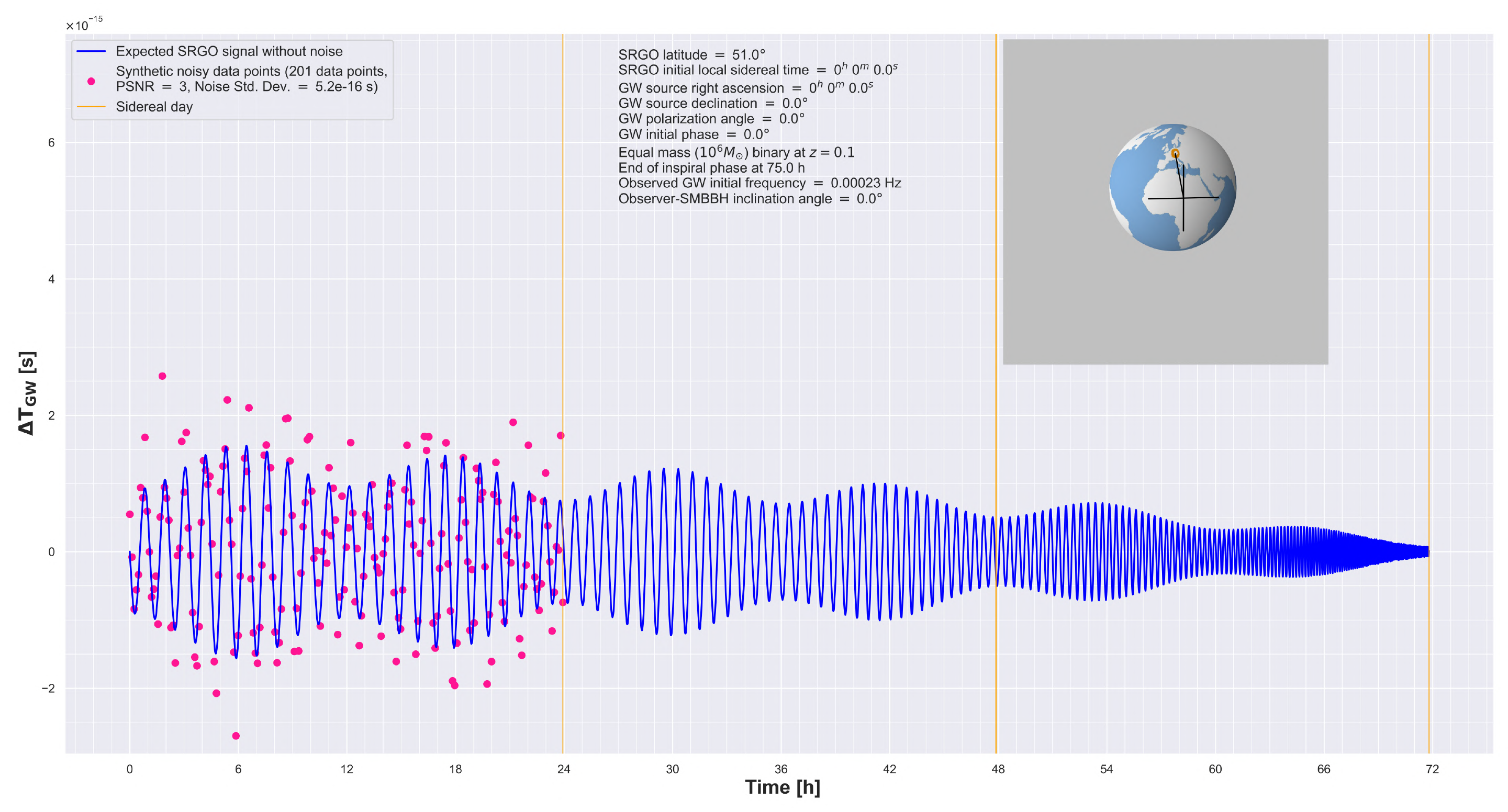}
\caption{The blue curve shows the expected response of an SRGO to a mHz GW, for the given parameters. The pink dots are a demonstration of discrete, noisy data points, created by adding artificial Gaussian noise (with a PSNR$ = 3$) to the SRGO response signal. The orange ring in the inset shows the SRGO position, while the black lines show the GW propagation direction and the $+$-polarization axes. For clarity regarding the camera angle in the inset, note that the black line along the GW propagation direction is parallel to Earth's equatorial plane, and passes through the center of the storage ring.}
\label{fig_response}
\end{figure*}

\begin{figure}[!ht]
\includegraphics[width=\linewidth]{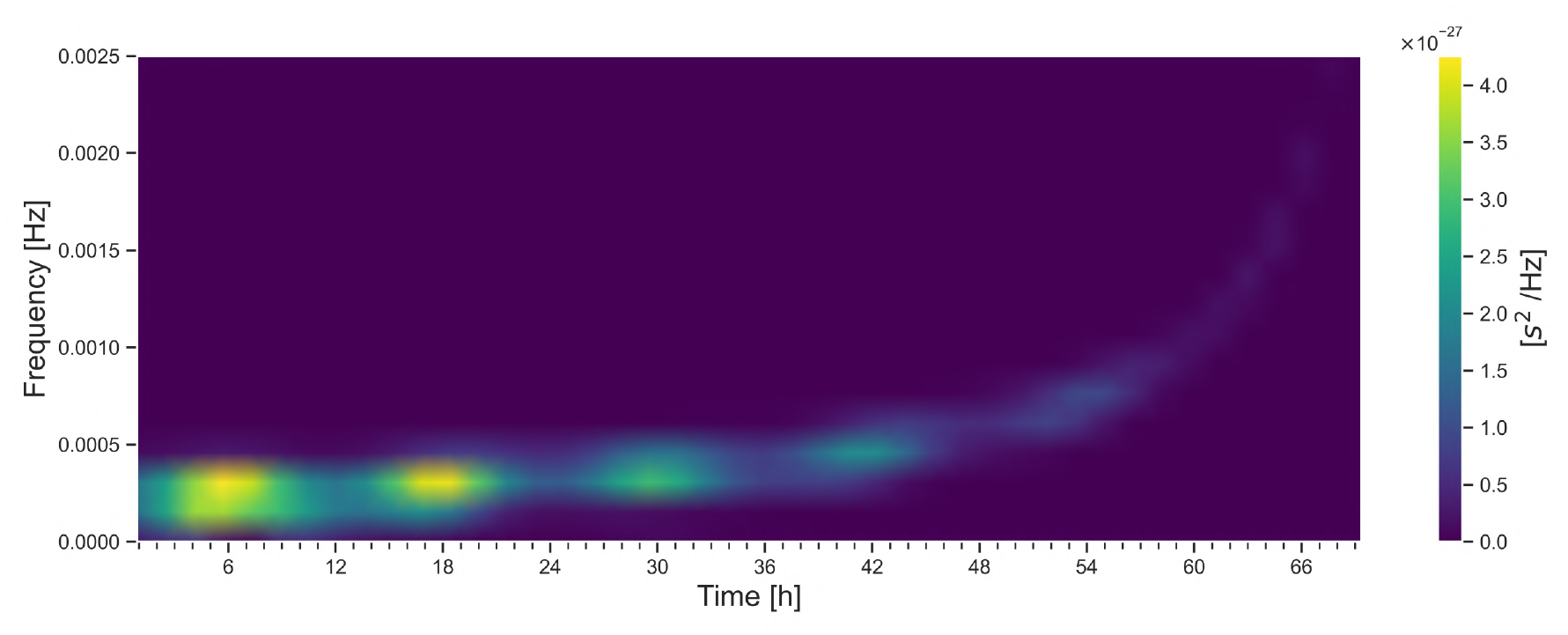}
\caption{Spectrogram of the SRGO response signal shown in Fig.~\ref{fig_response}, i.e.\ the power spectral density as a function of time (along the horizontal axis) and frequency (along the vertical axis) computed over a sliding time window.}
\label{fig_extras}
\end{figure}

According to the Nyquist-Shannon sampling theorem, the minimum sampling (data measurement) rate must be greater than twice the highest expected GW frequency, while the maximum possible sampling rate would correspond to the timing detector making one detection per revolution of the ion, i.e.\ every time it arrives at the detector. Given the parameter choices made above, our SRGO would have a maximum sampling rate of $f_{\rm sample} = 2.998$ MHz. While in reality one would prefer a sampling rate as close as possible to the maximum rate allowed by the instrument, in our simulations we choose significantly lower values of the sampling rate for the sake of computational speed.
However, we may always consider each simulated data point (with a given noise) to represent an appropriately weighted average of several `real' data points collected at a higher sampling rate and with higher noise. In our simulations, we always choose the total number of data points to be a power of two, as this allows for faster computation of the Fast Fourier Transform (FFT) done during the MCMC GW parameter estimation.

We assume that deterministic (i.e.\ systematic) sources of noise, such as gravity gradient noise, seismic activity, etc., are either technologically eliminated or accounted for by appropriately correcting the data, such that only stochastic (random) noise sources remain in the experiment. This residual noise is assumed to be Gaussian (an assumption also made by LIGO \cite{thrane_talbot_2019}), with a standard deviation, $\sigma_{\rm noise}$, between $10$ and $0.01$ times the peak GW signal. Values outside of this range of peak signal-to-noise ratio (PSNR) will give more or less trivial results, particularly when estimating the parameters of GW sources, and are hence not considered. A better noise model for the current study cannot be assumed until detailed studies of the noise sources have been carried out (Schmirander et al., in preparation).

\subsection{Numerical integration procedure}

In order to compute the response of our SRGO (Section~\ref{ring_model}) to a GW produced by a given source (Section~\ref{subsect_A}), we use Boole's quadrature rule \cite{Ubale2012} over a timestep corresponding to our chosen sampling rate to compute its contribution to the integral of Eq.~(\ref{dtcircular2}). Starting from the initial value of the integral ($= 0$), the contribution of each timestep is added to the integral, and its cumulative value is saved after each timestep. 
Thus, we numerically obtain the SRGO response signal as a time series.

\subsection{Markov Chain Monte Carlo (MCMC) fitting procedure}
\label{mcmc}

Below we will fit our GW source model (Section~\ref{subsect_A}) to a simulated SRGO response signal degraded by a given amount of noise in order to derive the relationship between the measurement noise and the precision with which the model parameters can be estimated, and to explore any degeneracies between these parameters. To this end, we will use an MCMC fitting procedure.

MCMC is a Bayesian inference tool for numerically obtaining the joint posterior probability distribution of unknown model parameters by directly drawing samples from the posterior. MCMC finds and explores the regions in parameter space that correspond to the maximum likelihood of the model being a fit for the given data and a prior probability distribution of the fitting parameters \cite{vanRavenzwaaij2016, 1909.12313, Luengo2020}.

We prefer MCMC methods over the conventional matched filtering algorithm \cite{Sathyaprakash2009} for thorough and efficient GW parameter estimation, because our GW model contains $9$ free parameters, and making a grid in a parameter space of such high dimensionality would require an unfeasibly long computation time. In general, we indeed fit for all $9$ model paramaters: the GW source masses $m_1, m_2$; the initial separation between the masses, $r_0$; the GW source inclination angle, $i$; the GW source redshift, $z$; the initial phase of the GW, $\delta_0$; the right ascension, $\alpha_{\rm src}$ and declination, $\delta_{\rm src}$, of the GW source; and the polarization angle, $\psi_{\rm eq}$ of the GW. The MCMC fitting is implemented using the PyMC3 python module \cite{Salvatier2016}.

We first create synthetic noisy data points by adding Gaussian noise to the SRGO response computed for a given set of parameters. The noisy data are then transformed to Fourier space and passed to the likelihood function.

We use flat priors for the unknown fitting parameters and a two-dimensional Gaussian noise `Whittle' likelihood function, as also done by LIGO for GW parameter estimation \cite{thrane_talbot_2019}. The priors corresponding to angular parameters are bounded between $-180 \degree$ and $180 \degree$, whereas priors corresponding to non-angular parameters are bounded between zero and twice their true parameter values for computational efficiency. 

We employ the Differential Evolution Markov Chain (DE-MC) algorithm \cite{Braak2006} for the MCMC chains, and run, per data set, 1000 parallel chains which draw 1250 samples each. Since we are purely interested in parameter estimation, and not in showing the convergence of chains to the region of maximum likelihood, we do not discard $25\%$ of the initial traces as the `burn-in' phase. Instead, we allow the chains to start from the true parameter values and then explore the likelihood around these locations. 

For a given set of parameters, we generate and fit not just $1$ but $10$ noisy data sets. We discard those fitting results where the true parameter values do not lie within the $3$-$\sigma$ ($99.7\%$) highest posterior density (HPD) region. The multiple joint posteriors provide us with the statistical variation of the joint posterior for each set of parameters. This is used to produce error bars on the sizes of the joint posteriors (`errors on the errors') plotted as a function of different controllable experiment parameters, as discussed in Section~\ref{sect_param_estim}.

\section{Results: SRGO response analysis}
\label{response_analysis}

In this section we discuss the general characteristics of the response signal of an SRGO (Section~\ref{signal_analysis}), its sensitivity curve (Section~\ref{subsec_sensi}), and the range over which various astrophysical sources of mHz GWs could potentially be observed (Section~\ref{obs_range}).

\subsection{Response signal analysis} 
\label{signal_analysis}

In Fig.~\ref{fig_response}, we show the response of an SRGO to a mHz GW from a supermassive binary black hole (SMBBH) inspiral, for an arbitrarily chosen set of parameters as indicated in the figure's legend. 

First, we note that the signal is oscillatory with a frequency equal to that of the GW, which increases with time due to the inspiral of the binary. Second, we find that the amplitude of the signal decreases with time, which, as we shall see below, is due to the increasing GW frequency. Therefore, unlike LIGO, the chirping of the GW strain amplitude is not reflected in the SRGO's response spectrogram, as shown in Fig.~\ref{fig_extras}. Finally, the overall decrease of the amplitude is modulated by an envelope which is due to the source's path across the SRGO's antenna pattern (the $F_+$ and $F_\times$ terms in Eq.~\ref{dtcircular2}) as the Earth rotates (cf.\ Fig.~\ref{fig_angles}).


\begin{figure*}
     \centering
     \begin{subfigure}{0.32\linewidth}
         \centering
         \includegraphics[width=\linewidth, keepaspectratio]{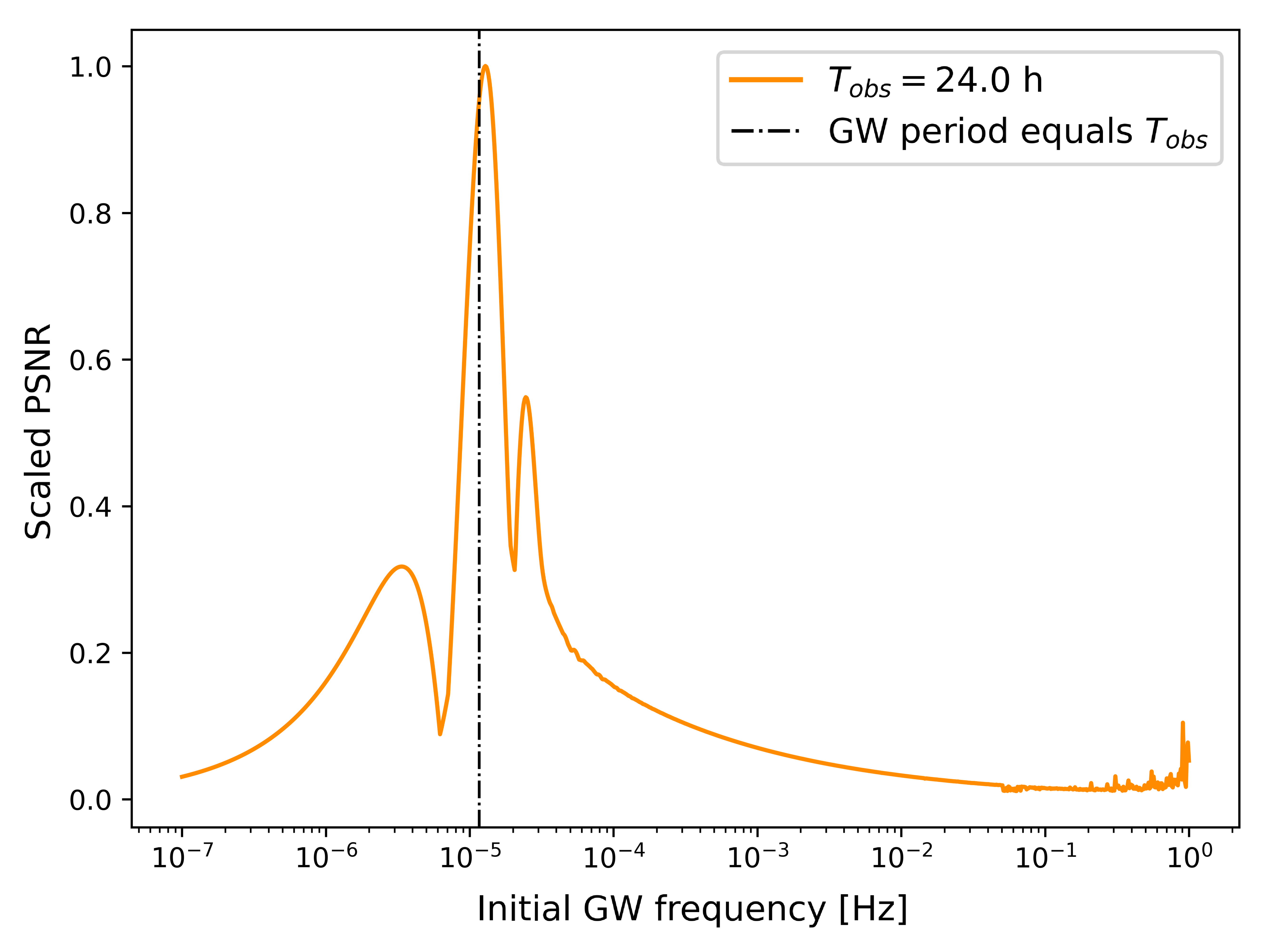}
         \caption{$T_{\rm obs}=24$ h}
         \label{subfigaa}
     \end{subfigure}
     \begin{subfigure}{0.32\linewidth}
         \centering
         \includegraphics[width=\linewidth, keepaspectratio]{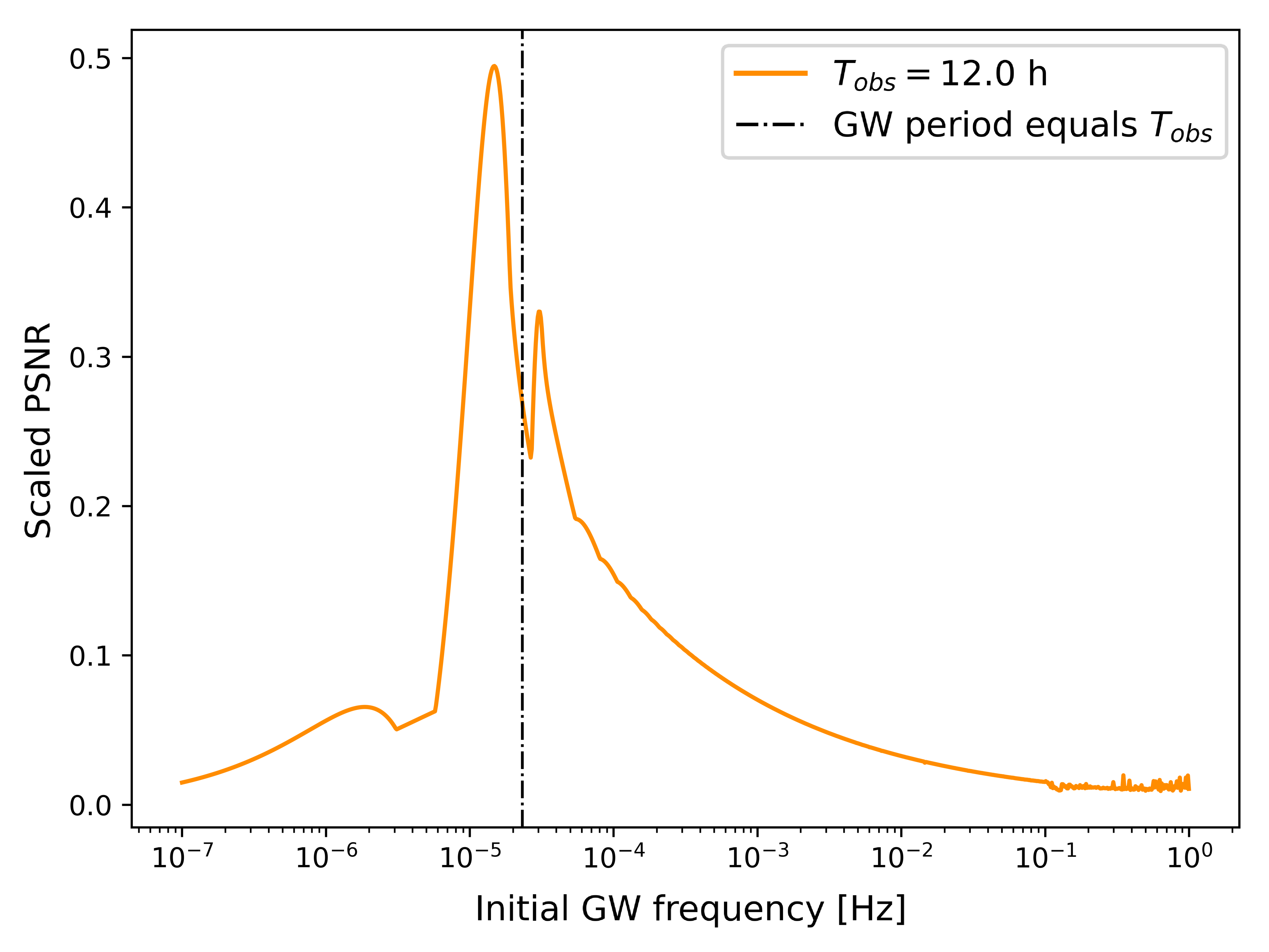}
         \caption{$T_{\rm obs}=12$ h}
         \label{subfigbb}
     \end{subfigure}
     \begin{subfigure}{0.32\linewidth}
         \centering
         \includegraphics[width=\linewidth, keepaspectratio]{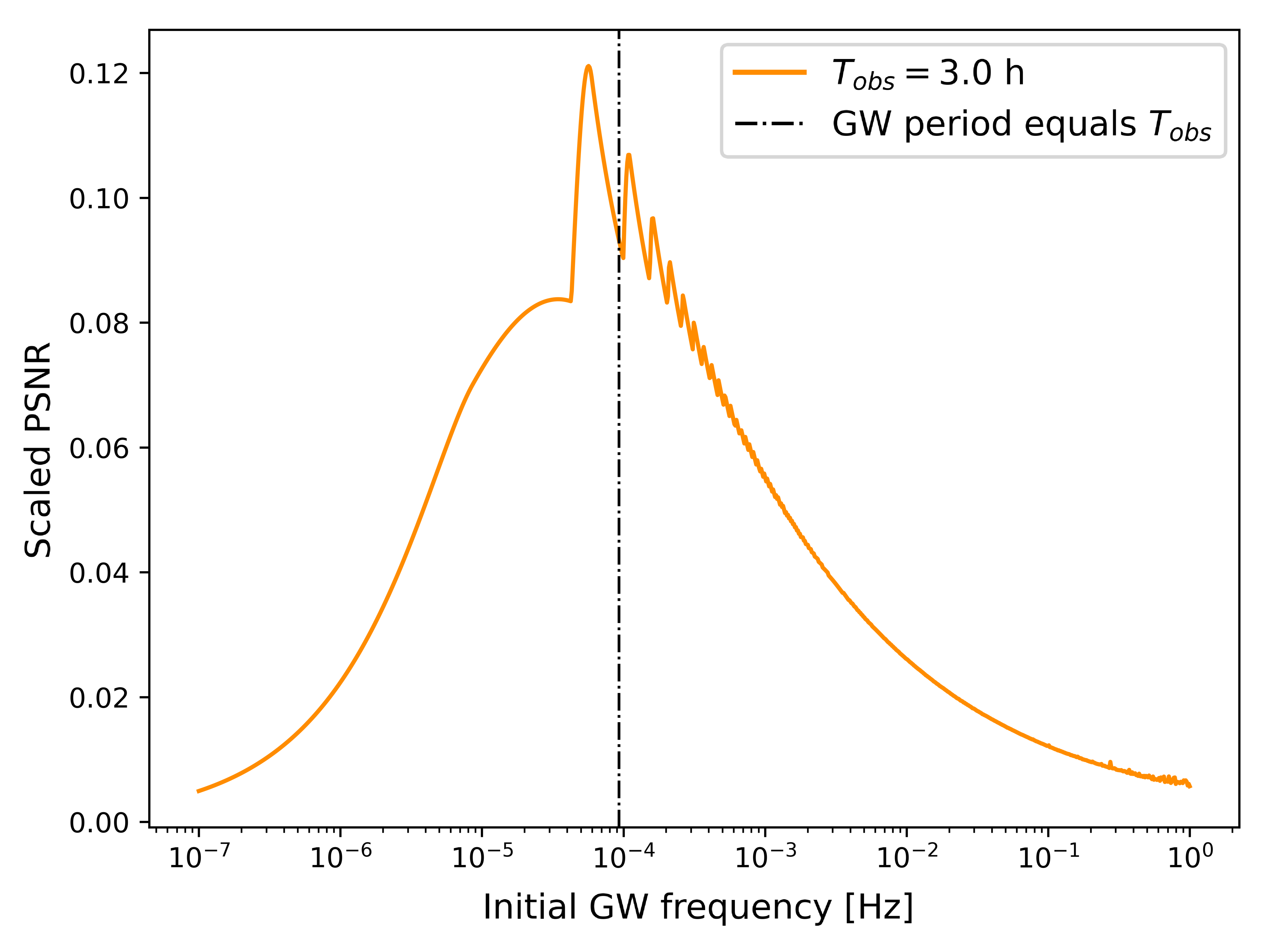}
         \caption{$T_{\rm obs}=3$ h}
         \label{subfigcc}
     \end{subfigure}
     \begin{subfigure}{0.32\linewidth}
         \centering
         \includegraphics[width=\linewidth, keepaspectratio]{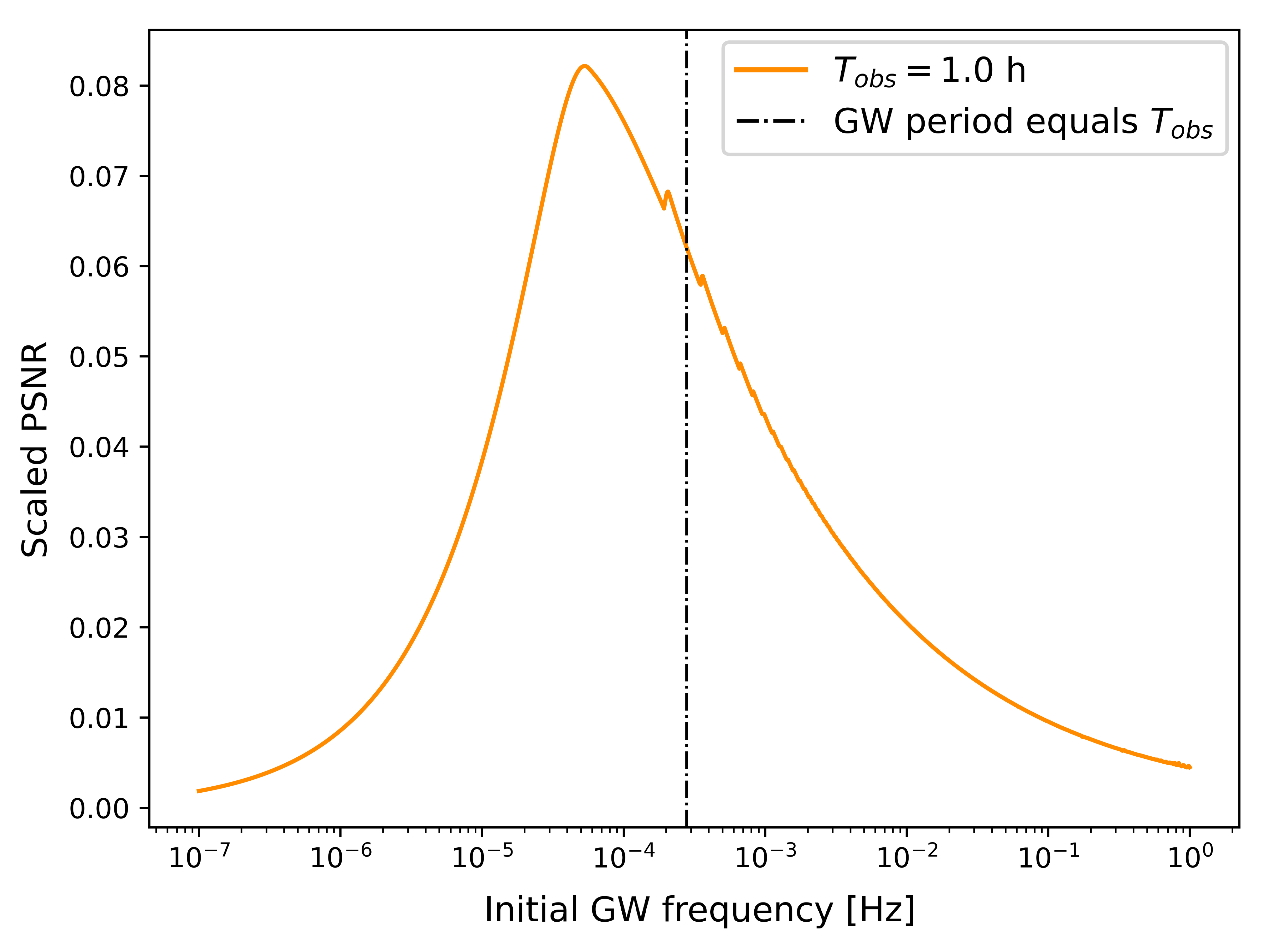}
         \caption{$T_{\rm obs}=1$ h}
         \label{subfigdd}
     \end{subfigure}
     \begin{subfigure}{0.32\linewidth}
         \centering
         \includegraphics[width=\linewidth, keepaspectratio]{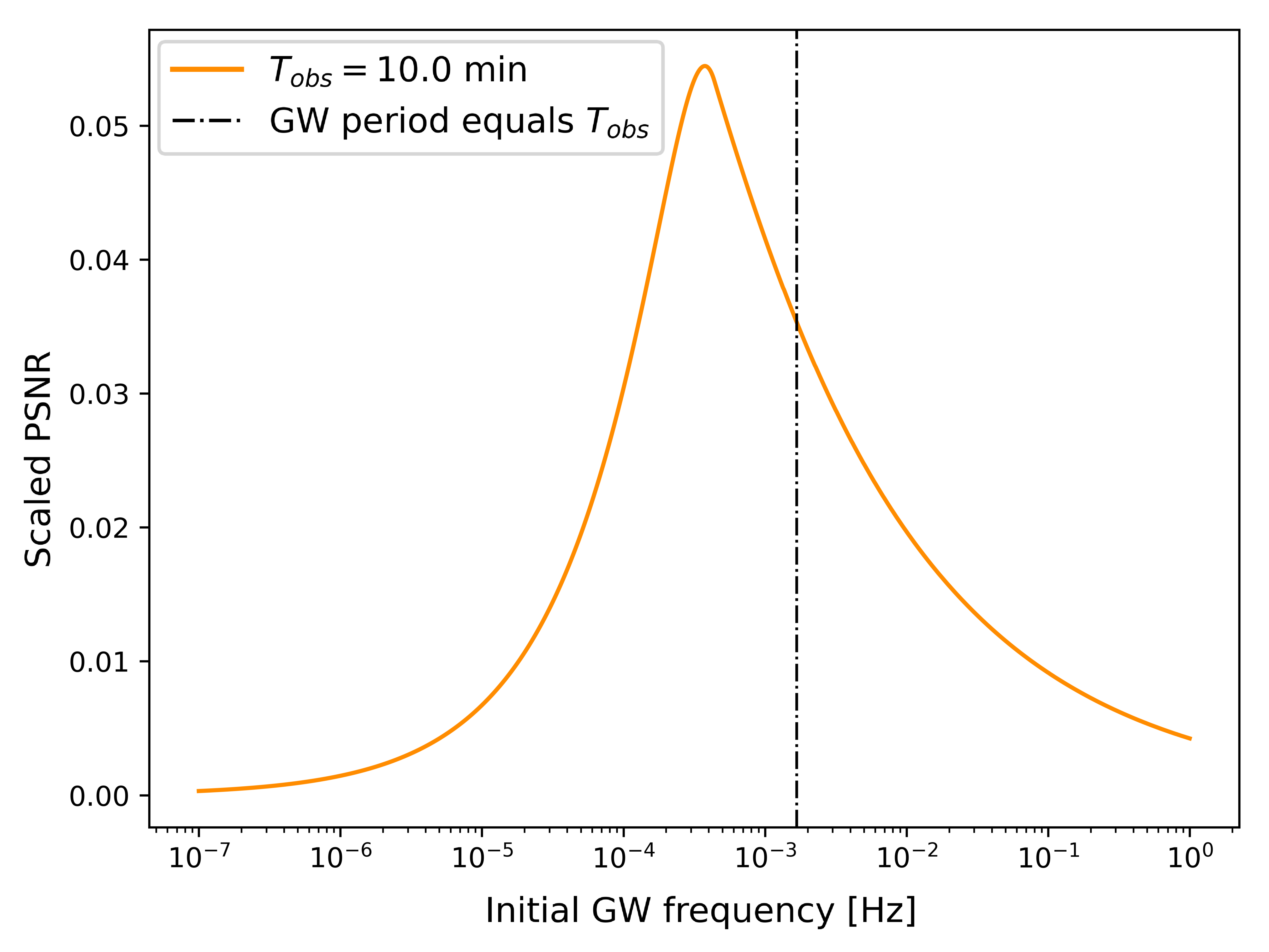}
         \caption{$T_{\rm obs}=10$ min}
         \label{subfigee}
     \end{subfigure}
     \begin{subfigure}{0.32\linewidth}
         \centering
         \includegraphics[width=\linewidth, keepaspectratio]{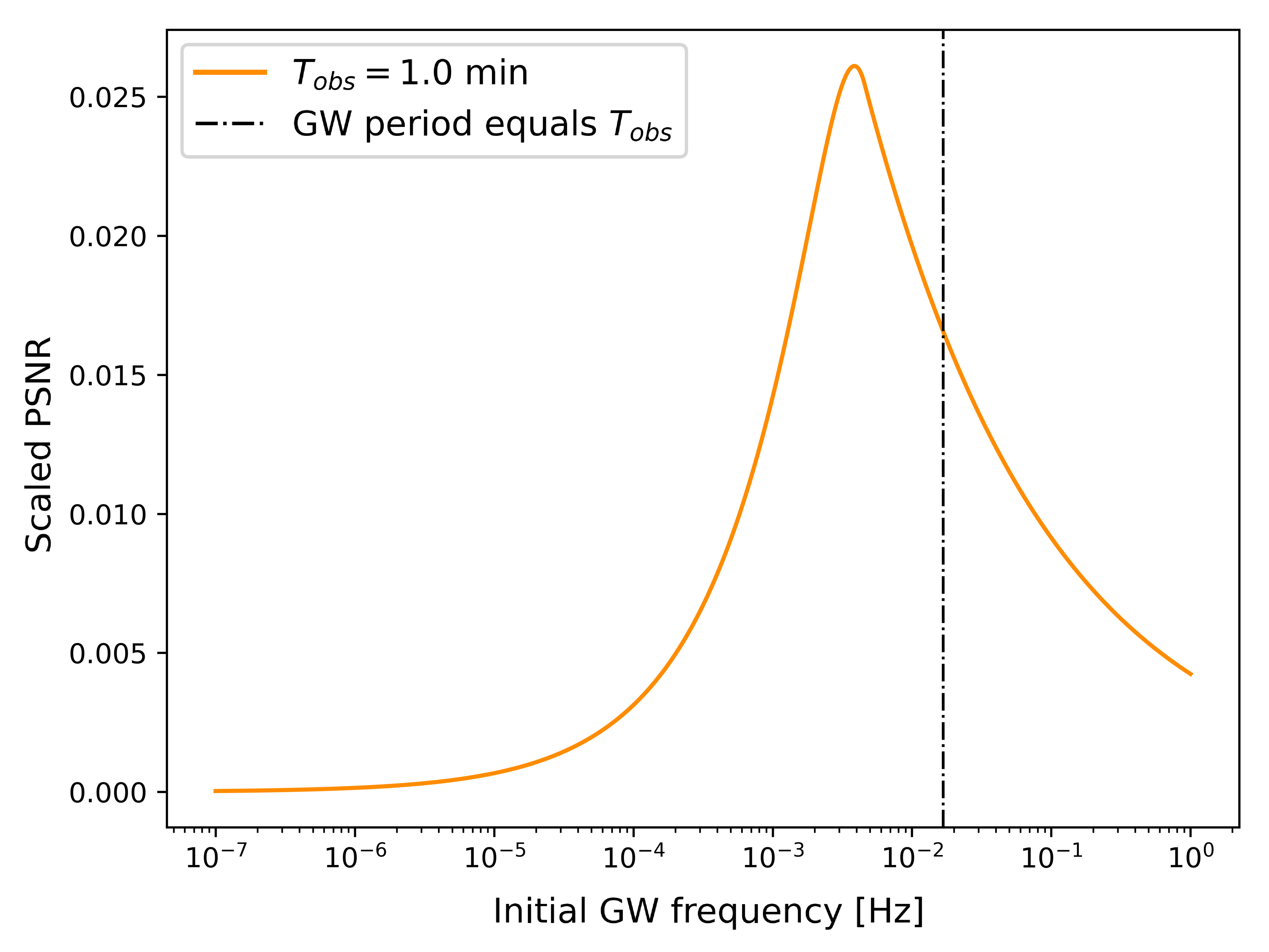}
         \caption{$T_{\rm obs}=1$ min}
         \label{subfigff}
     \end{subfigure}     
        \caption{Scaled values of the peak signal amplitude attained within $T_{\rm obs}$ as a function of the initial GW frequency (which is set by the initial separation of the binary), for different values of the observation time, $T_{\rm obs}$. All other parameters are equal to those shown in Fig.~\ref{fig_response}, except the black hole masses, which have been set to $3$~$M_\odot$ each, so that even for the highest initial GW frequency, the inspiral phase does not end before the the end total observation time. Note that, due to the scaling, the shape of the plots are independent of the binary mass. The scaling is done such that the largest peak among all plots has a value of unity.}
        \label{psnr_frq}
\end{figure*}

We now turn to a more detailed discussion of the dependence of the response signal amplitude on the GW frequency. In Fig.~\ref{psnr_frq} we plot a scaled version of the peak response signal amplitude obtained within a given observation time, $T_{\rm obs}$, as a function of the initial GW frequency, $f$, for different values of $T_{\rm obs}$.


We immediately notice that in all cases the curves peak near $f\,T_{\rm obs} \approx 1$, and decrease towards both higher and smaller frequencies.
We can understand this behaviour by analysing Eq.~\ref{dtcircular2}. For better clarity, let us temporarily ignore the contribution due to the envelope, contained in the antenna pattern terms $F_{+}$ and $F_{\times}$ (which corresponds to the situation in Fig.~\ref{subfigff}, where the observation time is much smaller than the Earth's rotation period). The response signal is basically a time integral of a periodic function that is the GW strain, and the value of this integral is maximised when integrating over some fraction of the GW period (depending on the initial phase), i.e.\ in the regime $f\,T_{\rm obs} \approx 1$. In this regime, if the GW period (or equivalently, the total observation time), is reduced, and even if the amplitude of the GW strain were to increase, the total area under the strain curve would actually decrease. The same reason also explains why, in the regime $f\,T_{\rm obs} \gg 1$, the peak signal decreases with increasing GW frequency. Note that in this regime, the peak signal amplitude would occur somewhere within the first GW period, ignoring the effect of the envelope.

The dynamically changing peaks and valleys in the series of plots can be somewhat elucidated: they are the result of the interplay between the GW period, initial GW phase, the orientation of SRGO relative to the GW source, Earth's rotation period and the observation period. The orientation of SRGO relative to the GW source and Earth's rotation period together determine the envelope seen in the response signal of Fig.~\ref{fig_response}. During the observation time, large peaks in these plots occur when peaks in the GW strain waveform align with the envelope peaks, and valleys occur when the GW strain peaks align with the envelope valleys (or vice versa, whichever produces a greater response signal). For a given observation time, as the initial GW frequency is increased, a number of successive GW strain peaks and valleys occur during the observation period, which may or may not align with the envelope peaks and valleys. An aligned GW strain peak (resulting in a local maxima in the plot curves), upon increasing the initial GW frequency, will become misaligned, resulting in a drop in the curve until the next peak starts becoming aligned, causing then a rise in the curve. This succeeding local maxima may be higher or lower than its predecessor, depending on the location of the global maxima, which occurs when the initial GW period is close to the observation time. This explains the spiky sections of the plot curves, prominently seen in Fig.~\ref{subfigcc}, but also present in the other plots. 

In Figs.~\ref{subfigaa}, \ref{subfigbb} and \ref{subfigcc}, the noisy regions in the right side tails of the plots have purely numerical origins, being caused by the abrupt halting of the code one timestep after the inspiral phase has been crossed. Therefore, they exist only at the ends of the right side tails (corresponding to high initial GW frequencies, meaning that the binary compact objects begin close to the end of their inspiral phases). As the observation time is reduced, the end of the inspiral phase would be reached within the observation time at higher initial frequencies. Therefore, the noisy regions shift towards higher frequencies and eventually disappear in Figs.~\ref{subfigdd}, \ref{subfigee} and \ref{subfigff}.

\begin{figure*}
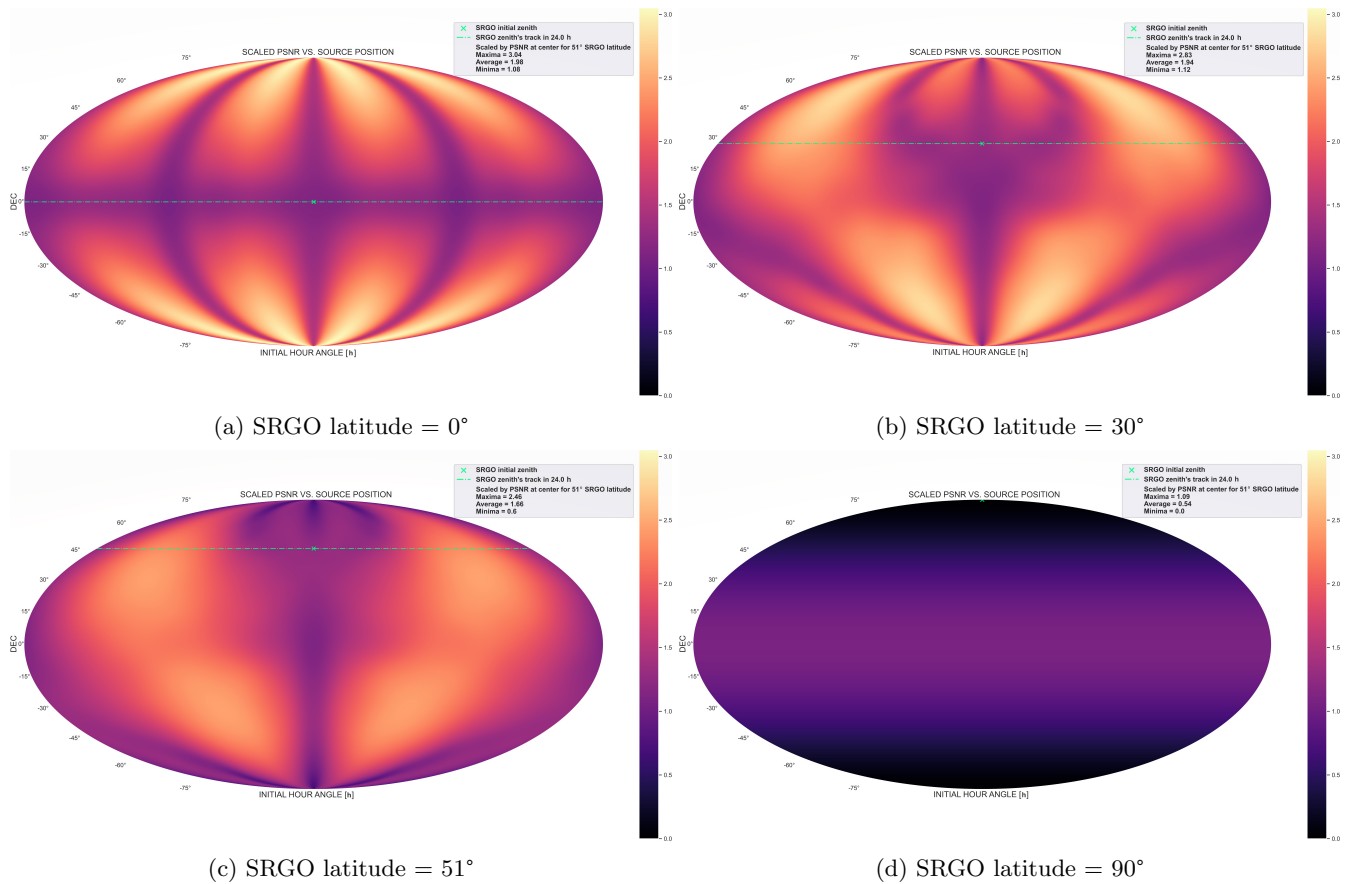

     \centering
     \begin{subfigure}{0.49\linewidth}
         \centering
         \includegraphics[width=\linewidth, keepaspectratio]{psnr_vs_srcpos1.pdf}
         \caption{SRGO latitude = $0 \degree$}
         \label{subfigA}
     \end{subfigure}
     \begin{subfigure}{0.49\linewidth}
         \centering
         \includegraphics[width=\linewidth, keepaspectratio]{psnr_vs_srcpos2.pdf}
         \caption{SRGO latitude = $30 \degree$}
         \label{subfigB}
     \end{subfigure}
     \begin{subfigure}{0.49\linewidth}
         \centering
         \includegraphics[width=\linewidth, keepaspectratio]{psnr_vs_srcpos3.pdf}
         \caption{SRGO latitude = $51 \degree$}
         \label{subfigC}
     \end{subfigure}
     \begin{subfigure}{0.49\linewidth}
         \centering
         \includegraphics[width=\linewidth, keepaspectratio]{psnr_vs_srcpos4.pdf}
         \caption{SRGO latitude = $90 \degree$}
         \label{subfigD}
     \end{subfigure}
        \caption{Scaled values of the peak signal-to-noise ratio as a function of the GW source's initial position in the sky relative to SRGO. All other parameters are equal to those shown in Fig.~\ref{fig_response}, except the initial separation between the masses which is 1 AU (corresponding to an initial GW period of $3.1$~h), and the observation time which is 1 day. The top-left, top-right, bottom-left and bottom-right figures respectively correspond to an SRGO latitude of $0 \degree$, $30 \degree$, $51 \degree$ and $90 \degree$. The scaling is done relative to the case corresponding to the center of the bottom-left plot.}
        \label{psnr_srcpos}
\end{figure*}

In Fig.~\ref{psnr_srcpos}, we plot the PSNR as a function of the source position in the sky, scaled such that the arbitrary case corresponding to Fig.~\ref{fig_response} has a PSNR of unity. While changing the source position, all other parameters remain the same as those in Fig.~\ref{fig_response}, but the initial SMBBH separation is 1 AU (corresponding to an initial GW period of $3.1$~h), and signal is computed over an observation time of 1 day. We repeat this for different latitudes of the SRGO location on Earth. 

We observe that the complex patterns in the plots are a modified projection of the SRGO antenna pattern on the sky, as expected from Eq.~\ref{dtcircular2}. Other parameters being fixed, for a given SRGO latitude and source declination, the horizontal variation is determined by the initial hour angle, the GW frequency, initial GW phase, GW polarization angle and Earth's rotation frequency. Maxima in the horizontal variation occur during the observation time when peaks in the GW strain waveform occur exactly at a time when there are peaks in the response signal envelope. Minima occur when the GW strain peaks align with the envelope valleys (or vice versa, whichever produces a greater response signal). We note that the pattern depends on the initial hour angle i.e. the right ascension of the GW source relative to the initial SRGO local sidereal time (and not on their absolute values). However, the same is not true between the SRGO latitude and source declination. That symmetry is broken by the Earth's spin axis, and therefore, the SRGO latitude variation produces different patterns in the plots. The bottom-right plot corresponds to SRGO being placed at the pole. For this case, if the GW source is also at one of the poles, then the orientation always remains face-on, and no response signal is produced. For all other cases, a non-zero response signal is produced over 24 hours, due to a changing orientation of the GW source relative to the ring.

Further, we observe that the plots show antipodal symmetry, since placing the ring and/or the GW source at antipodal positions, and/or having the ions circulating in the opposite direction, would all produce the same SRGO response. The plots also appear to have a left-right symmetry, because two GW sources at the same declination, initially located on either side of the ring and having the same hour angle magnitude, would produce the same value of the initial SRGO response. The response signals over a 24~h period would, however, not be exactly the same, breaking the symmetry. Although both sources would trace the same path across the antenna pattern, the phase difference between the GW and the path across the antenna pattern would be different, so that the peak signal would be attained at different times and at slightly different values. In addition, the symmetry is further broken by the time evolution of the GW frequency due to the inspiral, which in turn changes the amplitude of the response signal, as discussed above. This effect is particularly relevant when the rate of change of the GW frequency is large, which is not the case in the example shown in Fig.~\ref{psnr_srcpos}. Finally, there is also a latitudinal symmetry, as having SRGO located at equal and opposite latitudes would simply flip the plots about the horizontal axis. 

The SRGO latitude variation indicates that placing the ring near the equatorial latitudes on Earth would be more advantageous than placing the ring near the polar latitudes, offering an increase of the maximum PSNR by around 3 times, of the average PSNR by around 4 times, and of the minimum PSNR by a factor of unity from zero, between the polar and equatorial SRGO latitudes.

\subsection{SRGO sensitivity curve}
\label{subsec_sensi}

\begin{figure}[!ht]
\includegraphics[width=\linewidth, keepaspectratio]{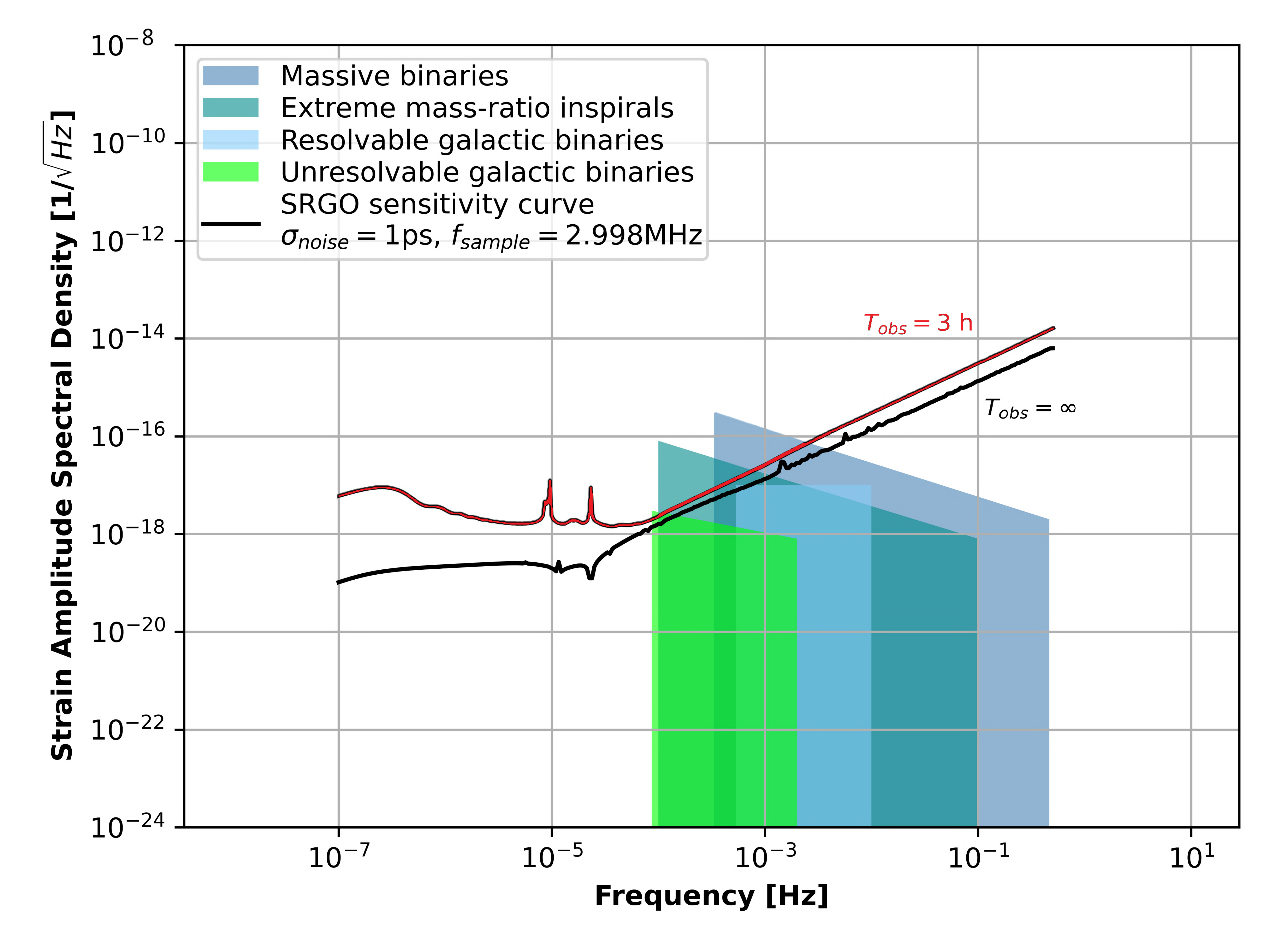}
\caption{The numerically computed sensitivity curve of an SRGO for the given parameter values, and averaged over all other parameter values. The red curve corresponds to a finite observation time of 3 h, while the black curve is the standard amplitude spectral density corresponding to an infinite observation time.}
\label{fig_sensi}
\end{figure}

The sensitivity curve of a GW detector may be defined as the curve in the plane of the strain amplitude spectral density versus frequency, where the signal-to-noise ratio is unity.

Conventionally, a GW sensitivity curve plot contains two elements: First, colored regions are shown corresponding to various astrophysical GW sources. These regions are computed solely from GW strain models of the sources, and their extent is determined by astronomical constraints applied to the model parameters. Second, a curve is overplotted over these regions, which is constructed by modeling the experimentally measured detector noise data as a strain, using the response model of the GW detector. Since this plot is in the frequency space, which is integrated over all time, the ratio of a point in the colored region to a point on the curve, at a given frequency, directly gives the integrated signal-to-noise ratio at that frequency. This allows us to directly read off the signal-to-noise ratio by eye from such a plot. However, for computing the sensitivity curve, an averaging over the antenna pattern of the GW detector is involved. For LIGO and other high frequency GW detectors, the averaging of the antenna pattern is time-independent and trivial, because the antenna pattern is practically stationary over the duration of the GW. For millihertz GWs, which is our region of interest for an SRGO, the antenna pattern is not stationary due to Earth's rotation, making such an averaging time-dependent and difficult.

Therefore, in our study, we plot the sensitivity curve in a different way. The colored regions in our plot are the same as in every conventional GW sensitivity plot, and taken from \cite{Moore_2015}. But the curve itself is not made by modeling the experimentally measured (or artificially generated) noise data as a strain. Instead, we compute the source GW strain that would be required to make a detection using SRGO, for chosen values of controllable experiment parameters such as the stochastic noise level, data sampling rate and total observation time. Hence, our sensitivity plot compares the expected GW strain (the colored regions) to the GW strain required to make a detection (the curve itself), as a function of GW frequency. Note that here too, the integrated signal-to-noise ratio can be read by eye from the plot, based on the definition of the sensitivity curve mentioned earlier.

In this study, we use our code to numerically compute the SRGO sensitivity curve. We start by equating the root mean square value of the response signal (taken over the total observation period) to the effective noise,

\begin{equation}
\label{sensi}
    \left[\Delta T_{\rm GW}\right]_{\rm rms} = \frac{\sigma_{\rm noise}}{\sqrt{f_{\rm sample} T_{\rm obs}}}, 
\end{equation}

where $T_{\rm obs}$ is the total observation time and $f_{\rm sample} = \frac{n_p v_0}{2\pi R}$ is the data sampling rate, with $n_p$ being the number of circulating test masses that are timed, which may be individual ions or bunches of ions. $\sigma_{\rm noise}$ can be the standard deviation of any stochastic noise, even the net stochastic noise in the system. In our study, we assume that the residual noise in the system is Gaussian, as mentioned in Sec.~\ref{ring_model}. Note that, although the condition Eq.~(\ref{sensi}) appears unconventional, but upon transposing $\sqrt{f_{\rm sample} T_{\rm obs}}$ over to the left, it is clear that this is a valid way of imposing the fundamental criterion for GW signal detectability, where some measure of the \textbf{integrated} signal, such as the root mean square of the signal multiplied by the square root of the observation time, should be equal to a measure of the noise level, such as $\sigma_{\rm noise}$. 

We then expand the left hand side using Eq.~(\ref{dtcircular2}), substituting $h_{+}$ and $h_{\times}$ from Eq.~(\ref{hplus}) and Eq.~(\ref{hcross}), respectively. Then we set $k=0$ in the GW frequency evolution to get continuous GWs (i.e. GWs with a constant frequency). Finally, we club together all the common time-independent terms within the integral, and call this quantity $h_0$. Hence, we have defined,
\begin{equation}
    h_0 = \frac{4}{d_L} \left(\frac{G\mathcal{M}}{c^2}\right)^\frac{5}{3} \left(\frac{\pi f}{c}\right)^\frac{2}{3}. 
\end{equation}

Rearranging Eq.~(\ref{sensi}), we thus obtain,
\begin{multline}
\label{sensi1}
    h_0 = \frac{4\sigma_{\rm noise}}{\sqrt{f_{\rm sample} T_{\rm obs}}} . \\
    \Biggl[ \int_{t_{0}}^{t_{0}+T}\biggl(F_{+}. \frac{1+\cos^2{(i)}}{2} \cos{\left(2 \pi f t + \delta_0\right)} \\
    + F_{\times}. \cos{(i)} \sin{\left(2 \pi f t + \delta_0\right)}\biggr)dt \Biggr]^{-1}_{\rm rms}.
\end{multline}

Hence, the quantity $h_0$ must satisfy the condition Eq.~(\ref{sensi1}) on the sensitivity curve. Now, we calculate and express the GW strain amplitude spectral density (ASD) in terms of $h_0$. It is defined as,

\begin{equation}
    ASD = \lim_{T_{\rm obs}\to\infty} \frac{|\tilde{h}(f)|}{\sqrt{T_{\rm obs}}}, 
\end{equation}

where the Fourier Transform of the GW strain, $\tilde{h}(f)$, can be calculated in terms of $h_0$ over the finite duration $T_{\rm obs}$, by first rewriting Eq.~(\ref{hplus}) as,

\begin{equation}
    h_{+} (t) = h_0 \frac{1+\cos^2{(i)}}{2} \cos{\left(2 \pi f t + \delta_0\right)}.
\end{equation}

Upon performing the calculations, we find,

\begin{multline}
\label{sensi2}
        ASD = \lim_{T_{\rm obs}\to\infty} \frac{h_0}{2} \frac{1+\cos^2{(i)}}{2} \sqrt{T_{\rm obs}} \\ 
        \sqrt{1 + \sinc^2{(2 \pi f T_{\rm obs})} + 2\sinc{(2 \pi f T_{\rm obs})}. \cos{(2\delta_0)} }.
\end{multline}

Finally, we substitute $h_0$ from Eq.~(\ref{sensi1}) and then apply the limit in Eq.~(\ref{sensi2}) to get,

\begin{multline}
\label{sensi3}
        ASD_{\infty} = \frac{2\sigma_{\rm noise}}{\sqrt{f_{\rm sample}}} \frac{1+\cos^2{(i)}}{2} \\        
        \Biggl[ \int_{t_{0}}^{t_{0}+T}\biggl(F_{+}. \frac{1+\cos^2{(i)}}{2} \cos{\left(2 \pi f t + \delta_0\right)} \\
        + F_{\times}. \cos{(i)} \sin{\left(2 \pi f t + \delta_0\right)}\biggr)dt \Biggr]^{-1}_{\rm rms}.
\end{multline}

Note that the root mean square in Eq.~(\ref{sensi3}) is over an infinite duration. But due to the periodicity of the GW signal, we can compute the root mean square over the lowest common multiple of the GW period and Earth's rotation period. For a finite observation time, we can skip the limit in Eq.~(\ref{sensi2}) to get,

\begin{multline}
\label{sensi4}
        ASD_{<\infty} = \frac{2\sigma_{\rm noise}}{\sqrt{f_{\rm sample}}} \frac{1+\cos^2{(i)}}{2} \\        
        \Biggl[ \int_{t_{0}}^{t_{0}+T}\biggl(F_{+}. \frac{1+\cos^2{(i)}}{2} \cos{\left(2 \pi f t + \delta_0\right)} \\
        + F_{\times}. \cos{(i)} \sin{\left(2 \pi f t + \delta_0\right)}\biggr)dt \Biggr]^{-1}_{\rm rms} \\ 
        \sqrt{1 + \sinc^2{(2 \pi f T_{\rm obs})} + 2\sinc{(2 \pi f T_{\rm obs})}. \cos{(2\delta_0)} }.
\end{multline} 

The GW strain amplitude spectral density is plotted as a function of the GW frequency, $f$, and it is numerically averaged over all the GW source parameters present in Eq.~(\ref{sensi4}). Note that, we would have obtained similar analytical results had we used $h_{\times}$ instead of $h_{+}$ in the above derivation, which would have lead to the same numerical result after averaging over the GW source parameters. The result shown in Fig.~\ref{fig_sensi} numerically confirms that an SRGO would be sensitive to mHz GWs by design. In the log scale, the SRGO sensitivity curve has a linear behaviour in the mHz frequency regime, as analytically predicted in paper~I \footnote{The `LHC-GW' sensitivity curve shown in Fig.~1 of paper~I was not a conventional one, since it was not normalized for a fixed observation time. It could not be directly compared with the sensitivity curves of other GW detectors, and was therefore shown in a separate plot. In this study, since we plot the strain amplitude spectral density, this issue is resolved.}. The sensitivity deteriorates at higher frequencies. This is because, as the GW period becomes smaller, the SRGO response amplitude also decreases, since the ions spend lesser time accumulating a timing deviation during every half-cycle of the GW. This aspect is discussed in the previous section and shown in Fig.~\ref{psnr_frq}. Therefore, a larger strain amplitude would be required to detect high frequency GWs. This would greatly exceed the predicted strain amplitudes from astrophysical sources in the decihertz or kilohertz ranges (i.e. for ``LIGO-like" sources).  

At very low frequencies, for a finite observation time (red curve in Fig.~\ref{fig_sensi}), the sensitivity curve rises. We may perform a thought experiment to analyse this situation: a zero frequency GW would be equivalent to an anisotropic spacetime having a constant distortion, and not necessarily a flat spacetime. In such a spacetime, if the instantaneous initial speed of the circulating test mass is measured and used to predict the expected future arrival times of the test mass at the timing detector, then in general the observed arrival times would deviate from the prediction, as the test mass traverses an anisotropic spacetime. This is why, if we input $f=0$ in Eq.~(\ref{dtcircular2}), we still get a finite value of the response signal. Therefore, unlike laser interferometers and atom interferometers (which use test masses that can only move linearly) which require both the temporal and spatial components of GW spacetime in order to probe it, an SRGO (which utilizes circulating test masses) would, in principle, be able to probe purely the spatial anisotropy of GW spacetime even at very low GW frequencies and finite observation times. However, for low frequency GWs, as the GW period far exceeds the total observation time, $T_{\rm obs}$, the peak response signal value would be comparatively small, as discussed in the previous section and shown in Fig.~\ref{psnr_frq}. That is why the SRGO sensitivity curve rises again at low frequencies. Also, for most resolvable astrophysical GW sources, the strain amplitude at near-zero frequencies would be near-zero. Further, we notice oscillations in the red curve at low frequencies. These correspond to the term containing sinc functions in Eq.~\ref{sensi4} which appears as a result of Fourier transforming the GW strain over a finite observation time.

However, for an infinite observation time (black curve in Fig.~\ref{fig_sensi}), in principle, no matter how small the GW frequency, one can observe several GW wavelengths. Since this regime corresponds to $f \, T_{obs} \gg 1$ as shown in Fig.~\ref{psnr_frq}, the signal would increase with decreasing frequency. This causes the black curve in Fig.~\ref{fig_sensi} to continue to decline at low frequencies, unlike the red curve.

In both curves of Fig.~\ref{fig_sensi}, we notice two significant spikes at GW frequencies corresponding roughly to a period of a full day and a half day. This suggests that when the GW period is equal to a full or half of Earth's rotation period, for certain parameter configurations, the signal amplitude is greatly boosted or reduced relative to the average. Therefore, while averaging over parameters to compute the sensitivity curve, such special configurations act as outliers and produce a spike in the average curve at certain frequencies. Moreover, in the black curve, which corresponds to an infinite observation time, we note that the curve changes its slope at around the region where the GW period equals half a day and one day, although the curve still continues to fall gradually with decreasing frequency. This is because, at this point, unlike before, the GW frequency determines the signal envelope instead of Earth's rotation frequency, thus changing the properties of the signal.

The red sensitivity curve of Fig.~\ref{fig_sensi} should corroborate with the curve in Fig.~\ref{subfigcc}, as both are computed for an observation time of 3 hours and we expect them to be inversely related. Although the general shape of the two curves agree with each other, their details are different, since the sensitivity curve has been computed by averaging over several parameters, whereas Fig.~\ref{subfigcc} corresponds to a fixed set of parameters. The GW frequency of the sensitivity curve minima matches the GW frequency of the maxima in Fig.~\ref{subfigcc}. In general, combining the insights from Figs.~\ref{psnr_frq} and \ref{fig_sensi}, we deduce that the minima of the sensitivity curve would occur at a GW frequency close to the inverse of the observation time, and that this minimum would be smaller for longer observation times. Based on the predicted astrophysical mHz GW sources, we can conclude that the minimum observation time for an SRGO experiment to be maximally sensitive to the entire mHz GW regime, would be of a few hours. This can be seen in Fig.~\ref{fig_sensi}, where the minima of the sensitivity curve lies close to the low-frequency edge of the predicted mHz GW regime.

Finally, we note from Eq.~\ref{sensi3} and Eq.~\ref{sensi4} that the absolute scale of sensitivity curves in Fig.~\ref{fig_sensi} of course scales with $\sigma_{\rm noise}/ \sqrt{f_{\rm sample}}$. This holds because we consider only stochastic noise to be present in the system, and any stochastic noise can be cut down by collecting more data points, due to the fundamental theorem in statistics that the variance of the mean goes as the inverse of the number of data points.

\subsection{SRGO observational range}
\label{obs_range}

\begin{figure}[!ht]
\includegraphics[width=\linewidth, keepaspectratio]{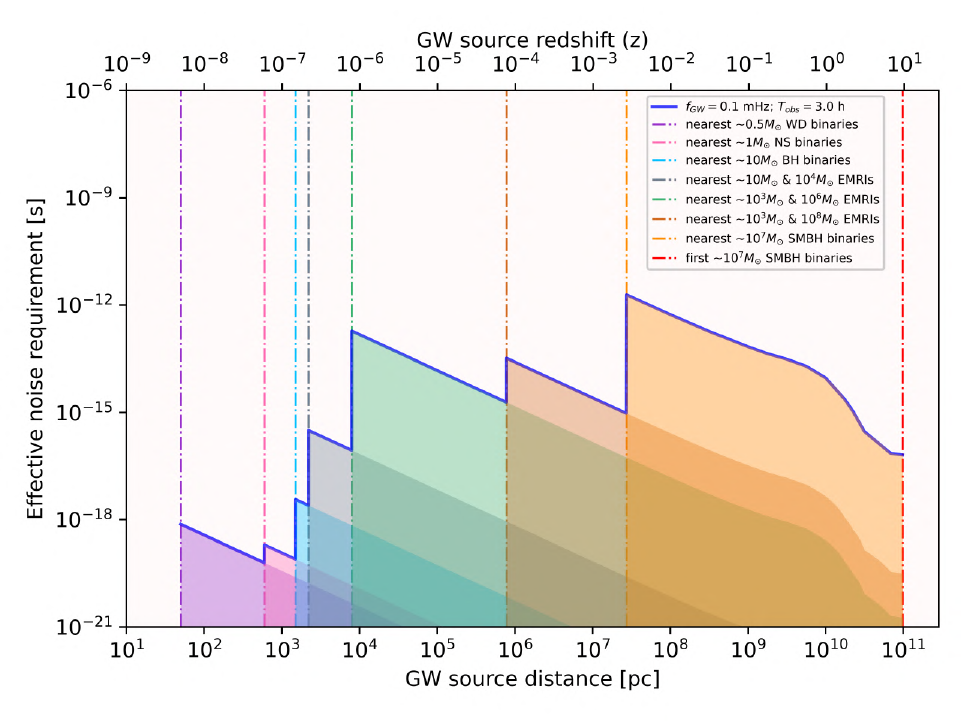}
\caption{The maximum effective noise allowed in an SRGO to make a detection, or equivalently, the largest SRGO response amplitude expected in the best-case (optimum parameter choice) scenarios which maximise the signal, due to GW sources at various distances. The colored dash-dotted lines indicate the nearest location of a particular type of source i.e. these GW sources are absent to the left of the colored line corresponding to them. All computations are done for a fixed observation time of 3 hours, and an initial GW frequency which is at the expected lower limit of the mHz regime, as this would maximise the SRGO response amplitude.}
\label{fig_range}
\end{figure}

In the previous sections, having simulated the response signal of an SRGO to astrophysical GWs and having computed the sensitivity curve, here we predict using our models, the required effective residual noise in the system to make a detection of various different types of GW sources. By `effective', we mean the residual stochastic noise divided by the square root of the number of data points taken over the observation period. We do this by computing the root mean square value of the strongest signal expected in an SRGO, using the principle of Eq.~(\ref{sensi}), discussed in the previous subsection. We use a realistic distribution of GW source types in the universe, combined with choosing a set of parameters for a given source that maximises the SRGO response.

From the ninth catalogue of spectroscopic binary orbits (SB9) \cite{sb9}, cross-referenced with the Gaia Data Release 3 \cite{gaia1, gaia2, gaia3}, we find that the nearest spectroscopic binaries within our galaxy, including white dwarf (WD) binaries with masses $ \sim 0.5 M_{\odot} $ and periods of a few days, are located at distances of a few tens of pc. Whereas, the nearest spectroscopic binaries with periods of a few hours are located at distances of several tens of pc. Hence, we choose a distance of 50 pc to mark the nearest WD binaries that would emit mHz GWs.

The nearest neutron star (NS) binaries \cite{Farrow_2019, Tauris_2017, cameron10.1093/mnrasl/sly003, Stovall_2018, Lynch_2018} with masses $\sim 1 M_{\odot}$ and periods of a few days, are located at distances of a few hundreds of pc \cite{Kaplan_2014}. Whereas, the nearest known double neutron star system with a period of a few hours is located at around 600 pc \cite{Lyne2004, PhysRevX.11.041050}. Hence, we choose this distance to mark the nearest NS binaries that would emit mHz GWs.

It is estimated that our Milky Way galaxy contains millions of stellar mass black holes \cite{mwbhs}. From binary black hole population simulations for Milky Way-like galaxies \cite{mwbbhs}, it is estimated that binary black holes may be present as close as 1 kpc from Earth, although most of them would be present 8 kpc away, near the galactic center. This also happens to agree with the recent unambiguous detection via astrometric microlensing, of an isolated stellar mass black hole \cite{Sahu_2022}, located at $1.58$ kpc from Earth. Hence, we choose this distance as an estimate of the nearest stellar mass binary black holes with masses $\sim 10 M_{\odot}$.

Intermediate mass black holes (IMBHs) of $10^2$ -- $10^5 M_{\odot}$ are expected to be found in globular clusters and massive star clusters, but would be more numerous within galactic bulges of large galaxies and within dwarf galaxies \cite{imbh1, imbh2}. In Milky Way-like galaxies, globular clusters containing IMBHs of $10^3$ -- $10^4 M_{\odot}$ are expected to be numerous at distances of 10 kpc from the galactic center, and these IMBHs can emit mHz GWs by merging with stellar mass black holes \cite{imbh3, imbh4}. Hence, we use the distance to the nearest known globular cluster ``M4", of 2.2 kpc, to estimate the whereabouts of the nearest $\sim 10 M_{\odot}$ \& $10^4 M_{\odot}$ extreme mass ratio inspirals (EMRIs).

A typical large galaxy can contain several ``wandering" SMBHs of $\sim 10^6 M_{\odot}$, spread out across the galactic halo, from near the galactic center to within the dwarf satellite galaxies and anywhere in between \cite{Ricarte_2021}. The Milky Way's central supermassive black hole (SMBH), Sgr A*, also happens to be of mass $\sim$$10^6\,M_{\odot}$ \cite{EHTC_2022}. According to \cite{Berti_2006}, the most promising mHz GW scenario in the IMBH -- light SMBH mass range, is of $\sim$$10^3\,M_{\odot}$ IMBHs merging with $\sim$$10^6\,M_{\odot}$ SMBHs. Lastly, the nearest large galaxy to us, M31 (Andromeda), contains a $\sim$$10^8\,M_{\odot}$ central SMBH \cite{Bender_2005}. For all these reasons, we choose a distance of 8\,kpc (distance from Earth to Milky Way's center, as well as to the closest dwarf galaxy, Canis Major) to represent the location of the nearest $\sim$$10^3\,M_{\odot}$ \& $\sim$$10^6\,M_{\odot}$ EMRIs. We choose 0.778\,Mpc (distance to M31) to represent the location of the nearest $\sim$$10^3\,M_{\odot}$ \& $\sim$$10^8\,M_{\odot}$ EMRIs.

The redshift evolution of the SMBH mass function \cite{smbh1, smbh2, smbh3} tells us that, on average, $10^7\,M_{\odot}$ SMBH mergers may be the most frequent. The nearest known inspiralling SMBHs due to a galaxy merger, are located at a distance of $27.4$\,Mpc \cite{smbh4}. Finally, according to
\cite{Colpi2019}, the first SMBH mergers happened at around $z=10$, when the first galaxies began to merge in the early universe. 

We use all this information to generate Fig.~\ref{fig_range}, which shows the maximum response signal, or equivalently, the upper limit on the effective residual noise allowed in an SRGO to detect a GW source at a given distance. From Fig.~\ref{fig_range}, we see that the largest SRGO response signals would correspond to mHz GWs from SMBH binaries in galaxy mergers. EMRIs involving a SMBH, typically the central SMBH of galaxies, or even wandering SMBHs interacting with smaller black holes, would also be significant sources. WD binaries, NS binaries and stellar mass BH binaries within our galaxy would not produce great responses, even if they were individually resolvable sources and located as close to Earth as possible. IMBH EMRIs within globular clusters in our galaxy may also produce reasonably large response amplitudes.

The linear behaviour of each section of the curve in Fig.~\ref{fig_range}, having a negative slope, comes from the inverse relation between the GW source luminosity distance and the GW strain (hence, also the response signal). To generate Fig.~\ref{fig_range}, we assume an observed initial GW frequency of $0.1$\,mHz (the expected lower limit of the mHz regime, which maximises the response signal). Since the GW is cosmologically redshifted over large distances, the emitted GW frequency (or equivalently, the orbital frequency of the binary) is an order of magnitude larger at high redshifts, between z=1 to z=10. Importantly, this causes the chirp rate of the frequency to also be significantly higher. Therefore, the signal amplitude reduces rapidly as the GW frequency increases (as discussed in Sec.~\ref{signal_analysis}), which explains the non-linear behaviour of the curve at high redshifts.

Fig.~\ref{fig_range} also tells us that an SRGO should aim for an effective residual stochastic noise of $\sim$$1$\,ps or better. Since the effective noise depends not only on the noise of an individual timing measurement, but also on the data sampling rate and observing time, the true residual stochastic noise may be greater than $\sim$$1$\,ps, and can effectively be reduced by collecting data at a higher sampling frequency during the observing run (see Sec.~\ref{subsec_sensi}). At this level of noise (or better), an SRGO could potentially detect mHz GW events involving SMBHs starting from within our galaxy, up to galaxy merger events at high redshifts.

Note that, while the sensitivity curve Fig.~\ref{fig_sensi} is a curve of unity SNR for the average signal as a function of GW frequency, Fig.~\ref{fig_range} is a curve of unity SNR for the maximum signal as a function of GW source distance, which also depends on the GW source type.

\section{Results: GW parameter estimation}
\label{sect_param_estim}

We have established a mathematical model to describe a basic SRGO, using which we have analysed its repsonse to GWs, obtained its sensitivity curve, and computed upper limits on the residual noise required to detect astrophysical mHz GW sources. In this section, we obtain first estimations on the capacity of an SRGO to constrain the physical parameters of a GW source. We do this by performing MCMC model fitting on synthetic noisy data generated with our model.  

The antenna pattern of a LIGO detector is largely omnidirectional (see Fig.~2 of \cite{SAULSON2013288}) and the same is true for an SRGO (see Fig.~2 of paper~I). Therefore, even with a high signal-to-noise ratio, a single stationary LIGO or SRGO, in principle, cannot pinpoint the position of the GW source in the sky. However, an Earth-based detector is not stationary, and continuously changes its orientation relative to an incoming GW, due to Earth's rotation. In principle, this should cause the GW source to sweep across the detector's antenna pattern and produce a corresponding characteristic envelope in its response, breaking the degeneracies of omnidirectionality and thus allowing the GW source to be pinpointed. But since LIGO is sensitive to GWs in the kHz frequency range that have short durations, it is unable to fully exploit the effect of Earth's rotation. Hence, three or more LIGO detectors working together are required to triangulate the GW source position via the relative time-delays between their detections from the same source.

On the other hand, since an ideal SRGO would be sensitive to GWs in the mHz frequency range (where the GW signal may last for hours, days, or much longer), it should be able to utilize Earth's rotation to pinpoint the GW source position. The GW source sky localization area, however, would obviously depend on the signal-to-noise ratio.  We demonstrate this with an example from our simulation results, where MCMC methods have been used to do GW parameter estimation on noisy data points that were created by adding Gaussian noise to the SRGO response signal.  

\begin{figure}
\includegraphics[width=\linewidth, keepaspectratio]{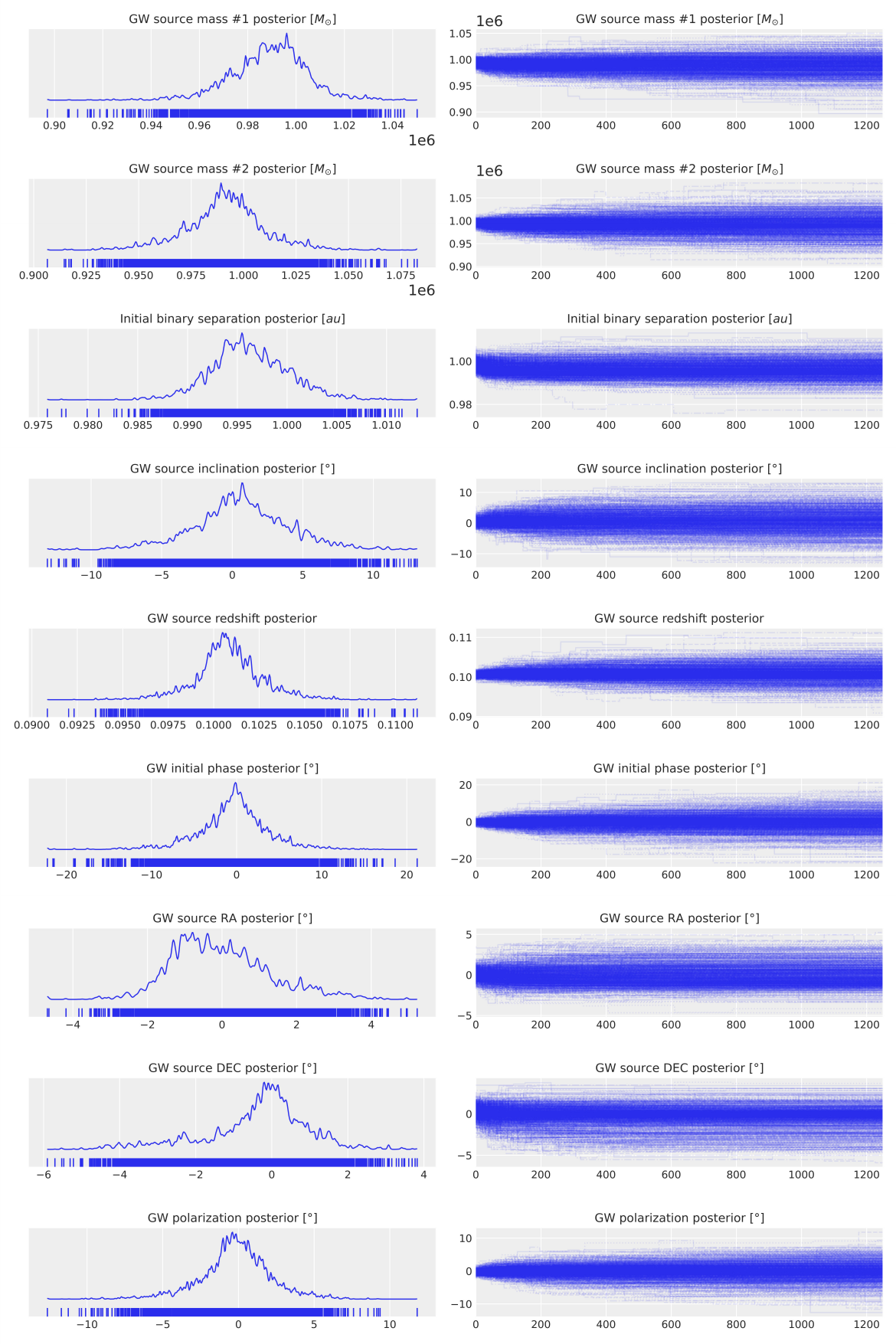}
\caption{On the left are the marginalized posteriors of the fitting parameters and on the right are the corresponding MCMC traces, consisting of 1000 parallel chains with 1250 samples each. The true parameter values for this case are the ones in Fig.~\ref{fig_response}, except the initial separation between the masses which is 1 AU (corresponding to an initial GW period of 3.1 h). 32 data points with artificial noise added (PSNR = 100) are taken over an observing time of 12 hours.}
\label{fig_trace}
\end{figure}

Fig.~\ref{fig_trace} is an example of the MCMC chain traces, and the diagonal elements of the $9 \times 9$ joint posterior corner-plot (shown in Appendix \ref{app_B}, Fig.~\ref{fig_last}). We see that in general, at sufficient PSNR values, starting from the correct parameter values, the MCMC chains explore around this location in the 9-dimensional parameter space.

\begin{figure}
\centering
\includegraphics[width=\linewidth, keepaspectratio]{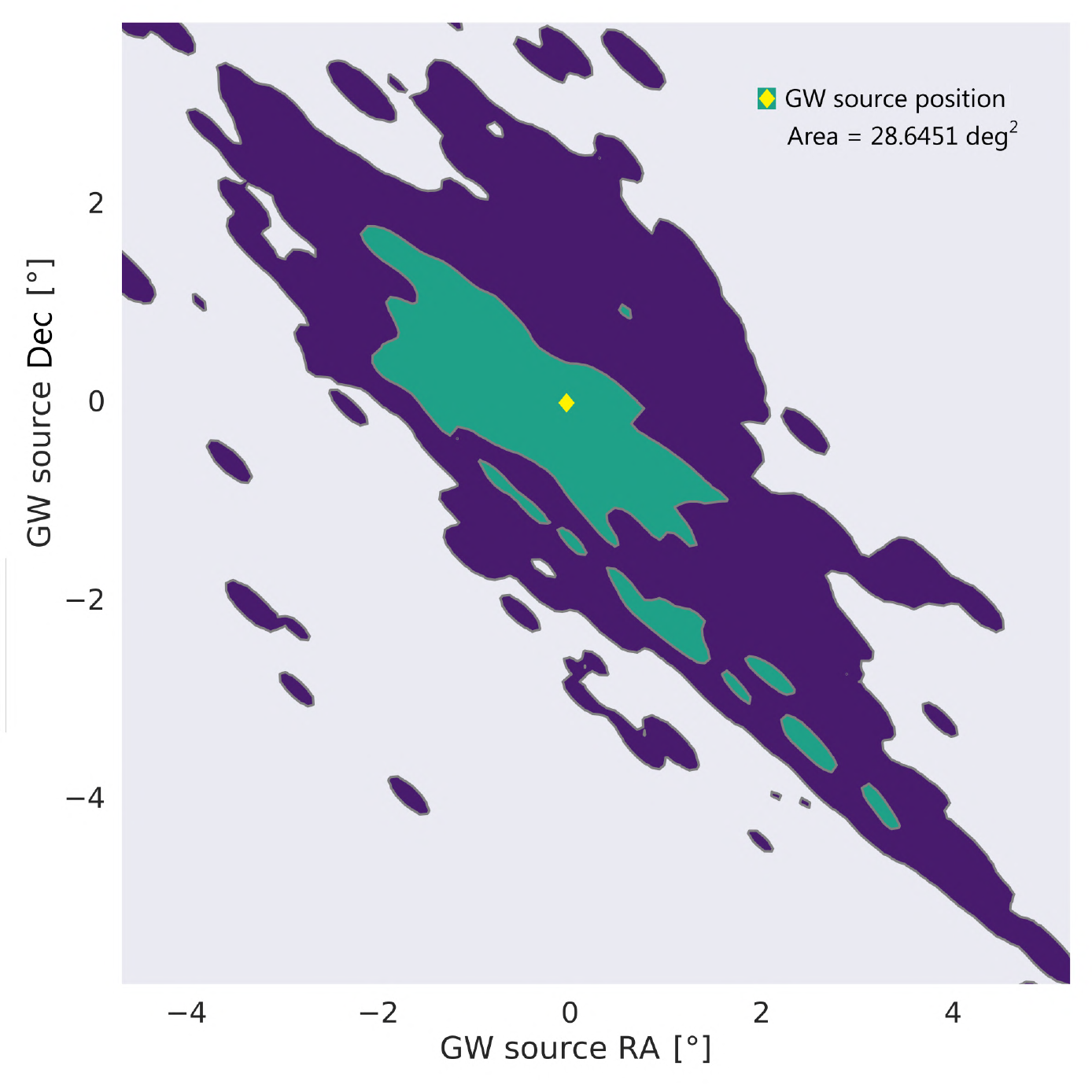}
\caption{The GW source sky localization, i.e.\ the 3-sigma ($99.7\%$, purple) and 1-sigma ($68\%$, green) HPD region on the joint posterior of the GW source's right ascension and declination. This figure corresponds to the case shown in Fig.~\ref{fig_trace}. This shows that a single SRGO can potentially use Earth's rotation to localize the GW source in the sky. The other joint posteriors corresponding to this case may be found in Appendix \ref{app_B}.}
\label{fig_localization}
\end{figure}

Fig.~\ref{fig_localization} corresponds to the same case as Fig.~\ref{fig_trace}, and it shows the sky localization (i.e.\ the joint posterior of the right ascension and declination of the GW source) for a case corresponding to 32 effective data points taken over 12 hours at a PSNR of 100. The sky localization area is calculated by computing the area on the sky (in $\deg^2$) of the posterior region.


Due to the specifics of the MCMC setup, described in Sec.~\ref{mcmc}, we miss the antipodal sky localization region, which should exist because placing the ring and/or the GW source at antipodal positions, and/or having the ions circulating in the opposite direction, would all produce the same SRGO response. Although we provide flat priors and allow the MCMC chains to explore over the full range of the angular parameters, the chains seem to explore only around the true parameter values. This is of course due to the nature of the DE-MC algorithm used, which is known for converging quickly to a solution in the parameter space and staying around it. An antipodal region would double the sky localization area, hence, we manually do so for all measurements of sky localization area.

We observe that the joint posteriors we obtain are in many cases not smooth (see Sec.~\ref{app_B} for the entire set of joint posteriors corresponding to Fig.~\ref{fig_localization}). However, a non-smooth posterior can still be statistically well-behaved and provide meaningful information about the parameters of interest. The smoothness or lack thereof in a posterior distribution does not necessarily indicate the quality or reliability of the results. If the model is highly non-linear or complex, it might naturally lead to non-smooth posteriors. The choice of MCMC sampling algorithm can also play a role. Different algorithms have different exploration and convergence properties, which can affect the smoothness of the obtained posteriors. The `blobbiness' observed in our posteriors comes from a combination of the complex model and the DE-MC algorithm. We chose the DE-MC because not only did it offer faster computation times, but it also gave more reliable results compared to other algorithms. 

Following this brief overview, in the next sub-sections, we shall explore the SRGO parameter estimation results in greater depth.

\begin{figure*}
     \centering
     \begin{subfigure}{0.49\linewidth}
         \centering
         \includegraphics[width=\linewidth, keepaspectratio]{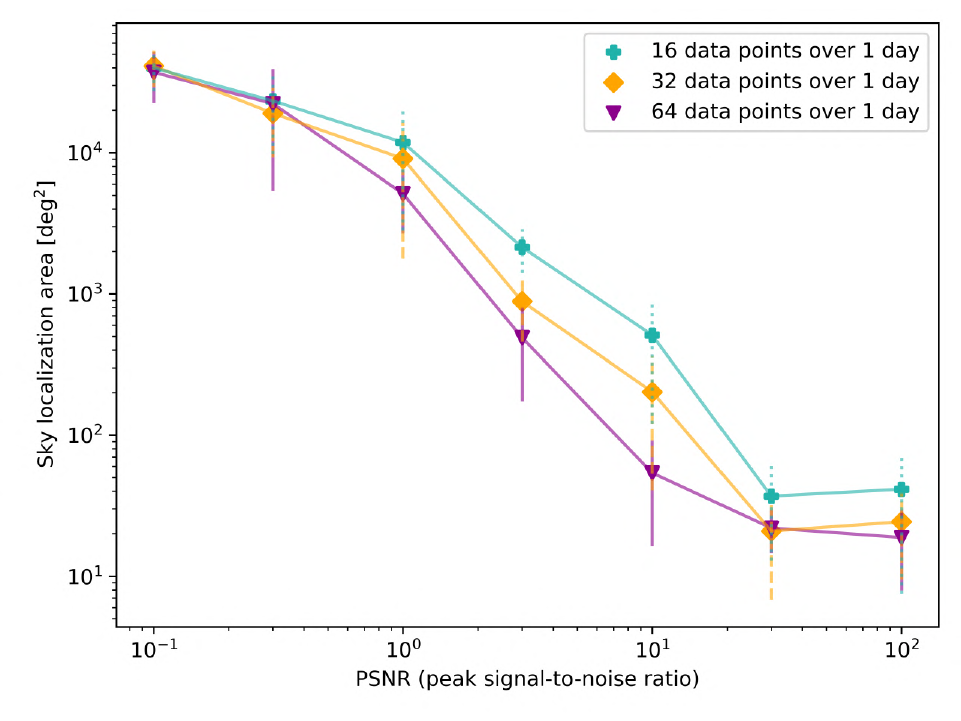}
         \caption{Sky localization area vs.\ PSNR}
         \label{subfig-i}
     \end{subfigure}
     \begin{subfigure}{0.49\linewidth}
         \centering
         \includegraphics[width=\linewidth, keepaspectratio]{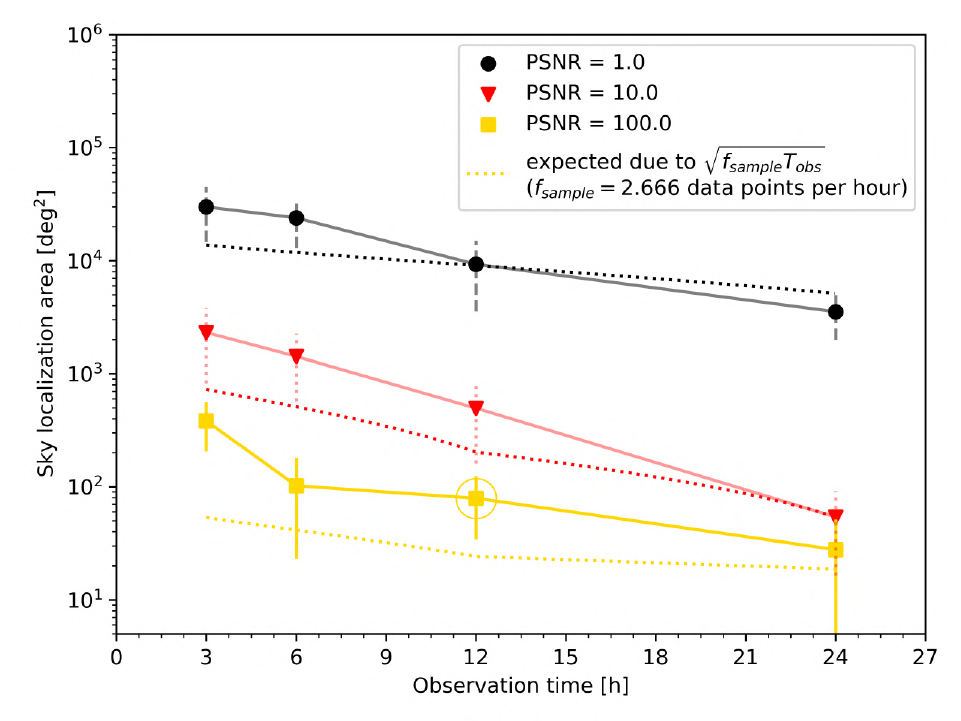}
         \caption{Sky localization area vs.\ $T_{\rm obs}$}
         \label{subfig-ii}
     \end{subfigure}
     \begin{subfigure}{0.49\linewidth}
         \centering
         \includegraphics[width=\linewidth, keepaspectratio]{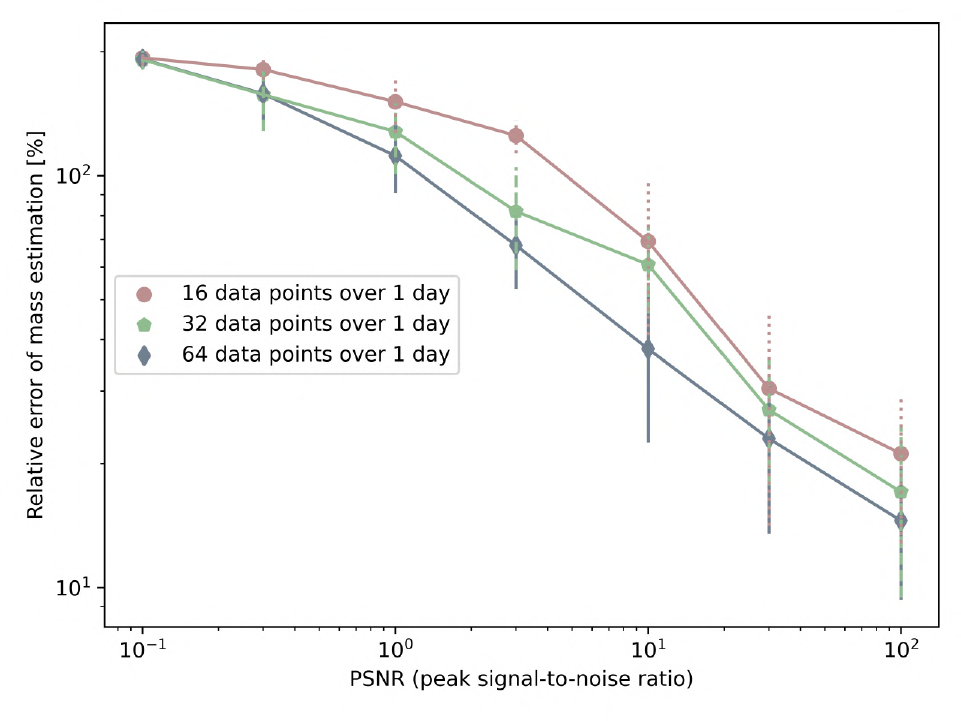}
         \caption{Mass estimation error vs.\ PSNR}
         \label{subfig-iii}
     \end{subfigure}
     \begin{subfigure}{0.49\linewidth}
         \centering
         \includegraphics[width=\linewidth, keepaspectratio]{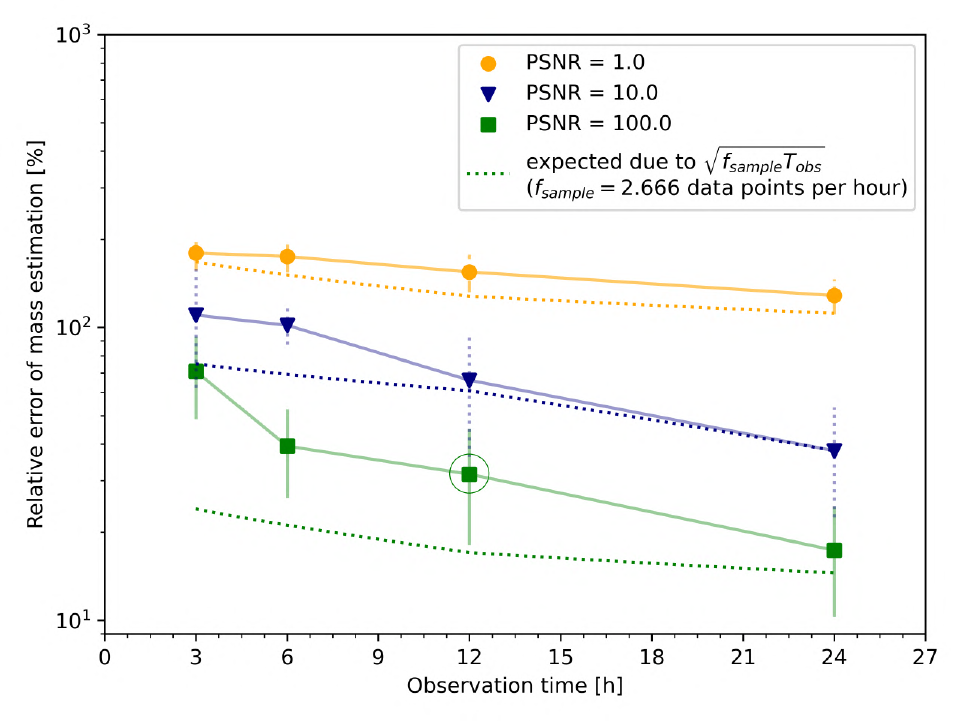}
         \caption{Mass estimation error vs.\ $T_{\rm obs}$}
         \label{subfig-iv}
     \end{subfigure}
     \begin{subfigure}{0.49\linewidth}
         \centering
         \includegraphics[width=\linewidth, keepaspectratio]{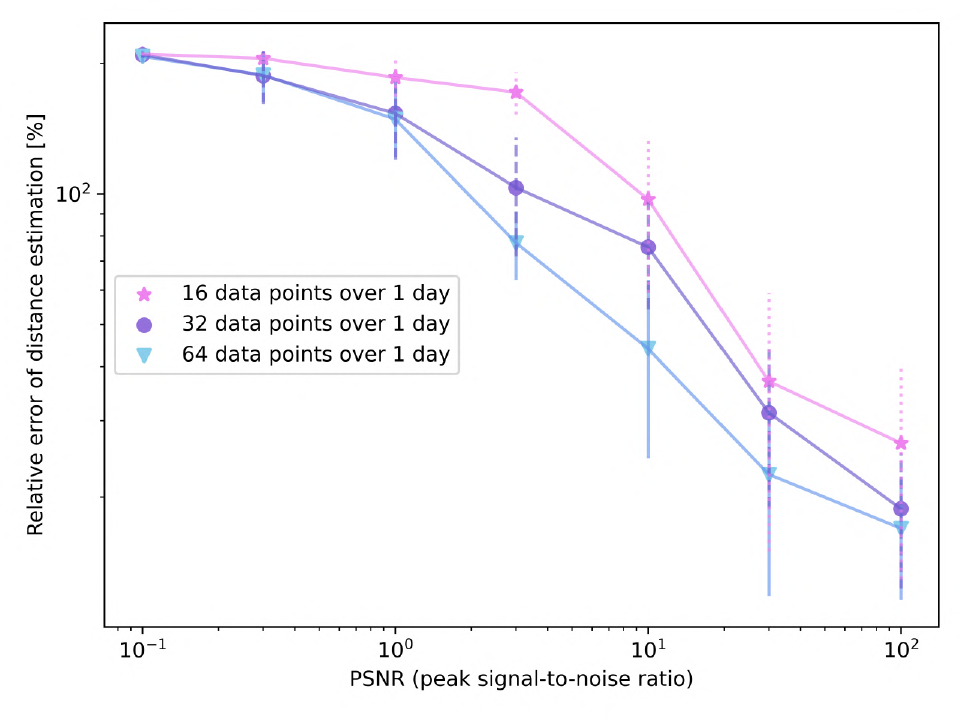}
         \caption{Distance estimation error vs.\ PSNR}
         \label{subfig-v}
     \end{subfigure}
     \begin{subfigure}{0.49\linewidth}
         \centering
         \includegraphics[width=\linewidth, keepaspectratio]{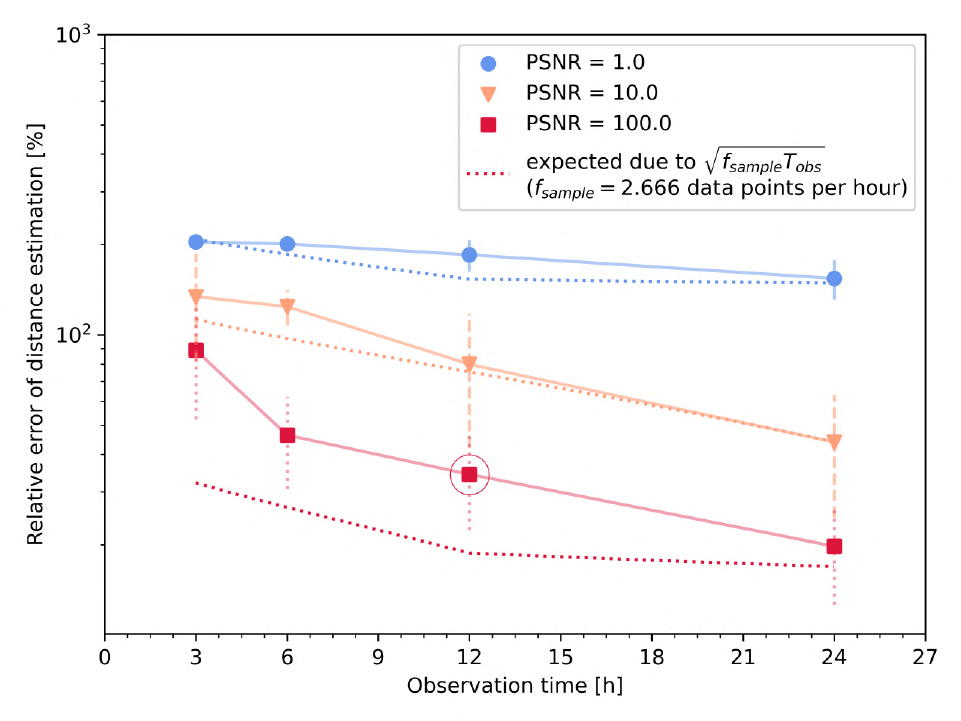}
         \caption{Distance estimation error vs.\ $T_{\rm obs}$}
         \label{subfig-vi}
     \end{subfigure}     
        \caption{The parameter estimations of three parameters are shown: GW source sky localization area, relative errors of distance and mass estimation. The left column shows their variation with PSNR for different data sampling rates. The right column shows their variation with observation time at a fixed data sampling rate, for different values of PSNR. The circled scatter points on the right panels correspond to the case shown in Fig.~\ref{fig_trace}, Fig.~\ref{fig_localization} and App.~\ref{app_B}.}
        \label{param_estimation}
\end{figure*}

\subsection{Variation of parameter estimation errors}
\label{estim_variation}

We explore the constraints derived from parameter estimation, by computing the variation of the posteriors as functions of some controllable experiment parameters (such as the observation time, data sampling rate and PSNR), for 5 out of the 9 fitting parameters in our model. These are: the GW source component masses, $m_1$ and $m_2$ ; the GW source redshift, $z$ ; and the GW source sky position, $\alpha_{\rm src}$ and $\delta_{\rm src}$. The results for $\alpha_{\rm src}$ and $\delta_{\rm src}$ jointly correspond to the sky localization area, shown in Figs.~\ref{subfig-i} and \ref{subfig-ii}. By symmetry, the results for $m_1$ and $m_2$ are the same, and correspond to Figs.~\ref{subfig-iii} and \ref{subfig-iv}. The results for $z$ have been translated into the GW source luminosity distance, and correspond to Figs.~\ref{subfig-v} and \ref{subfig-vi}. We choose these 5 parameters as they are astrophysically the most relevant ones.

We generate 16, 32 and 64 data points in this study for a multitude of reasons: first, as explained in Sec.~\ref{mcmc}, we use powers of two as this allows for faster computation of the FFT. Furthermore, in our computations, 16 data points taken over one day (or $2.666$ data points per hour) happens to be the lower limit for the Shannon-Nyquist condition to hold for the case corresponding to Fig.~\ref{fig_response}, which is used to generate results in this section. Hence, this would give us an upper limit on the constraints that can be derived from parameter estimation for a given PSNR and observation time. Also, although in reality it would be possible for a timing detector to make several million measurements within an observation time of hours to days, we use only a small number of data points for computational efficiency. This is justified because a smaller number of data points at a given PSNR may be interpreted as binning a large number of data points that correspond to a lower true PSNR, thus giving the same effective PSNR.

Note that, for results pertaining to parameter estimation, we chose to work with the PSNR (peak signal-to-noise ratio) rather than an integrated SNR, such as that used for the sensitivity curve in Sec.~{\ref{subsec_sensi}}. Integrated SNR is a useful measure for probing the limits of detectability, but large values can be obtained for a given signal and noise level, simply by taking measurements over longer observation times. The PSNR, on the other hand, is independent of the observation time. For a given noise level, the PSNR necessarily reflects whether the signal is strong or weak overall, unlike the integrated SNR. However, if the observation time is fixed, then both kinds of SNR effectively convey the same information.

In Fig.~\ref{subfig-i}, we see that, for a given data sampling rate and observation time, the sky localization area decreases with decreasing noise, and saturates at around a few tens of $\deg^2$ for high PSNR. This is due to parameter degeneracies that cannot be broken further, unless multiple SRGOs are utilized or better models are utilized that, for instance, account for higher-order harmonic modes of GWs \cite{2208.13351}. The error bars show the statistical variation of the parameter estimation (``error on the error''), and they increase with increasing noise. This is likely due to the DE-MC algorithm: at low PSNR, it would generally cause the chains to spread over a wide range on the parameter space, but they would sometimes tend to converge to some spurious value on the parameter space, thus increasing the standard deviation of the parameter estimation values obtained from multiple data sets generated for the same case.

Furthermore, at a given PSNR, the sky localization improves upon increasing the data sampling rate. This is because the effective noise is inversely proportional to the square root of the total number of data points, or in other words, the square root of the data sampling rate times the observing time. The parameter estimation becomes unreliable at high levels of noise, with the sky localization area covering almost the entire sky for PSNR values lower than $\sim 0.1$. An effective PSNR of $\sim 80$ seems to be the threshold for a single SRGO to achieve its best possible sky localization. As stated in the previous sections, `effective' implies dividing the noise (or multiplying the PSNR) by $\sqrt{f_{\rm sample} T_{\rm obs}}$, i.e.\ the square-root of the number of data points. 

For a given PSNR in Fig.~\ref{subfig-i}, upon interpolating over the three data points corresponding to different number of samples, and then extrapolating this, we plot model curves (dotted lines) for different PSNR values in Fig.~\ref{subfig-ii}, which show the expected variation of the sky localization due to a change in the total number of samples taken over different observation times, at a fixed sampling rate. This would correspond to a realistic scenario, where the sampling rate of the SRGO would be fixed, and changing the total observation time would change the number of data samples measured. However, when we overplot values (scatter points) obtained by doing an MCMC fit to simulated noisy data, we observe a deviation from the extrapolated model curves. This deviation is greater for higher PSNR values and for smaller observation times. In fact, this deviation shows that changing the observing time does not just have the effect of changing the number of data samples, but further, the effect of Earth's rotation can be exploited to a greater extent to better break some parameter degeneracies. At small observation times, this effect is inefficient, and hence the scatter points perform worse than the model fit. Note that the model fit agrees with the scatter point at an observation time of 24 hours, since that was the fixed observation time in Fig.~\ref{subfig-i} which was used to create the model curves.  

An exception to this trend may occur when the Shannon-Nyquist condition is violated, i.e.\ for a fixed number of data points, a smaller observation time may result in better parameter estimation if increasing the observation time (reducing the data sampling rate) results in aliasing. This scenario would typically not be relevant for a realistic SRGO experiment, where the data sampling rate would be orders of magnitude higher than the GW frequency. Finally, beyond 24 hours observing time, the Earth's rotation cannot break any more parameter degeneracies in principle, and therefore even at high effective PSNR values in Figs.~\ref{subfig-i} and \ref{subfig-ii}, the sky localization tends to saturate at around 20 $\deg^2$. 

In Figs.~\ref{subfig-iii}, \ref{subfig-iv}, \ref{subfig-v} and \ref{subfig-vi}, we observe similar trends for the GW source mass and distance estimation as observed for the sky localization. The relative errors of mass and distance estimation saturate at $200\%$ for low PSNR values only because of the bounded flat priors that we use in the MCMC algorithm, mentioned in Sec.~\ref{mcmc}. Even at the highest PSNR value, the errors on the mass and distance estimation remain finite, at $\sim$$20$\%. For the same reasons mentioned previously, comparing values from a given curve in Fig.~\ref{subfig-iv} (\ref{subfig-vi}) with the corresponding values at the same PSNR in Fig.~\ref{subfig-iii} (\ref{subfig-v}), we notice the trend that for the same effective PSNR, increasing the observation time improves parameter estimation. However, unlike the sky localization, the errors on the mass and distance estimations do not yet saturate at the highest PSNR values probed, but are expected to eventually saturate at even higher PSNR values.

Gravitational wave interferometers can measure the binary masses to a higher accuracy than obtained in our results, as calculated in \cite{PhysRevD.49.2658}. For the same SNR, the lower accuracy of mass estimation in our results may be attributed to using only the quadrupole moment contribution in our GW strain model, compared to the post-Newtonian model used in \cite{PhysRevD.49.2658}, which of course improves parameter estimation accuracy due to the higher order terms. Also, in our computation, we have only considered the inspiral phase of the GW source, avoiding the merger and ringdown phases, which generally provide further information for accurate parameter estimation \cite{PhysRevLett.70.2984}.

\subsection{Parameter degeneracies}
\label{param_degen}

In Appendix \ref{app_B}, we show an example of the 36 joint posterior pair-plots for our 9 model parameters. These correspond to a case where 32 noisy data points were taken over 12 hours, at a PSNR of 100. The true parameter values for this case are the same as in Fig.~\ref{fig_response}, except that the initial binary separation is 1~AU (corresponding to an initial GW period of 3.1~h). At such a high PSNR, the joint posterior correlations show the degeneracies between the parameters. Here, we try to explain the observed correlations based on the model details described in Sec.~\ref{subsect_A}:

The joint posterior of the two binary masses (Fig.~\ref{ssubfig1}) shows an anti-correlation, because upon increasing one of the masses, the other must be decreased to have the same chirp mass. The two binary masses are also positively correlated with the initial binary separation (Figs.~\ref{ssubfig2} and \ref{ssubfig9}), since increasing the mass increases the GW strain amplitude and also affects the frequency evolution, which can be countered by increasing the initial binary separation. The positive correlation between the masses and the GW source redshift (Figs.~\ref{ssubfig8} and \ref{ssubfig11}) has already been explored in detail in Sec.~\ref{obs_range}. It exists because increasing the redshift decreases the response signal, which can be countered by increasing the mass of the binary. Instead of increasing the binary mass, we can also counter this by increasing the initial binary separation, which reduces the GW frequency, thus increasing the response signal, as discussed in the beginning of Sec.~\ref{signal_analysis}. That is why the joint posterior between the source redshift and the initial binary separation also shows a positive correlation (Fig.~\ref{ssubfig17}).

An interesting joint posterior to analyze is the source redshift, $z$ vs.\ the source inclination angle, $i$ (Fig.~\ref{ssubfig22}). The degeneracy between the source distance, $d_L$, and inclination angle is well known in GW astronomy. Therefore, we expect them to be anti-correlated, since increasing the source distance decreases the GW strain amplitude, but decreasing the inclination angle can counter this. However, in Fig.~\ref{ssubfig22} we consider the correlation of $i$ with $z$, rather than with the luminosity distance, and $z$ also affects the chirp mass and GW frequency. Therefore, in our results, the $z-i$ joint posterior shows no correlation, since these two parameters control very different aspects of the response signal. Further, a small change in $z$ corresponds to a large change in $d_L$. If a posterior of $d_L$ and $i$ were computed, for an approximately constant $z$, then the expected degeneracy may be easily seen. 

Another noteworthy joint posterior is that of the GW polarization angle vs.\ the GW initial phase (Fig.~\ref{ssubfig33}), which shows a strong linear correlation. This can be derived analytically for our special case of $i=0$. For this case, the response signal Eq.~(\ref{dtcircular2}) can be re-written as follows, after substituting all the terms and collecting the common terms into a factor $h_0$,

\begin{multline}
   \Delta T_{\rm GW} = \int_{t_{0}}^{t_{0}+T}h_0\sin^2{\theta}\Bigl(\cos{\left(2\psi\right)}\cos{\left(2\pi ft + \delta_0\right)} \\ + \sin{\left(2\psi\right)}\sin{\left(2\pi ft + \delta_0\right)}\Bigr)dt \\
   = \int_{t_{0}}^{t_{0}+T}h_0\sin^2{\theta}\cos{\left(2\pi ft + \delta_0 - 2\psi\right)}dt.
\end{multline}

The above equation shows that, $i=0$ corresponds to circularly polarized GWs observed with an SRGO, which may be interpreted as GWs with a constant phase, $\delta_0$, whose polarization axes are rotating in a circle with an `effective' GW polarization angle, $\psi_{\rm eff} = \psi - \pi ft$. Other parameters being fixed, if we Taylor-expand $\psi$ as a function of $\psi_{\rm eq}$ around its true value, then for small deviations of $\psi_{\rm eq}$ from its true value, $\psi$ will be linear in $\psi_{\rm eq}$. Hence, the above equation explains why the joint posterior of $\psi_{\rm eq}$ and $\delta_0$ shows a linear correlation. Note that all values within the parentheses of the above equation are in radians, while the values shown in Fig.~\ref{ssubfig33} are in degrees, but these values are consistent with the Taylor series assumption that deviations of $\psi_{\rm eq}$ from its true value, in radians, is small.

Thus, for circularly polarized GWs, changing the initial phase of the GW can produce the same signal as changing the GW polarization angle. If the Earth were stationary, this would be the same as beginning the observation run at a different time. 


\section{Discussion}
\label{discuss}

\textit{Why do we choose ultrarelativistic test mass particles?}

It is currently unclear, but using ultrarelativistic ions might be beneficial (Schmirander et al., in preparation). This is because a single ion with sufficiently high energy would, in principle, by its sheer momentum, appreciably attenuate the effect of many stochastic and deterministic noise sources that would directly perturb the ion (some of which were explored in paper~I). This would also allow a coasting ion to deflect or deviate less from its ideal orbit, and last longer in a non-ideal vacuum, thus potentially allowing for longer SRGO observation runs.

\textit{What are the limitations and caveats of this study?}

One of the primary limitations of this study is the simplistic static Gaussian noise model for the residual system noise, based on the assumption that our hypothetical storage ring facility is capable of attenuating most of the (yet un-studied) noise sources, similar to LIGO. We cannot yet model noise sources in detail, especially their frequency and time dependence, until a thorough study is conducted (Schmirander et al., in preparation).

Next, our GW waveform models do not account for the spins of the compact objects in the binary, the eccentricity of their orbit, and other parameters corresponding to realistic binary systems. While our model contains 9 unknown fitting parameters, realistic GW models contain around 15 to 17. However, as a first step towards establishing a novel experiment concept, for the sake of consistency and ease of analysis, it is better to use a realistic toy model for making order-of-magnitude estimations, rather than to use complex and detailed models from the very beginning, which can make analysis quite difficult in a topic that has not been explored to such an extent prior to this study. Although our GW source and ring models are simple, they are realistic enough to provide correct orders of magnitude of the estimates. Perhaps, incorporating detailed GW waveforms and storage ring models is the next logical step in this series of works. 

We also do not model the merger and ringdown phases, and cover only the inspiral phase of the binaries. However, because the response signal tends to decrease with increasing GW frequency (Fig.~\ref{psnr_frq}), the merger and ringdown phases would likely not produce an SRGO response signal as large as the one during the inspiral phase. 

Furthermore, our GW source models, which are derived from post-Newtonain (PN) analysis, cannot accurately model EMRIs (extreme and intermediate mass ratio inspirals). In Sec.~\ref{obs_range}, Fig.~\ref{fig_range}, some GW sources that are actually EMRIs, have been estimated with our post-Newtonian GW waveform model, which is not optimal. But since we are interested in order-of-magnitude estimates and since we do not expect the difference between our model and an EMRI GW waveform to be orders-of-magnitude greater, we regard this as a justifiable simplification. This is supported by \cite{PhysRevD.88.024038}, where it can be seen that the simple PN waveform models are accurate enough to model EMRIs for small observation times of a few hours or days, as considered in our study. 

Many of our results have been generated by averaging over as many parameters as possible, so that the conclusions interpreted from them may remain accurate and general. However, some of our conclusions are extrapolated based on results for a specific and arbitrary combination of parameters, corresponding to Fig.~\ref{fig_response} (with some variations which are described in the sections pertaining to each result). This was done for cases where averaging over parameters was very difficult or computationally expensive. These include results in Sects.~\ref{signal_analysis} and \ref{sect_param_estim}. However, we do not expect the parameter-averaged results to be different in order-of-magnitude for these cases, and hence expect them to be sufficient for first estimates and general conclusions. For instance, the results in Sec.~\ref{signal_analysis}, which are based on scaled values of the PSNR, are intrinsically independent of some parameters, such as the GW source mass and distance. Moreover, we can make estimates of how some of the results would change for a different set of parameters. For example, the results of Sec.~\ref{sect_param_estim}, for a different set of true parameters, can be estimated by combining the results of Sects.~\ref{signal_analysis} and \ref{estim_variation}, which should at least be correct in order of magnitude.

Lastly, we have not yet accounted in the SRGO response formulation, the effect of the GW on the storage ring magnetic field, which may possibly boost the response signal. This would be included in future works (Schmirander et al, in preparation).

\textit{How to measure the instantaneous initial ion speed, $v_i$?} 

Two timing detectors placed close by would detect a passing ion with a delay. Dividing the known distance covered by the ion with this timing delay would give us $v_i$. This measurement could be made more accurate by repeating this procedure over the first several revolutions and then taking an average value. However, performing this procedure with a single timing detector (i.e. dividing the orbit circumference by the time interval between two successive detections of the same ion by a single detector) would be less accurate, because although the time-varying quantities would change negligibly during a single ion revolution, but the ion would still be affected by the anisotropy of the spacetime. Hence, compared to the former procedure, this way would give us a slightly worse substitute for the quantity that we wish to measure.

\textit{How to measure $\Delta T_{\rm GW}$?} 

Using $v_i$ and the circumference of the ion orbit, we can predict the expected arrival times of the ion to the timing detector. These must then be subtracted from the actual ion arrival times that are measured by the timing detector. The result will constitute the discrete noisy data points $\Delta T_{\rm GW}$, which when plotted against the expected arrival times, will look like Fig.~\ref{fig_response}. This is why the second term within the integral of Eq.~(\ref{dtcircular2}) differs from that of Eq.~(\ref{dtcircular}), when we measure $v_i = v_0 \left(1 + \frac{h_{\theta \phi \psi} (0)}{2} \right)$ instead of $v_0$. Since the speed is used to predict the times when the ion would arrive at the detector, a different speed would change the predicted ion arrival times, and thus, also the signal (which is the observed arrival time minus the predicted arrival time of the ion).

\textit{Do GWs affect the atomic clock of the timing detector?}

Since the storage ring ion clock and the atomic/optical clock of the timing detector would be located next to each other, they would both be affected in the same way due to the temporal component of the GW metric (or any other spacetime metric). Therefore, in principle, the temporal component of the spacetime metric cannot be measured by a comparison between the ion clock and atomic clock geodesics (the working principle of SRGO). However, since the location of the atomic clock would be stationary in the reference frame, while the ion would revolve in an anisotropic GW spacetime, the spatial components of the GW metric would affect the storage ring ion clock differently as compared to the atomic/optical clock. This difference would result in  the response signal that can, in principle, be measured by an SRGO. This is the reason why, as explained in Sec.~\ref{obs_range}, laser and atom interferometer GW detectors cannot probe the anisotropy of a static distorted spacetime (such as very low frequency GW spacetimes over short observation times), even in principle. Whereas, this would possible in principle with an SRGO, even in the absence of Earth's rotation. However, practically, this might never be tested because of the stochastic gravitational wave background (SGWB), which exists due to an overlap of a large number of unresolved and incoherent astrophysical GW sources at low frequencies \cite{Caprini_2018, Christensen_2019, galaxies10010034}.

\textit{Why did we choose MCMC methods over Fisher Information for parameter estimation?}

The Fisher Information Matrix (FIM) can be described as the inverse of the covariance matrix of some distribution. It may also be interpreted as the curvature of the log-likelihood graph. The FIM can be calculated analytically, requiring only the model that generates the response signal. This makes the FIM a fast and simple method of obtaining the precision of the parameter estimation pipeline without actually having to make a measurement of artificial noisy data. However, the FIM does have limitations: It assumes a model with linearly correlated parameters, a detector with Gaussian noise, and a high SNR. It has been shown that for a non-spinning binary GW source model with 9 unknown parameters such as ours, at total binary mass higher than $10.0 M_{\odot}$, the standard deviation predicted by the FIM does not agree with the standard deviation of a fully calculated posterior by MCMC methods \cite{PhysRevD.88.084013}.

Further, an MCMC fit gives us the added advantage of studying correlations between parameters, and getting a qualitative impression of how well-behaved a fit is.

\textit{What are the implications for conducting an SRGO experiment at the FCC (Future Circular Collider)?}

FCC \cite{2203.06520} is a proposed circular particle accelerator which will be able to accelerate ultrarelativistic ions at even higher energies than the LHC. This could increase the natural attenuation of any stochastic noise sources directly acting on the ions, due to the ions having a higher relativistic mass or momentum. However, the proposed 100 km circumference of the FCC would have implications for noise levels from sources such as seismic noise, gravity gradient noise and others, which unlike the expected SRGO response signal, would likely be sensitive to the ring size. Currently, it is unclear whether a larger or smaller ring size would be more suitable for an SRGO experiment. It is hoped that, upon detailed computational modeling of the noise sources, an optimal configuration within the parameter space can be found, which reveals the optimal ring size (Schmirander et al., in preparation).

However, another pertinent point regarding the FCC in the context of this study is that, innovation in particle accelerator technology through big projects such as the FCC would indeed help the future realisation of an ideal SRGO facility. As we recall from Sec.~\ref{ring_model}, the ideal SRGO assumed in this study currently seems to be beyond the frontier of modern technology, and work on the FCC may help push that frontier in ways that serendipitously unlock solutions for realising an SRGO.

\textit{What are the potential implications for multi-messenger astronomy?}

The yet undetected mHz GW events are also predicted to be associated with the emission of electromagnetic radiation and neutrinos \cite{2019arXiv190706482B, 2211.13759, Baker2019Multimessenger, Eracleous2019Arena}. For transient astrophysical events that correspond to high frequency GWs such as those detected by LIGO, the usual case for multi-messenger observations is of the event first being detected by the omnidirectional GW detectors, which then perform fast parameter estimations and send out real-time alerts to other observatories, providing the estimated GW source component types, masses, spins and importantly, the sky localization region. An effort is then made to quickly and simultaneously observe the GW event via the other messenger signals, using the alert information. However, for mHz GW events, fast alert response would be of lesser concern, because most of these events would be long-lasting. Therefore, improving parameter estimation, especially the sky localization, would be most important for multi-messenger studies of mHz GW events. Other than improving detector sensitivities, this is best achieved by collaboration between multiple mHz GW detectors. It is estimated that a proposed mHz GW detector such as LISA, by itself, would not be good enough to pinpoint the host galaxies of mHz GW sources \cite{Ruan2020, 2211.13759}. On this front, it is clear that the successful realization of an SRGO would greatly complement other mHz GW detectors such as LISA, and improve the GW alerts for multi-messenger observations. 

Assuming that a mHz GW event is detected simultaneously by LISA and SRGO, and further assuming that the realized SRGO has effectively the same capabilities as the hypothetical system considered in Sec.~\ref{ring_model} of this study, then we can make a rough estimation of the improvement in the GW source sky localization due to a combination of SRGO and LISA. The LISA sky localization for massive black holes is estimated to be $1 - 100 \deg^2$, and LISA would be lagging the Earth orbit by $20 \degree$ \cite{2017arXiv170200786A}. In the optimistic case, assuming that a single SRGO on Earth manages to localize the same GW event between $4 - 40 \deg^2$ as obtained in Sec.~\ref{estim_variation}, then by combining this data via simple 3D geometry, we can roughly estimate that the improved sky localization may be as good as sub--$\deg^2$, and as bad as a few tens of $\deg^2$. Overall, this would be a very good improvement, and it could be made even better by having multiple SRGOs at different locations on Earth. 

For atom interferometers sensitive to the dHz frequency range, suggestions for further improvement to parameter estimation beyond the effect of Earth’s rotation have been made, incorporating the effect of Earth’s revolution \cite{PhysRevD.97.024052}. In our case, we did not consider this effect due to the knowledge that maintaining a coasting beam in a storage ring beyond a few hours is currently very difficult. In principle, if a year-long beam could be sustained, then further improvement to the parameter estimation by a single detector would be possible.

\section{Summary and Conclusion}
\label{summary}

In Sec.~\ref{intro}, we discuss previous studies on storage rings as GW detectors, highlighting what they missed, and explaining the novelty of our idea. We provide comparisons and analogies between an SRGO and other known GW detection techniques. We also discuss references that support our findings and throw light on potential ways for realizing an SRGO. In Sec.~\ref{review}, we provide a review of the theory behind an SRGO, and revise important formulae to display them in a better format. Sec.~\ref{models} describes the mathematical models and numerical procedures of our simulation code. 

In Sec.~\ref{signal_analysis}, we study the variation of the response signal with the experiment parameters, obtaining useful physical insights about how an SRGO works. Our results suggest that the response signal would be maximised by placing an SRGO at equatorial latitudes on Earth and by having long observation times. In Sec.~\ref{subsec_sensi}, we numerically obtain the SRGO sensitivity curve, which shows that an SRGO would be sensitive to the mHz GW regime provided that a minimum observation time (run time of the storage ring) of at least a few hours can be achieved in an SRGO experiment. In Sec.~\ref{obs_range}, we find that a typical SRGO would require an effective residual noise lower than $\sim 1$ps to detect astrophysical mHz GW sources. At this level of noise, an SRGO could potentially detect mHz GW events involving supermassive black holes starting from within our galaxy, up to galaxy merger events at high redshifts.

The results of Sec.~\ref{sect_param_estim} prove that even a single SRGO can, in principle, perform accurate GW parameter estimation, being able to provide a closed region on a sky map for the GW source localization, which would improve with increasing PSNR. In Sec.~\ref{estim_variation}, we find that an effective PSNR (i.e. true PSNR times the square root of the total number of data points) of at least $\sim 80$ would be required to constrain parameters effectively with a single SRGO, which may be achieved by a combination of noise reduction and increasing the data measurement rate. At this effective PSNR or higher, a single SRGO would be capable of constraining the GW source parameters (such as the sky localization area, relative distance and mass estimations, etc.) to within a few tens of percent of their true values. In Sec.~\ref{param_degen}, we obtain more physical insights by studying the parameter degeneracies of an SRGO experiment.

Finally, in Sec.~\ref{discuss}, we discuss the limitations of this study; justify some approaches we have taken in this study; answer fundamental questions about the working principle of an SRGO; and discuss future implications of realizing an SRGO. 

In conclusion, SRGO seems promising as a near-future Earth-based GW detector sensitive to the yet undetected mHz GWs. It could complement space-based detectors such as LISA, or even make detections prior to the launch of LISA, assuming that rapid technological development during this decade allows a functional SRGO to be built. The main effort required in this direction would be detailed studies, techniques and technologies to handle noise sources; finding the optimum operation mode of a storage ring for an SRGO experiment; techniques and technologies for the timing data readout. Further studies of single ion storage rings and improvement in vacuum technology would also help.

\begin{acknowledgments}
We acknowledge Saloni Priya, Thorben Schmirander, Roman Schnabel, Mikhail Korobko, Florian Grüner, Wolfgang Hillert, and Velizar Miltchev for fruitful discussions. This research was supported by the Deutsche Forschungsgemeinschaft (DFG, German Research Foundation) under Germany’s Excellence Strategy – EXC 2121 Quantum Universe – 390833306. This work has made use of data from the European Space Agency (ESA) mission
{\it Gaia} (\url{https://www.cosmos.esa.int/gaia}), processed by the {\it Gaia}
Data Processing and Analysis Consortium (DPAC,
\url{https://www.cosmos.esa.int/web/gaia/dpac/consortium}). Funding for the DPAC
has been provided by national institutions, in particular the institutions
participating in the {\it Gaia} Multilateral Agreement. Finally, we acknowledge the arXiv community for some useful correspondence which improved the paper since its first publication on the platform.
\end{acknowledgments}

\appendix
\section{Contribution to $\Delta T_{\rm GW}$ from beam orbit shape distortions}
\label{app_A}

Consider a circular ion beam of radius $R$, which gets distorted into, say, an ellipse with axes $R \pm \Delta R$, where $\frac{\Delta R}{R} = h$ represents the GW strain amplitude, which is much smaller than unity.

The perimeter of a near-circular ellipse is approximated to an excellent accuracy by Ramanujan's formula \cite{math/0506384},

\begin{equation}
    C_{\rm ellipse} = \pi(a+b)\left( 1 + \frac{3\lambda^2}{10 + \sqrt{4-3\lambda^2}} \right).
\end{equation}

Here, $a = R + \Delta R$, $b = R - \Delta R$, $\lambda = \frac{(a-b)}{(a+b)} = h$. The error in Ramanujan's approximation is $\mathcal{O}(h^{10})$. 
Over many revolutions, the ion circulation time deviation will be proportional to a time integral over the difference between the perimeters of the distorted and ideal orbit shapes, 

\begin{equation}
\Delta T_{\rm orbit} \propto \int_{t_{0}}^{t_{0}+T} \left( C_{\rm ellipse} - 2\pi R \right) dt \propto h^2 .    
\end{equation}

This result is, in general, also true for more complex beam orbit shape distortions caused by other sources (such as seismic activity), as long as the corresponding quantity equivalent to $h$ is small.

\section{Corner plot shown as individual joint posterior plots}
\label{app_B}

Due to space constraints, we show in Fig.~\ref{fig_last}, the 36 individual joint posterior pair-plots corresponding to the non-diagonal elements of the $9 \times 9$ corner plot. The diagonal elements of the corner plot have been shown in Fig.~\ref{fig_trace}. In each pair-plot, we show the 1-sigma ($68\%$, green) and 3-sigma ($99.7\%$, purple) highest posterior density (HPD) regions. The true parameter values for this case correspond to those in Fig.~\ref{fig_response}, except the initial binary separation which is 1 AU (corresponding to an initial GW period of 3.1 h). 32 data points have been taken over an observing time of 12 hours at a PSNR of 100. The results of Fig.~\ref{fig_last} are discussed in Sec.~\ref{param_degen}.

\begin{figure*}[h]
     \begin{subfigure}{0.245\linewidth}
         \includegraphics[width=\linewidth, keepaspectratio]{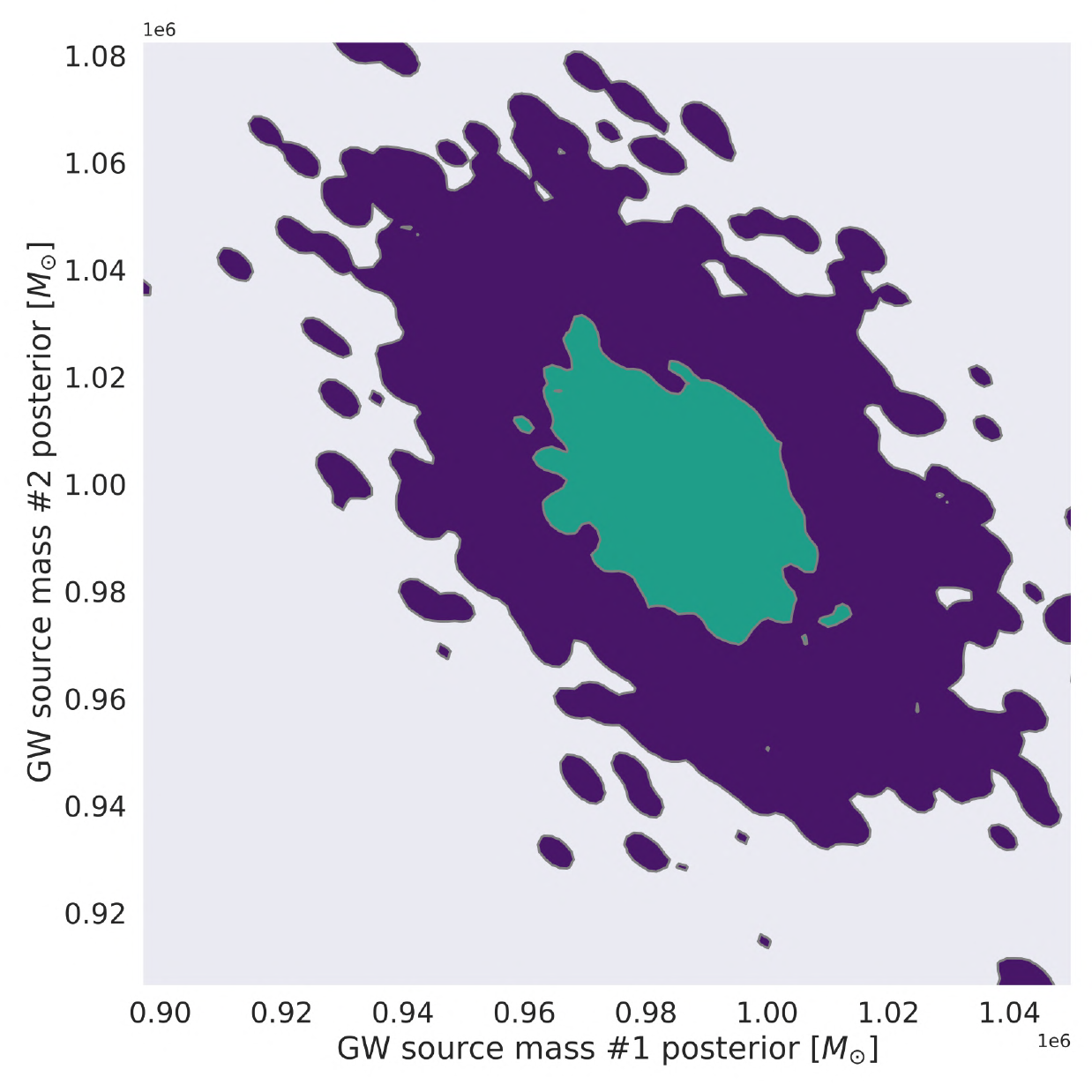}
         \caption{$m_1$ and $m_2$}
         \label{ssubfig1}
     \end{subfigure}
     \begin{subfigure}{0.245\linewidth}
         \includegraphics[width=\linewidth, keepaspectratio]{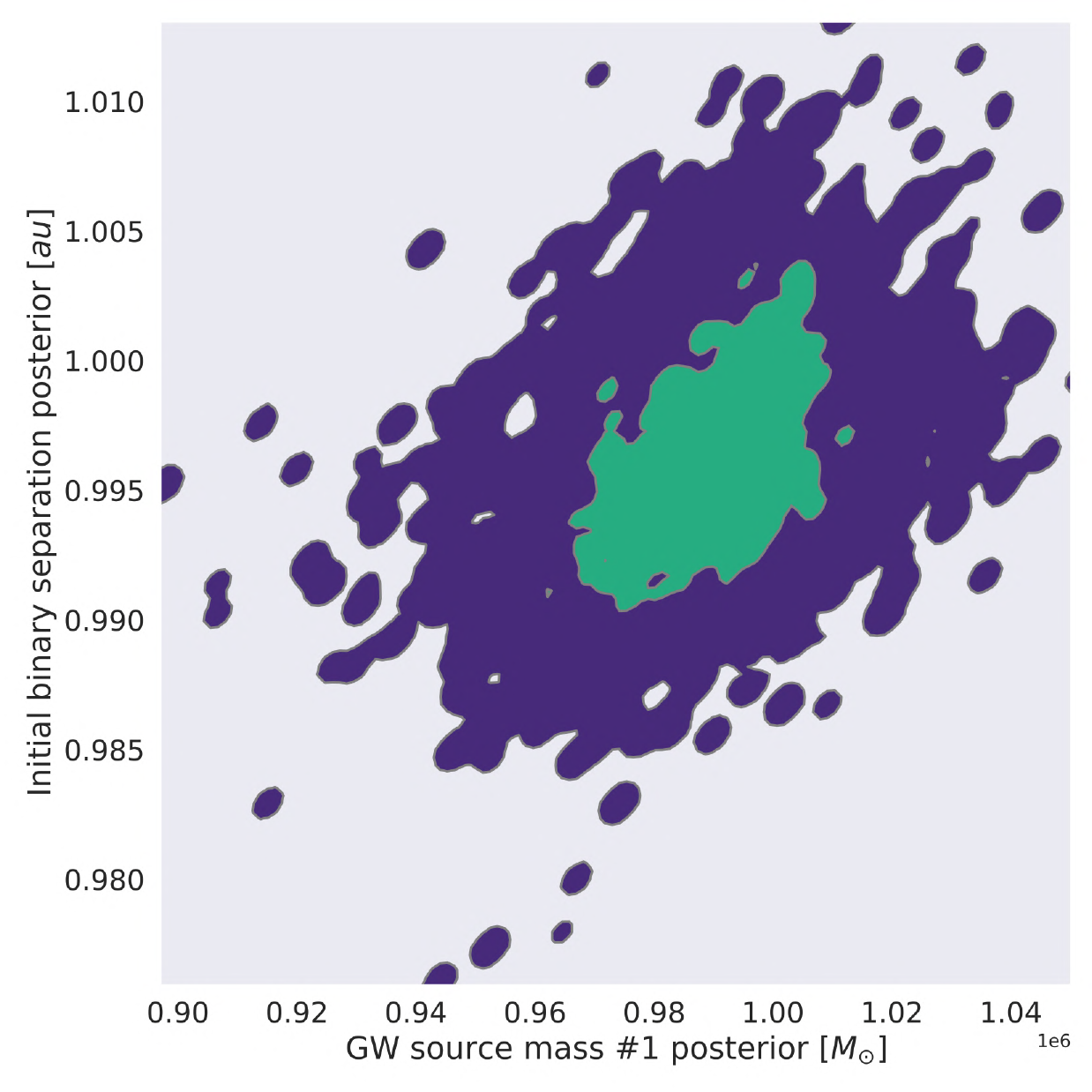}
         \caption{$m_1$ and $r_0$}
         \label{ssubfig2}
     \end{subfigure}
     \begin{subfigure}{0.245\linewidth}
         \includegraphics[width=\linewidth, keepaspectratio]{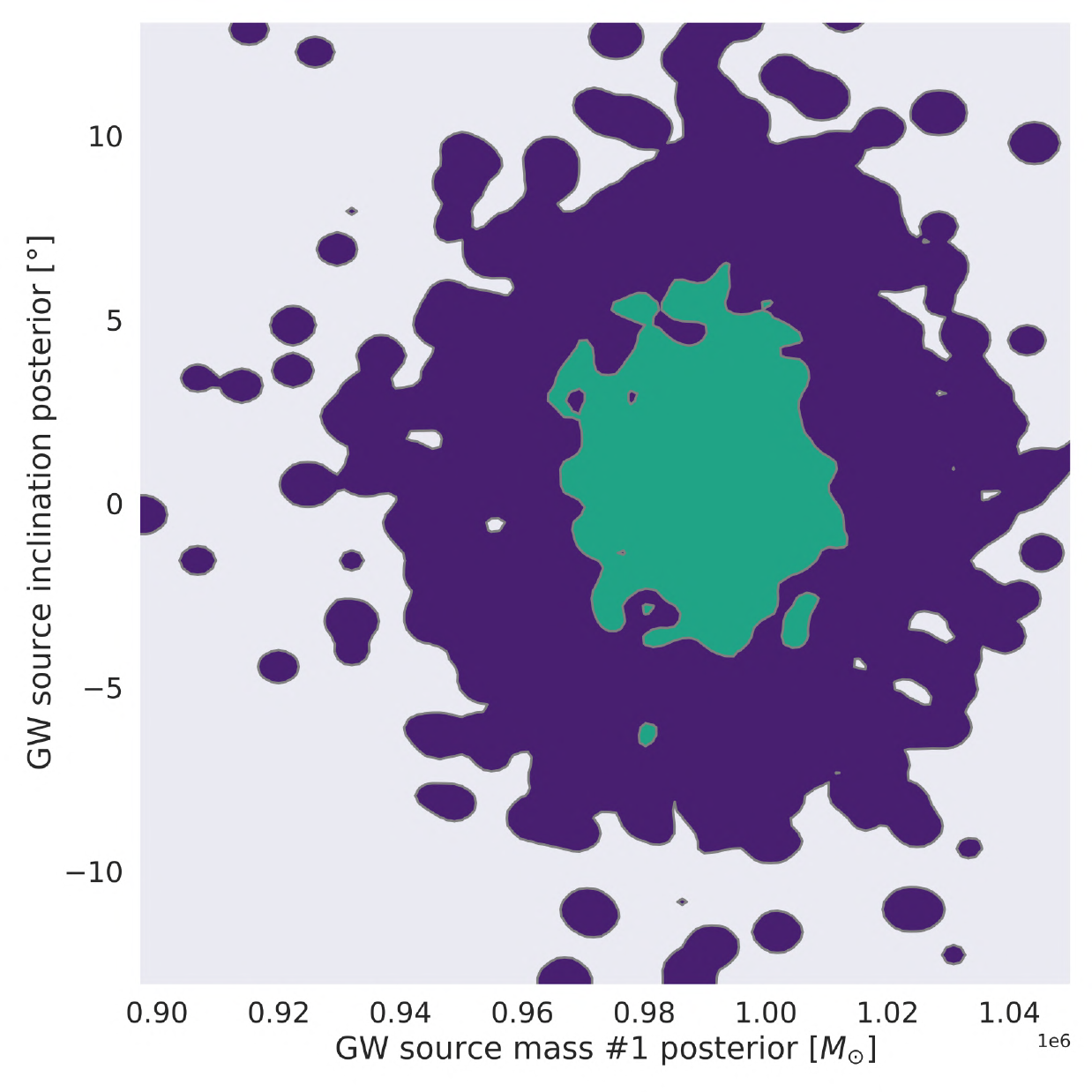}
         \caption{$m_1$ and $i$}
         \label{ssubfig3}
     \end{subfigure}
     \begin{subfigure}{0.245\linewidth}
         \includegraphics[width=\linewidth, keepaspectratio]{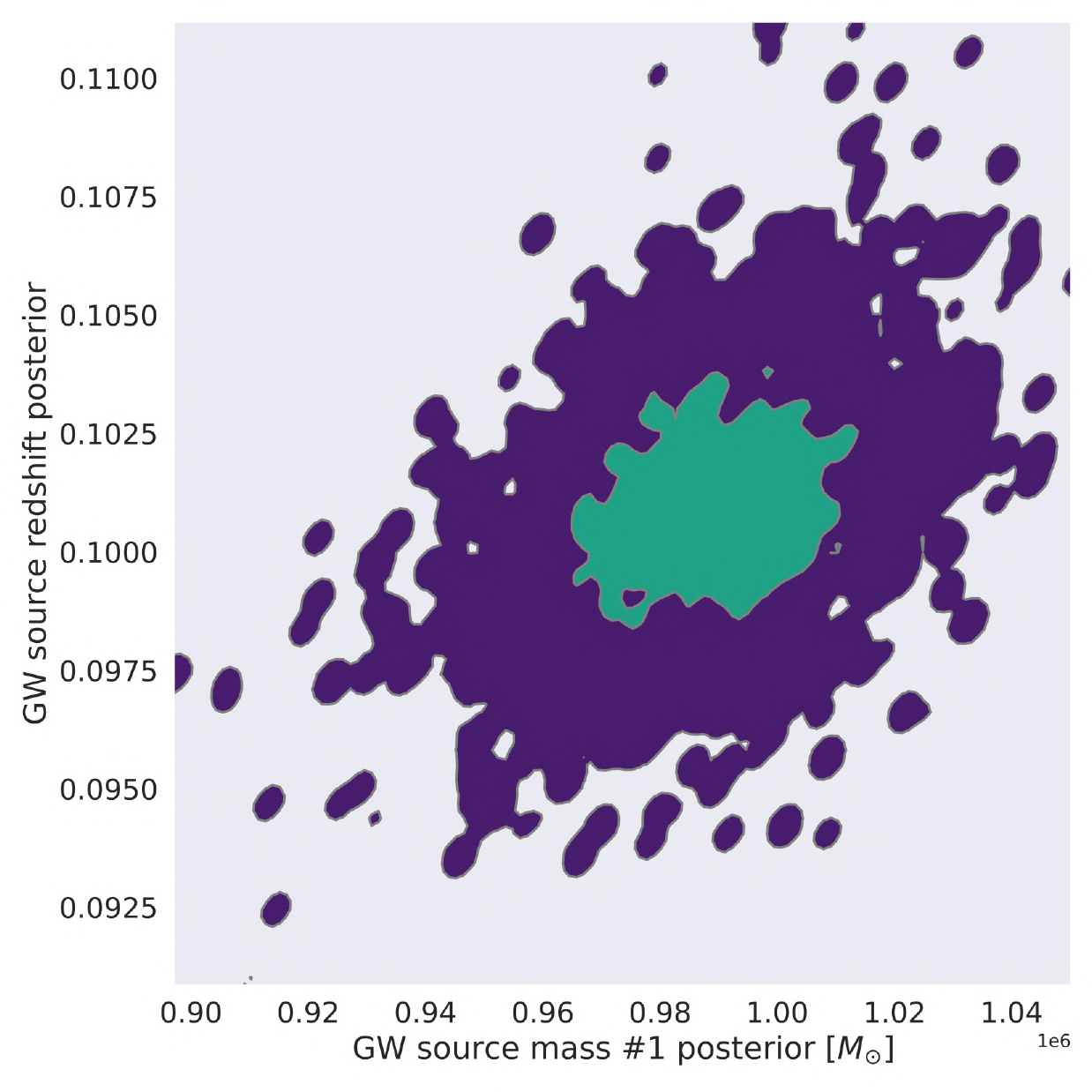}
         \caption{$m_1$ and $z$}
         \label{ssubfig4}
     \end{subfigure}
     \begin{subfigure}{0.245\linewidth}
         \includegraphics[width=\linewidth, keepaspectratio]{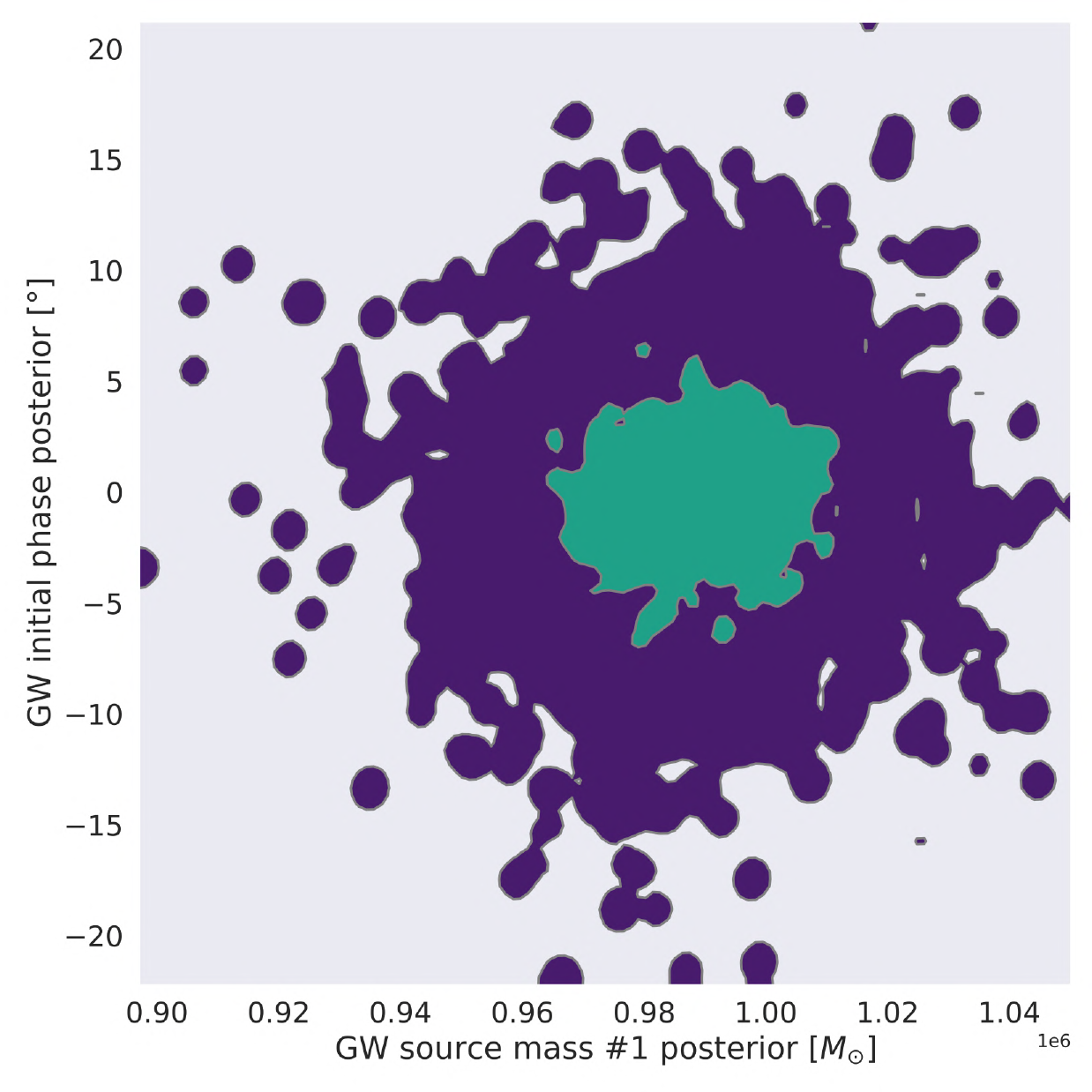}
         \caption{$m_1$ and $\delta_0$}
         \label{ssubfig5}
     \end{subfigure}
     \begin{subfigure}{0.245\linewidth}
         \includegraphics[width=\linewidth, keepaspectratio]{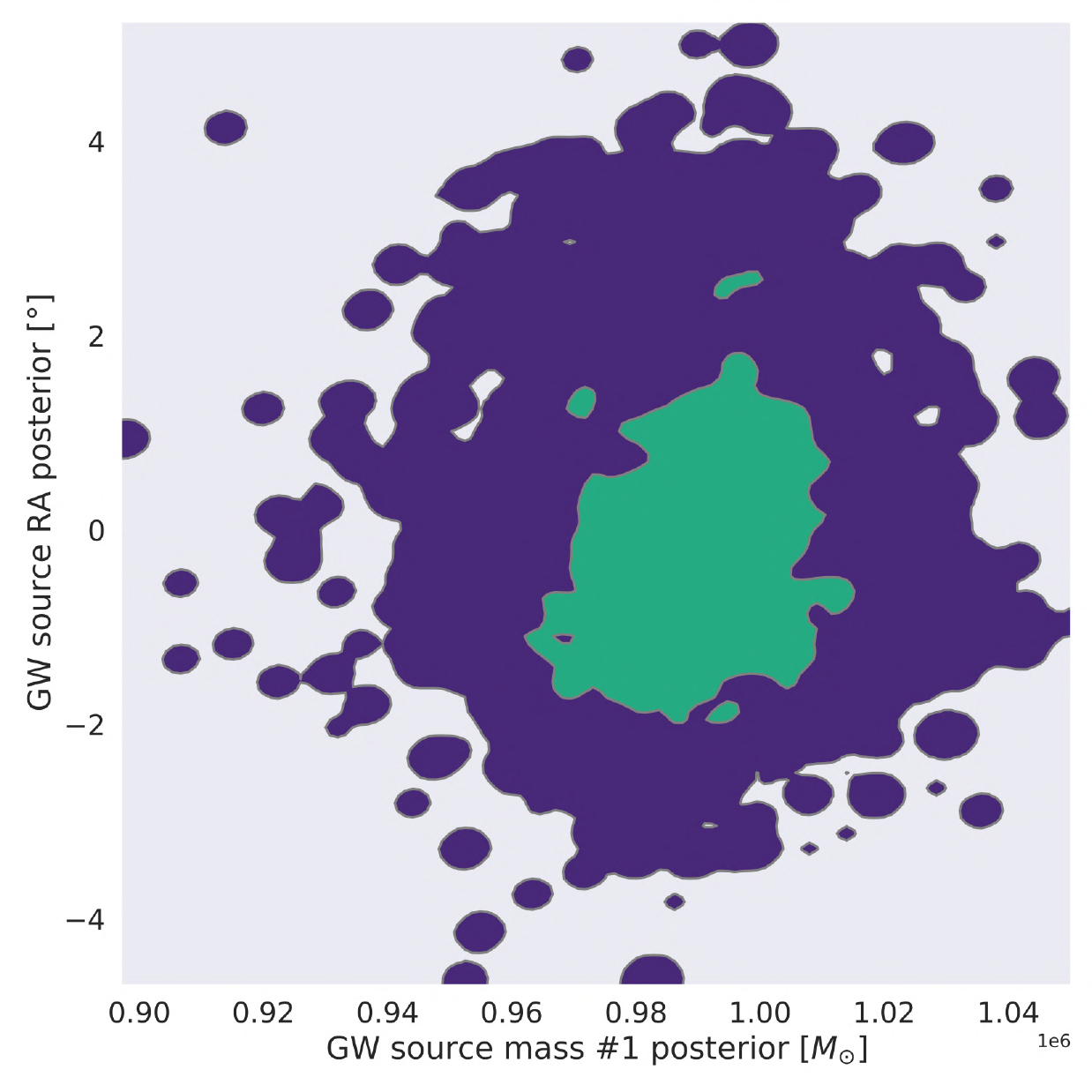}
         \caption{$m_1$ and $\alpha_{src}$}
         \label{ssubfig6}
     \end{subfigure}
     \begin{subfigure}{0.245\linewidth}
         \includegraphics[width=\linewidth, keepaspectratio]{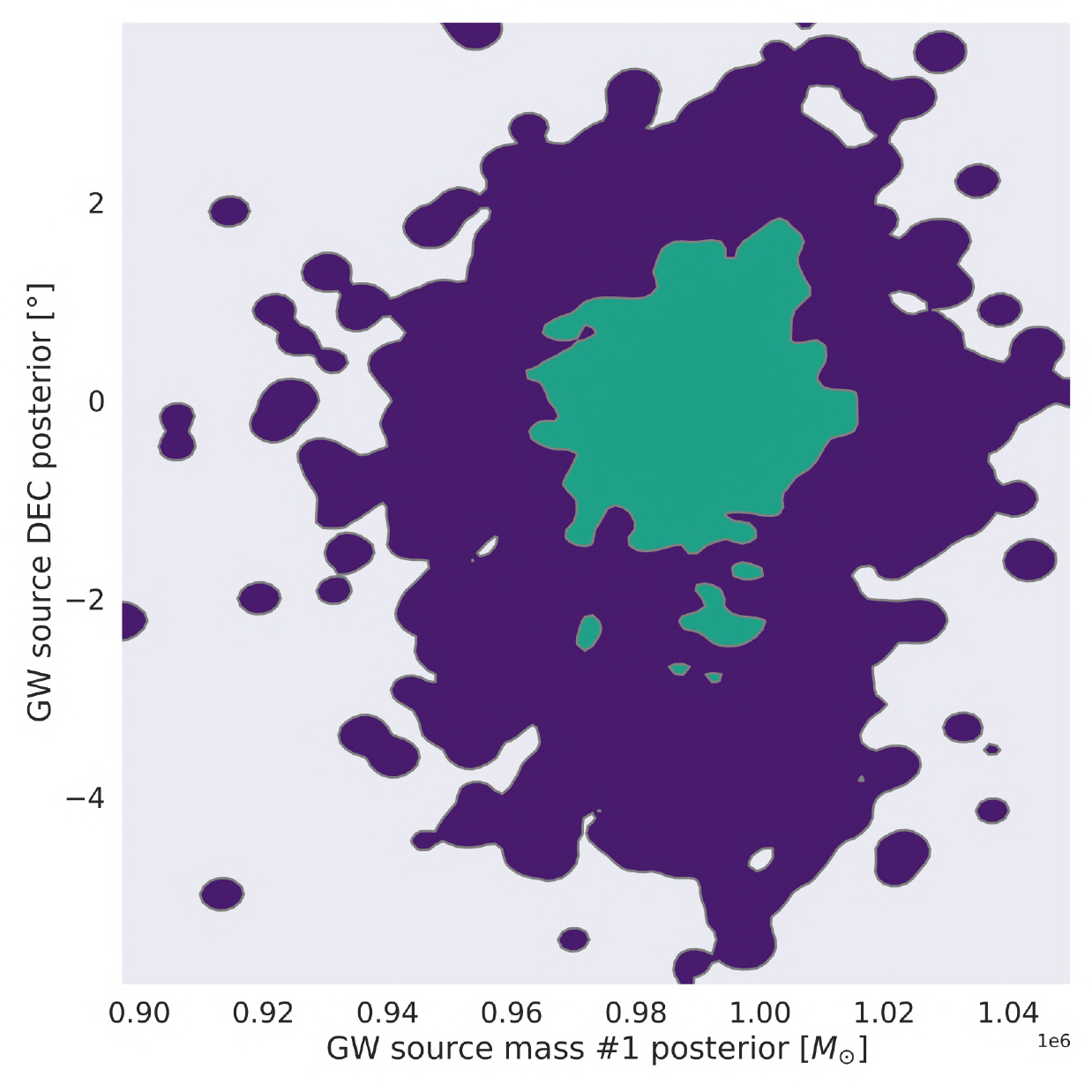}
         \caption{$m_1$ and $\delta_{src}$}
         \label{ssubfig7}
     \end{subfigure}
     \begin{subfigure}{0.245\linewidth}
         \includegraphics[width=\linewidth, keepaspectratio]{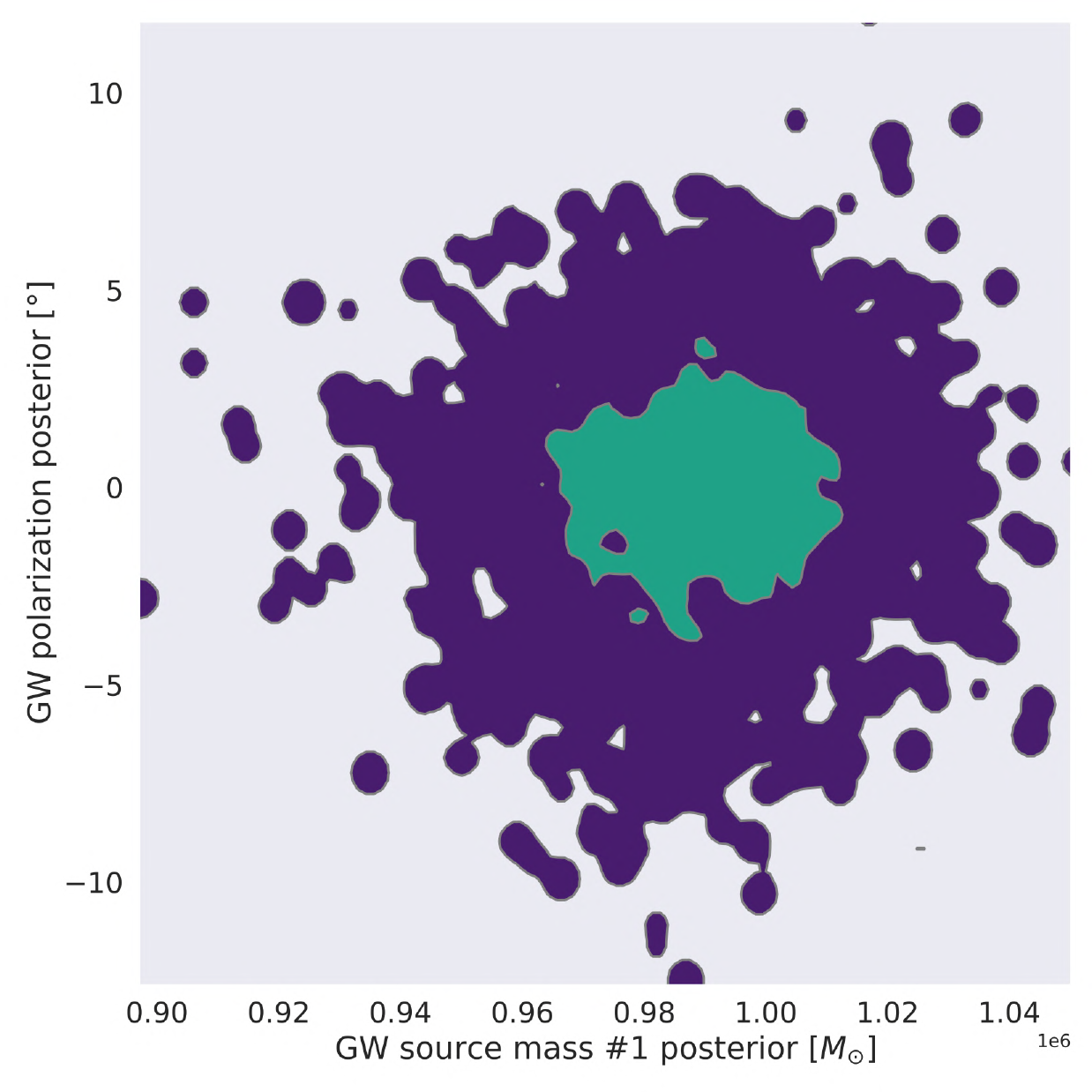}
         \caption{$m_1$ and $\psi_{eq}$}
         \label{ssubfig8}
     \end{subfigure}
     \begin{subfigure}{0.245\linewidth}
         \includegraphics[width=\linewidth, keepaspectratio]{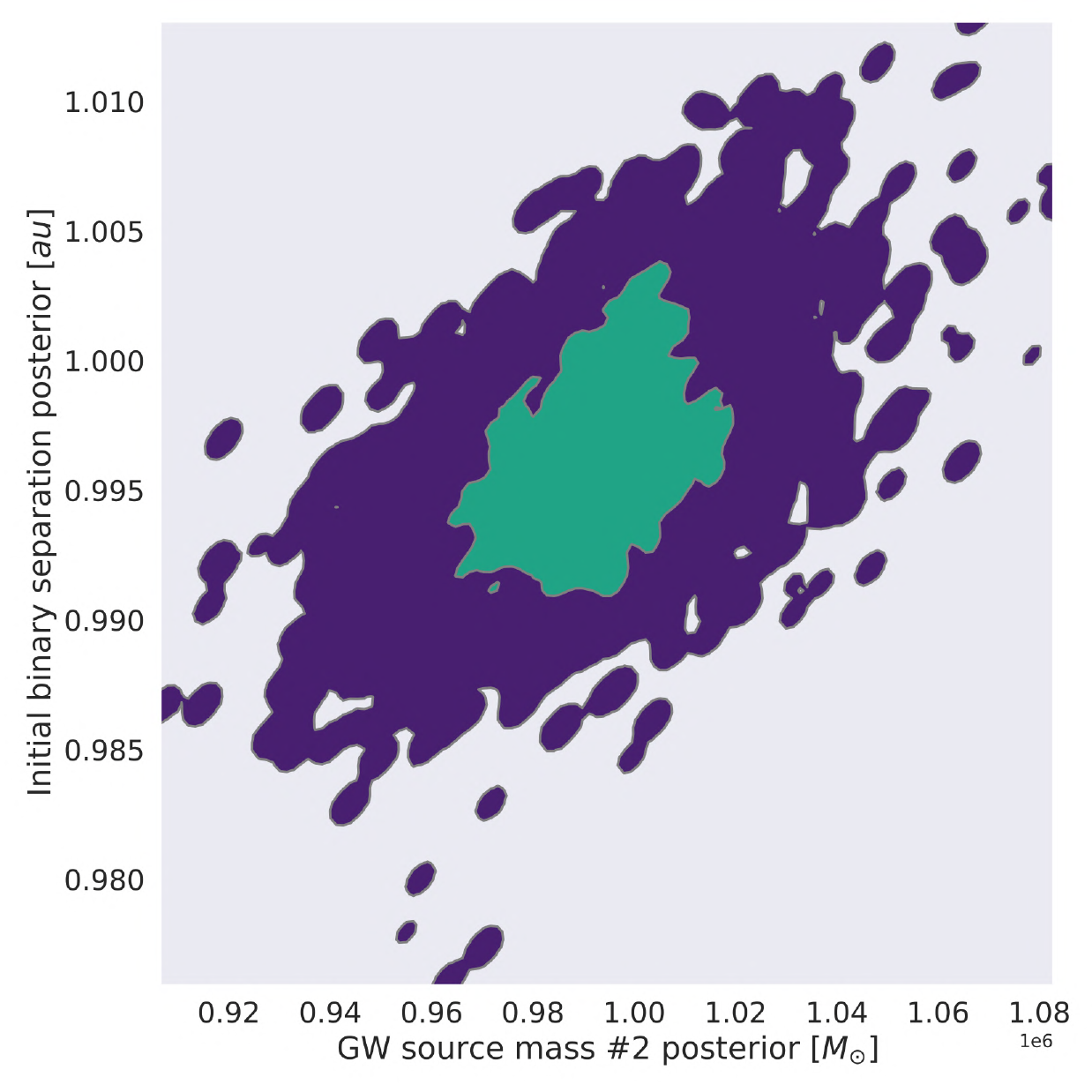}
         \caption{$m_2$ and $r_0$}
         \label{ssubfig9}
     \end{subfigure}
     \begin{subfigure}{0.245\linewidth}
         \includegraphics[width=\linewidth, keepaspectratio]{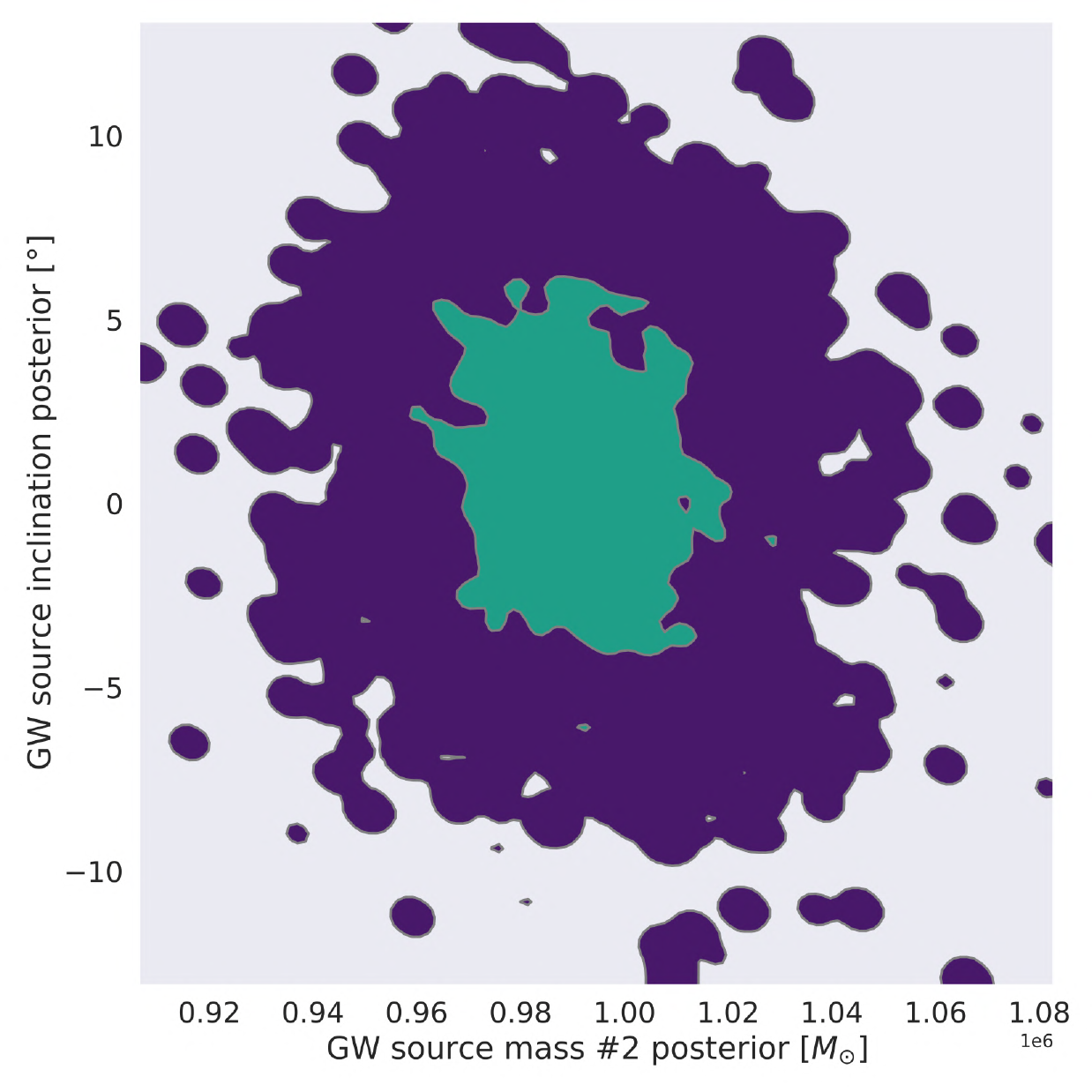}
         \caption{$m_2$ and $i$}
         \label{ssubfig10}
     \end{subfigure}
     \begin{subfigure}{0.245\linewidth}
         \includegraphics[width=\linewidth, keepaspectratio]{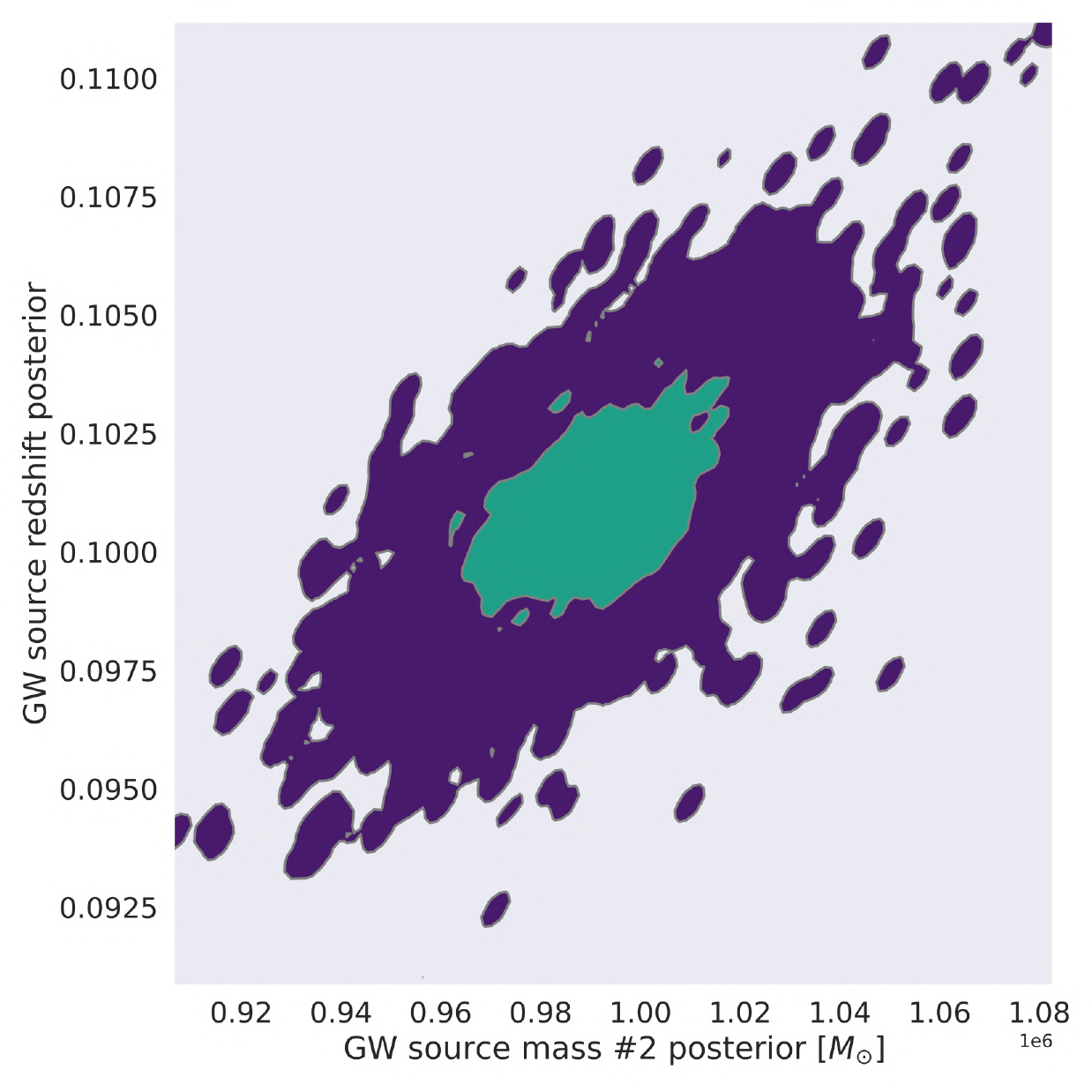}
         \caption{$m_2$ and $z$}
         \label{ssubfig11}
     \end{subfigure}
     \begin{subfigure}{0.245\linewidth}
         \includegraphics[width=\linewidth, keepaspectratio]{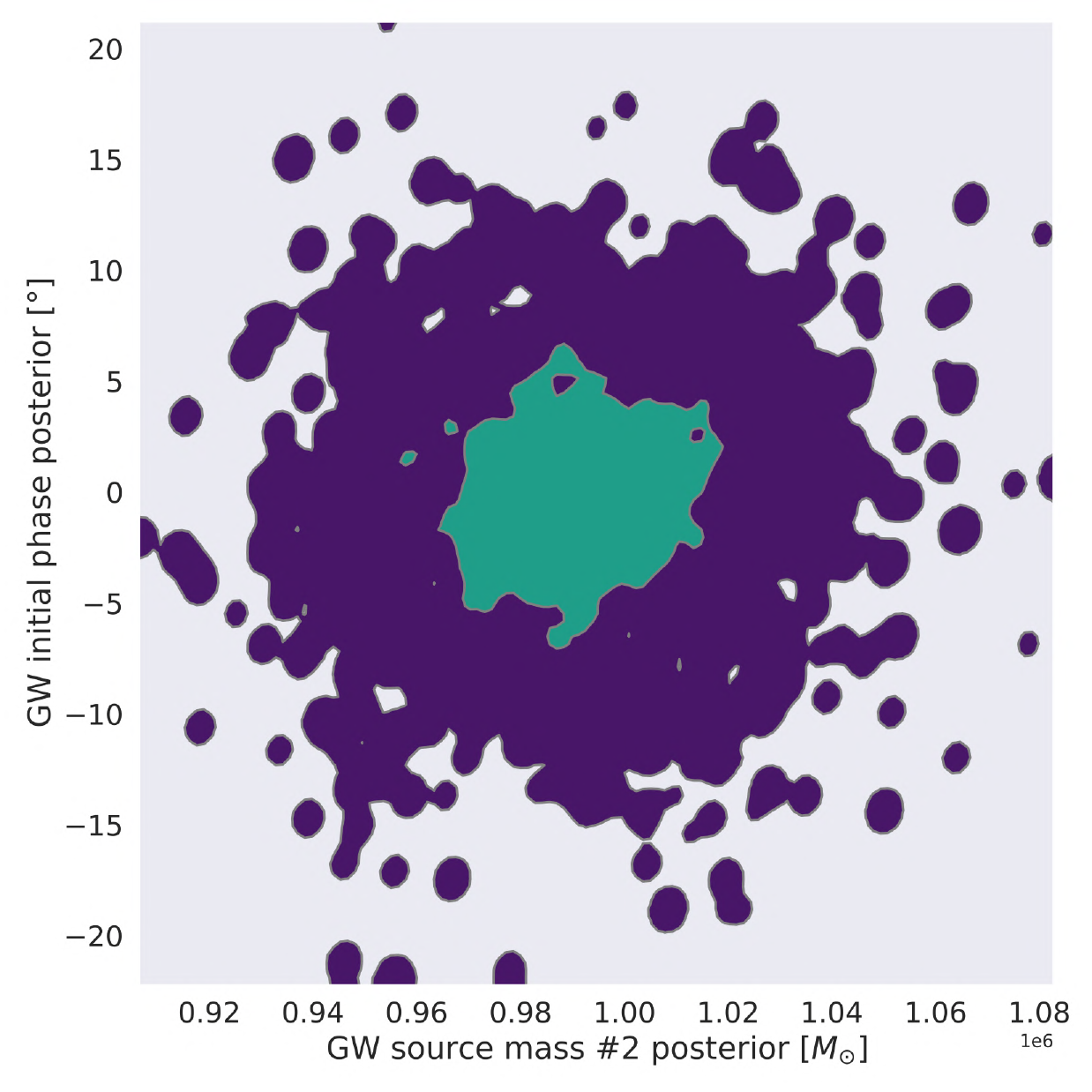}
         \caption{$m_2$ and $\delta_0$}
         \label{ssubfig12}
     \end{subfigure}
     \begin{subfigure}{0.245\linewidth}
         \includegraphics[width=\linewidth, keepaspectratio]{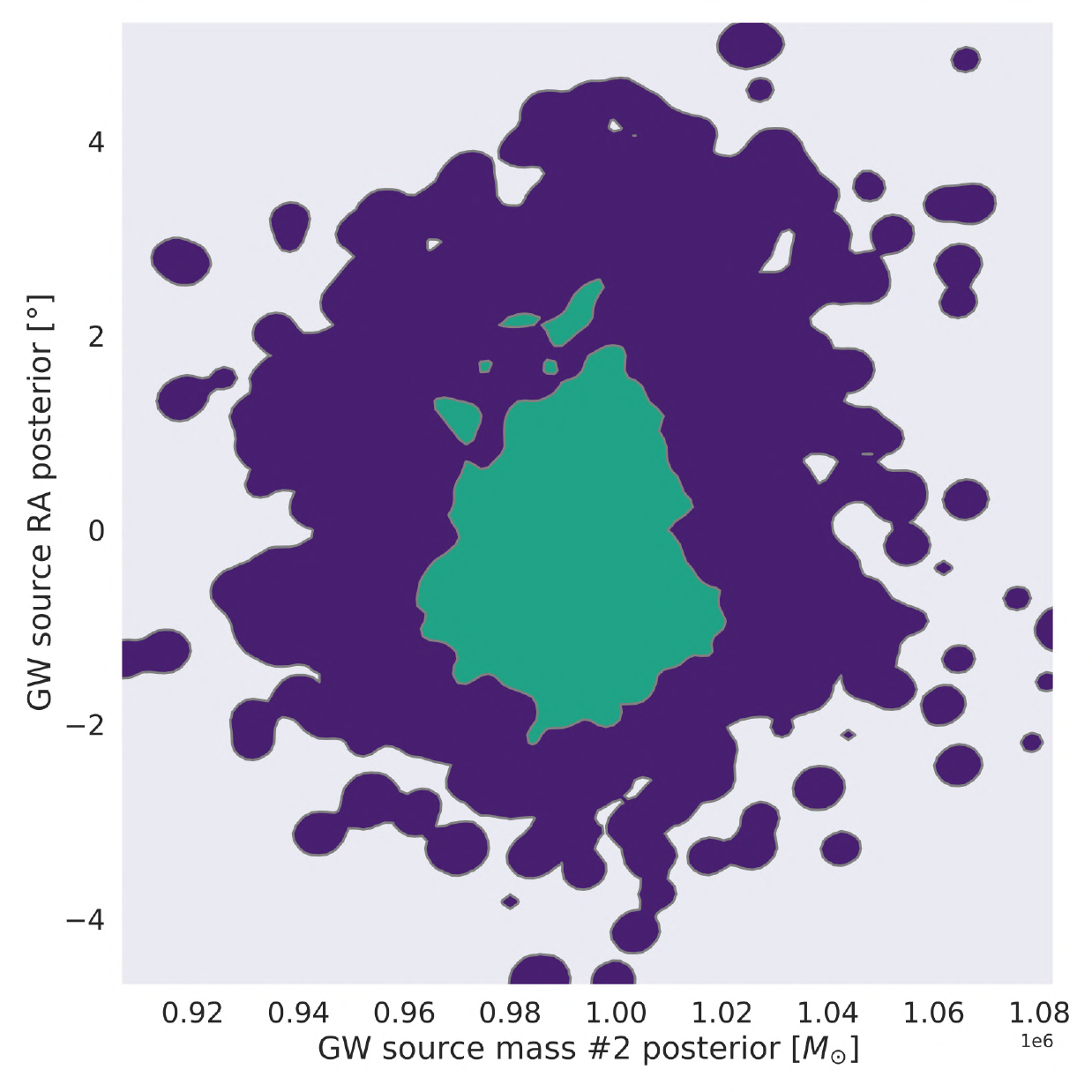}
         \caption{$m_2$ and $\alpha_{src}$}
         \label{ssubfig13}
     \end{subfigure}
     \begin{subfigure}{0.245\linewidth}
         \includegraphics[width=\linewidth, keepaspectratio]{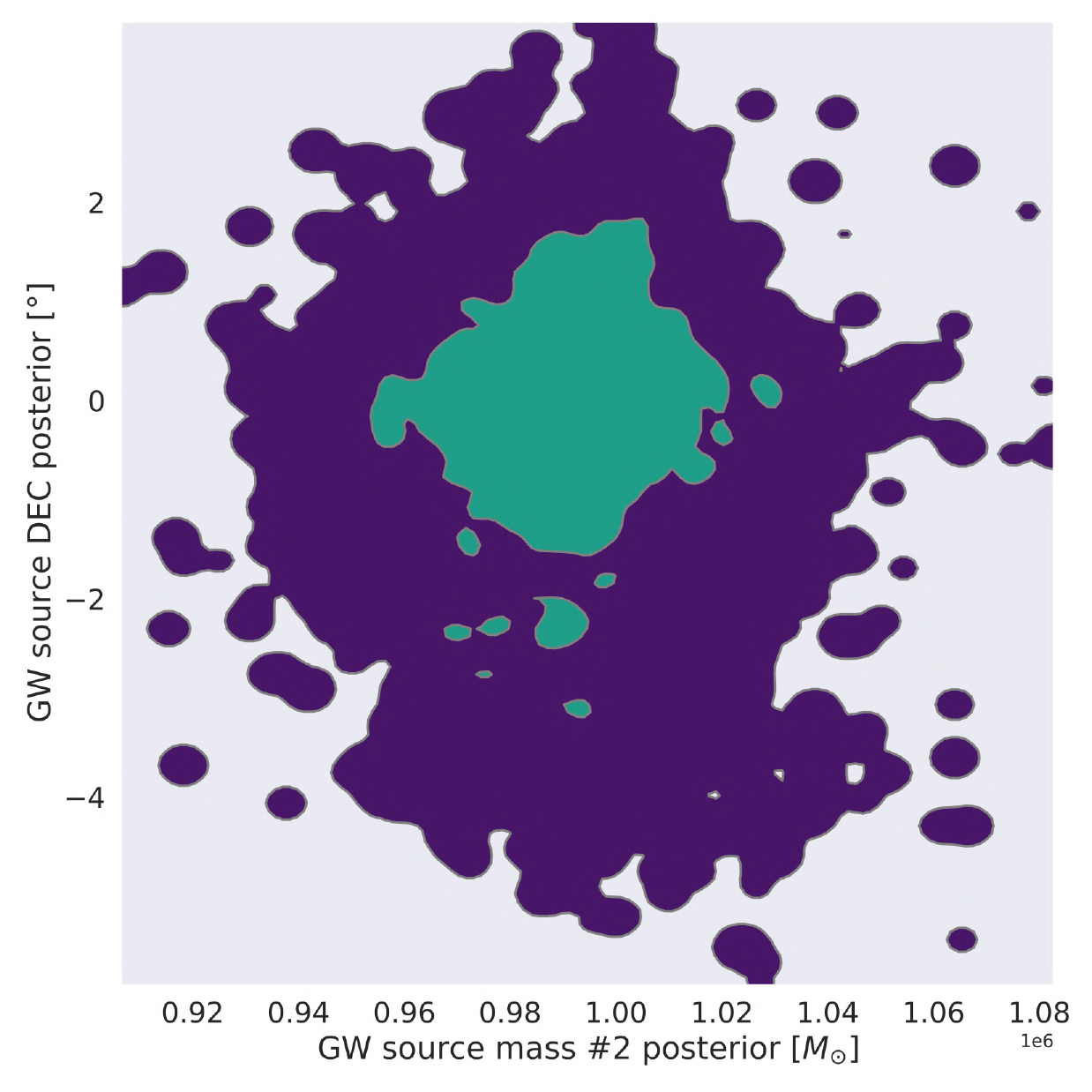}
         \caption{$m_2$ and $\delta_{src}$}
         \label{ssubfig14}
     \end{subfigure}
     \begin{subfigure}{0.245\linewidth}
         \includegraphics[width=\linewidth, keepaspectratio]{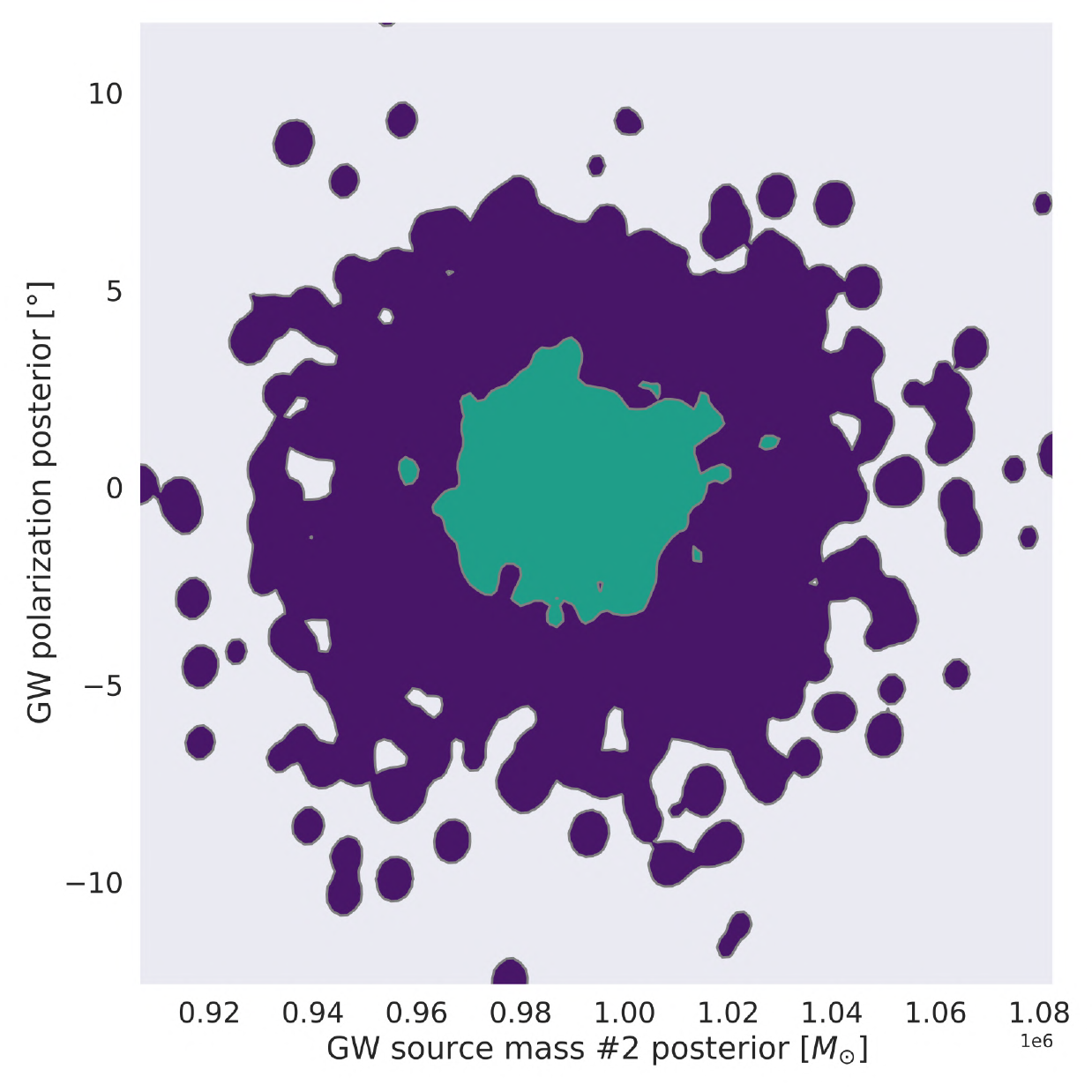}
         \caption{$m_2$ and $\psi_{eq}$}
         \label{ssubfig15}
     \end{subfigure}
     \begin{subfigure}{0.245\linewidth}
         \includegraphics[width=\linewidth, keepaspectratio]{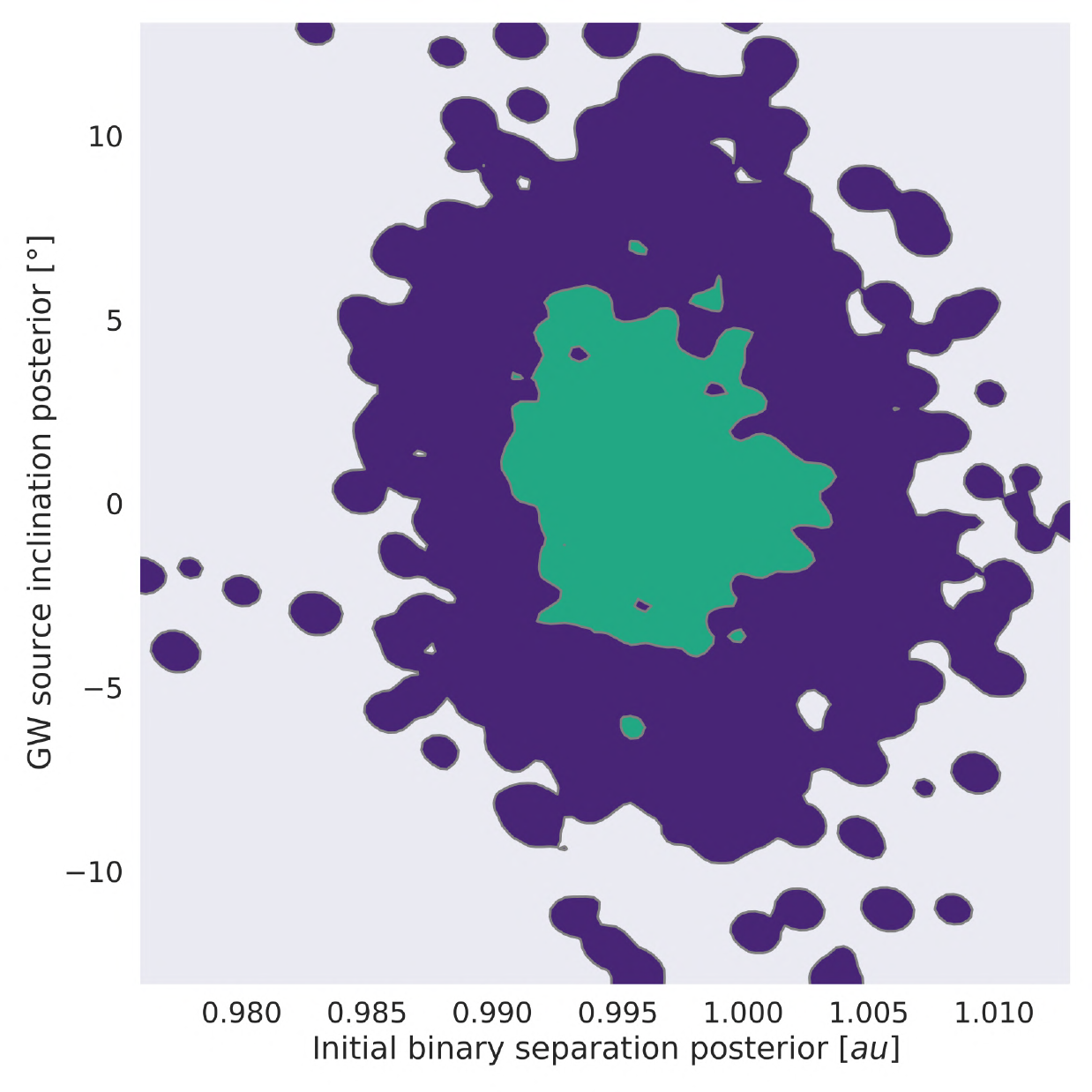}
         \caption{$r_0$ and $i$}
         \label{ssubfig16}
     \end{subfigure}
     \begin{subfigure}{0.245\linewidth}
         \includegraphics[width=\linewidth, keepaspectratio]{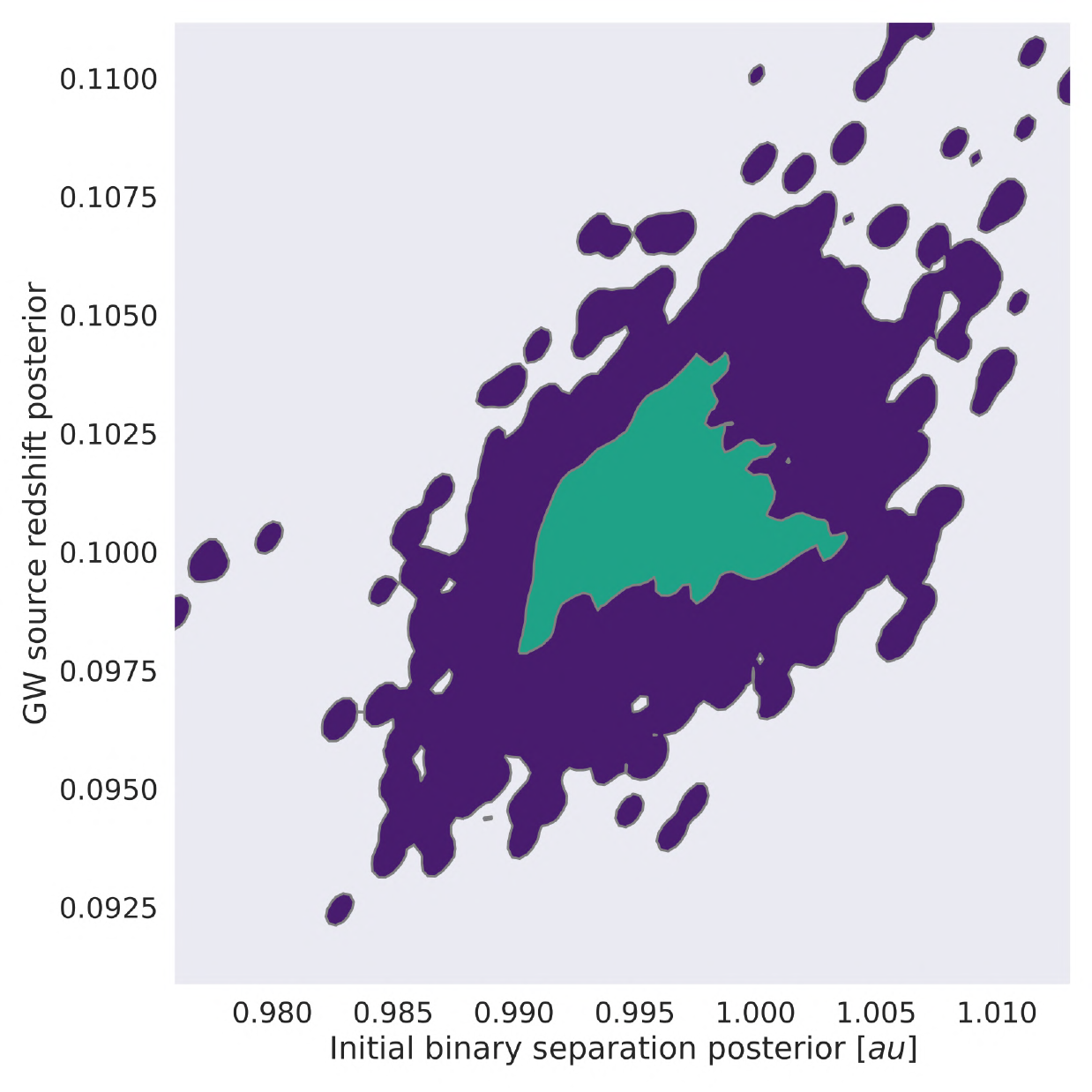}
         \caption{$r_0$ and $z$}
         \label{ssubfig17}
     \end{subfigure}
     \begin{subfigure}{0.245\linewidth}
         \includegraphics[width=\linewidth, keepaspectratio]{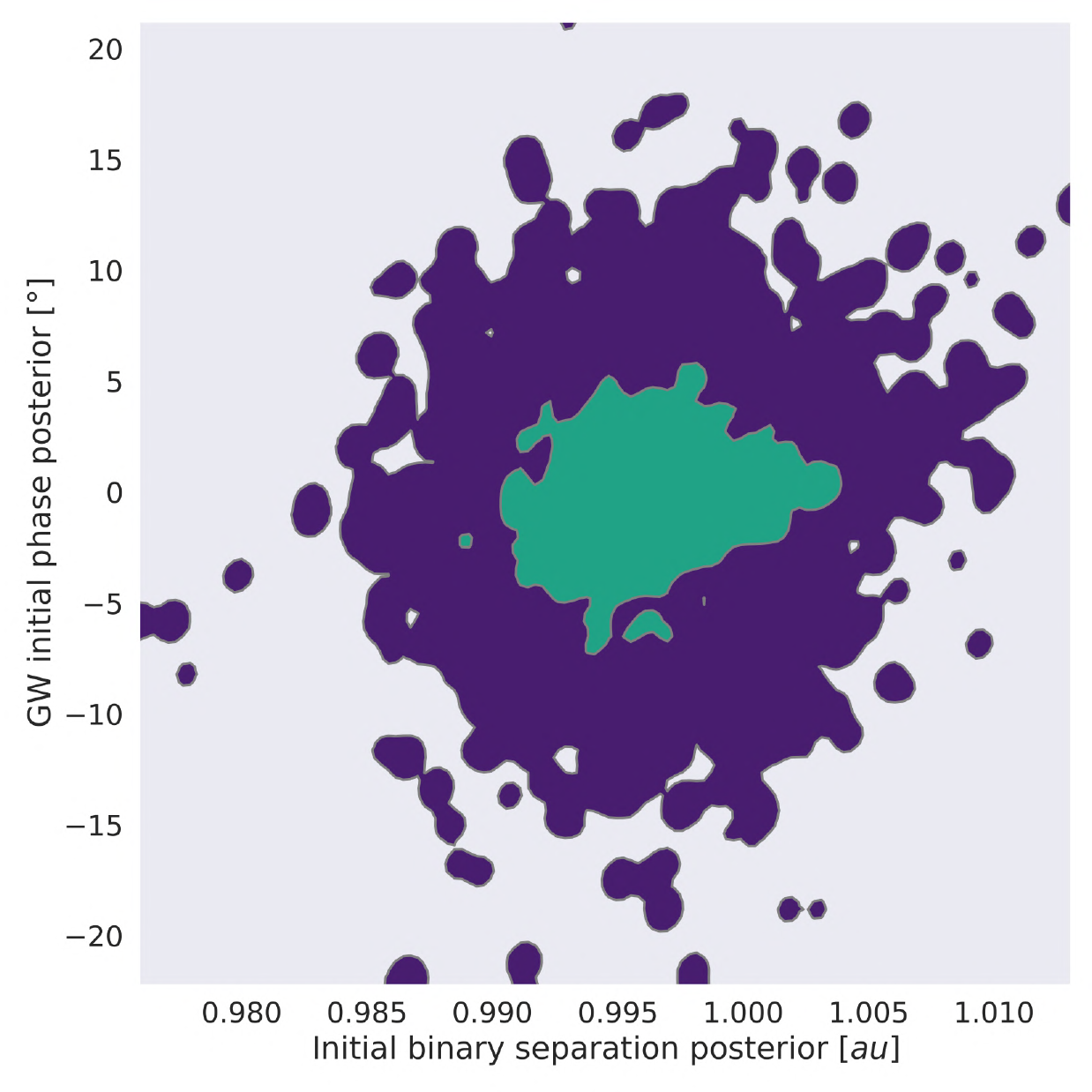}
         \caption{$r_0$ and $\delta_0$}
         \label{ssubfig18}
     \end{subfigure}
     \begin{subfigure}{0.245\linewidth}
         \includegraphics[width=\linewidth, keepaspectratio]{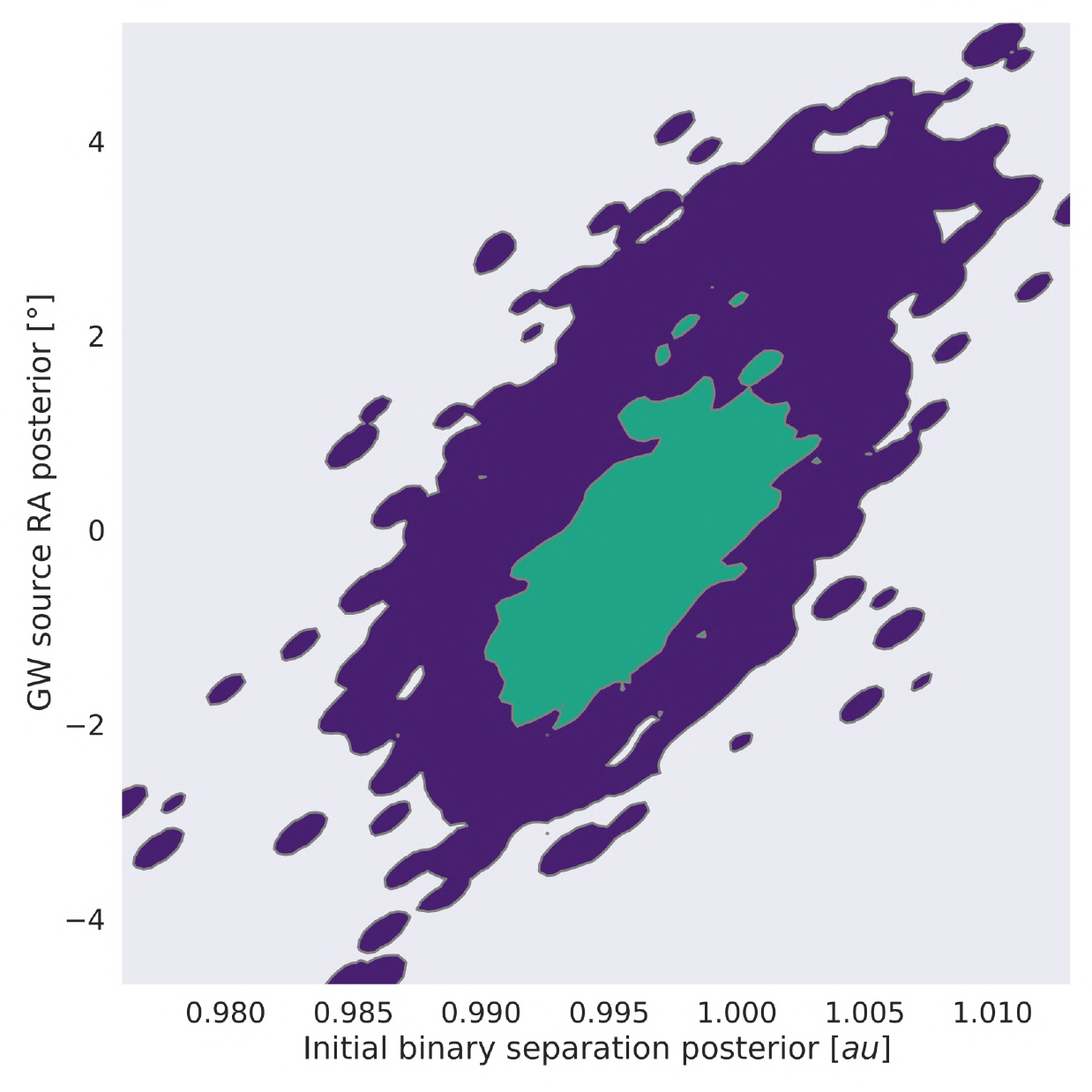}
         \caption{$r_0$ and $\alpha_{src}$}
         \label{ssubfig19}
     \end{subfigure}
     \begin{subfigure}{0.245\linewidth}
         \includegraphics[width=\linewidth, keepaspectratio]{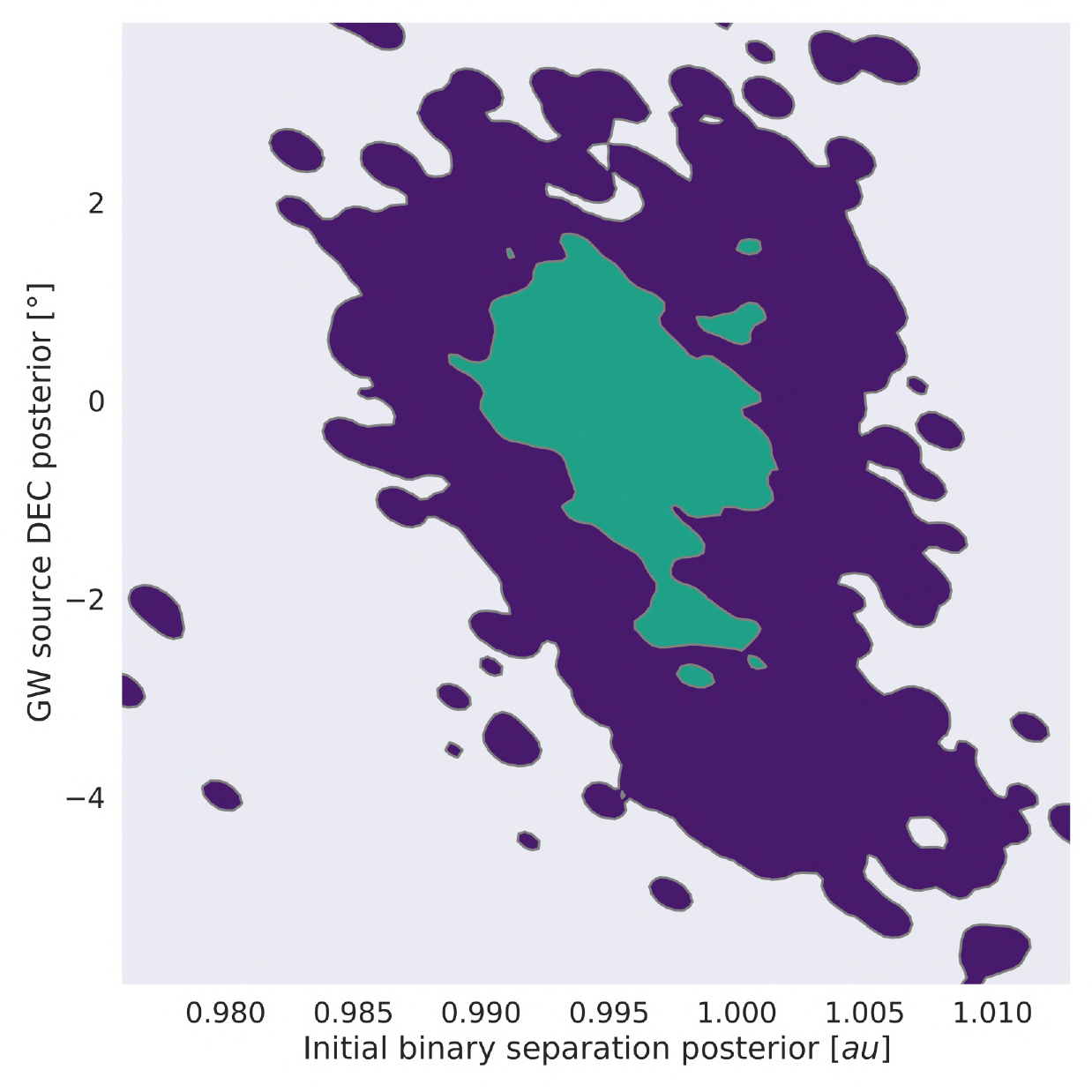}
         \caption{$r_0$ and $\delta_{src}$}
         \label{ssubfig20}
     \end{subfigure}     
\end{figure*}

\begin{figure*}[h]
\ContinuedFloat
     \begin{subfigure}{0.245\linewidth}
         \includegraphics[width=\linewidth, keepaspectratio]{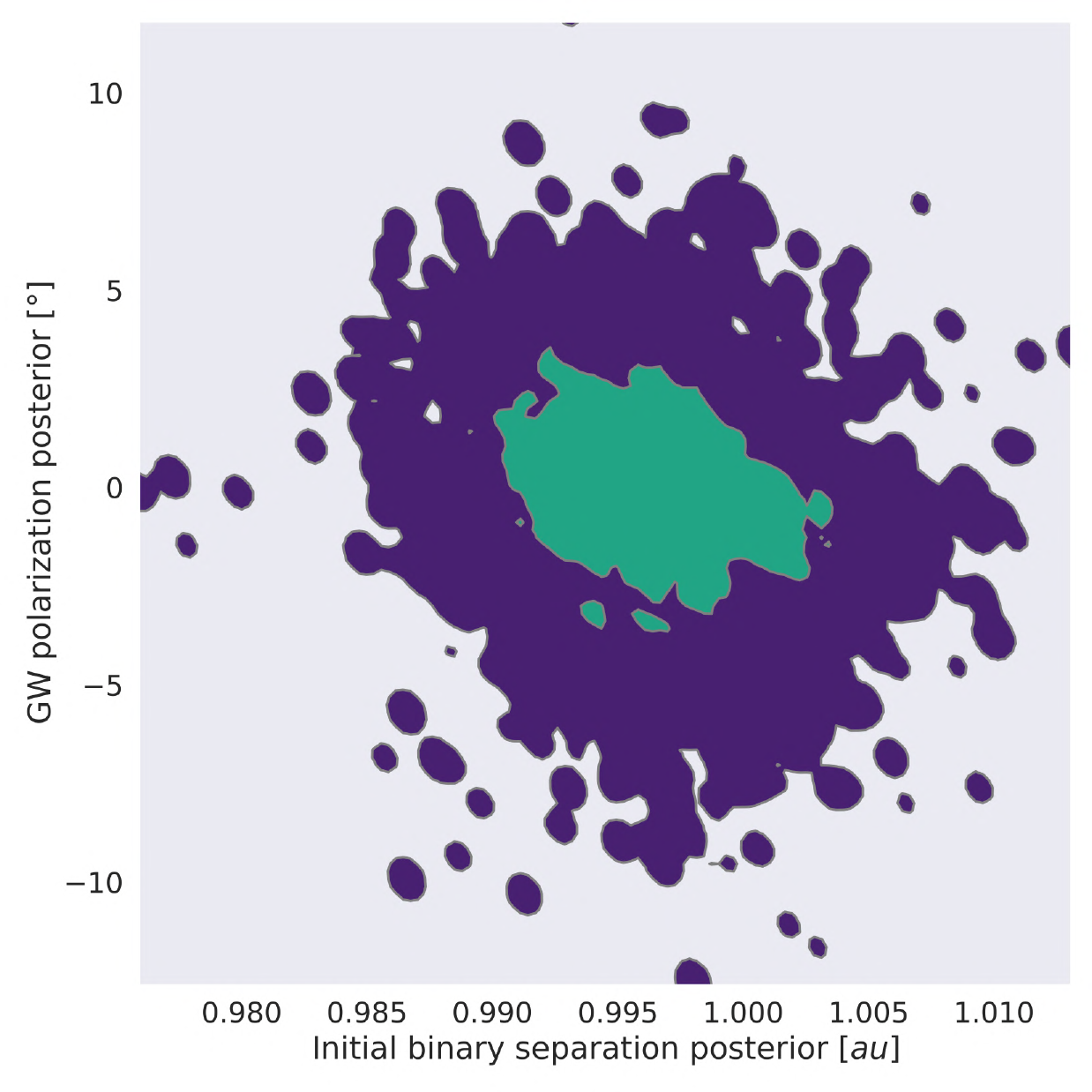}
         \caption{$r_0$ and $\psi_{eq}$}
         \label{ssubfig21}
     \end{subfigure}
     \begin{subfigure}{0.245\linewidth}
         \includegraphics[width=\linewidth, keepaspectratio]{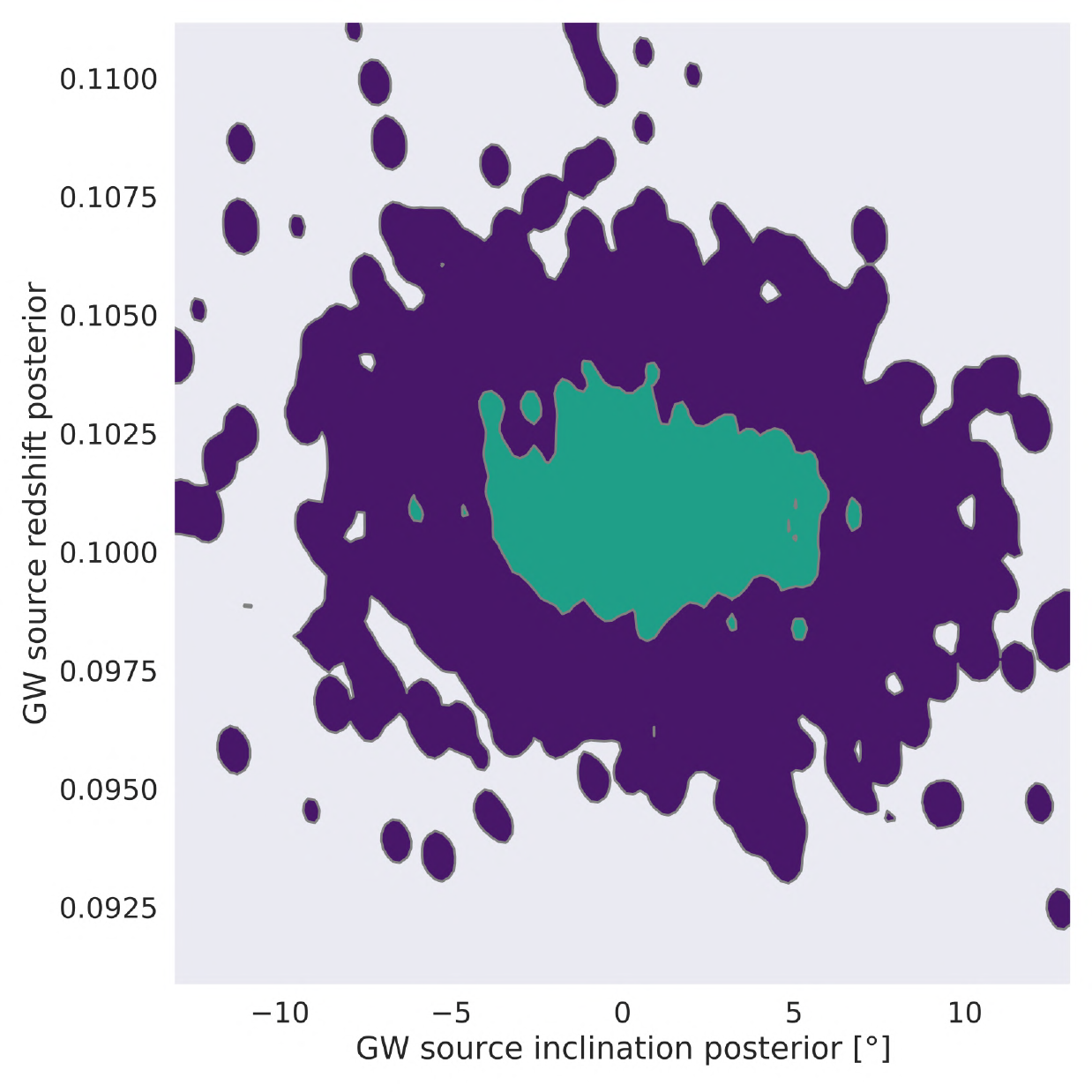}
         \caption{$i$ and $z$}
         \label{ssubfig22}
     \end{subfigure}
     \begin{subfigure}{0.245\linewidth}
         \includegraphics[width=\linewidth, keepaspectratio]{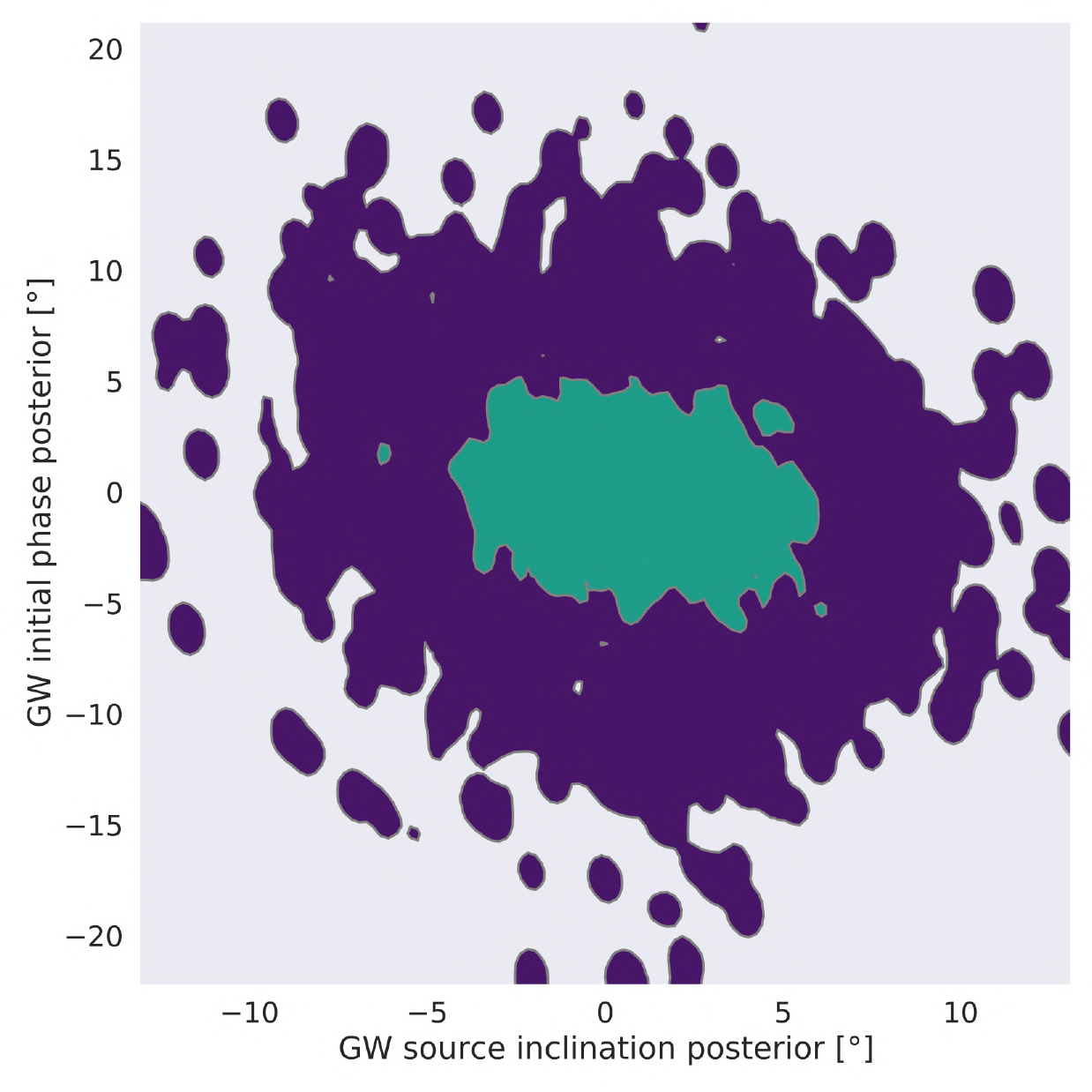}
         \caption{$i$ and $\delta_0$}
         \label{ssubfig23}
     \end{subfigure}
     \begin{subfigure}{0.245\linewidth}
         \includegraphics[width=\linewidth, keepaspectratio]{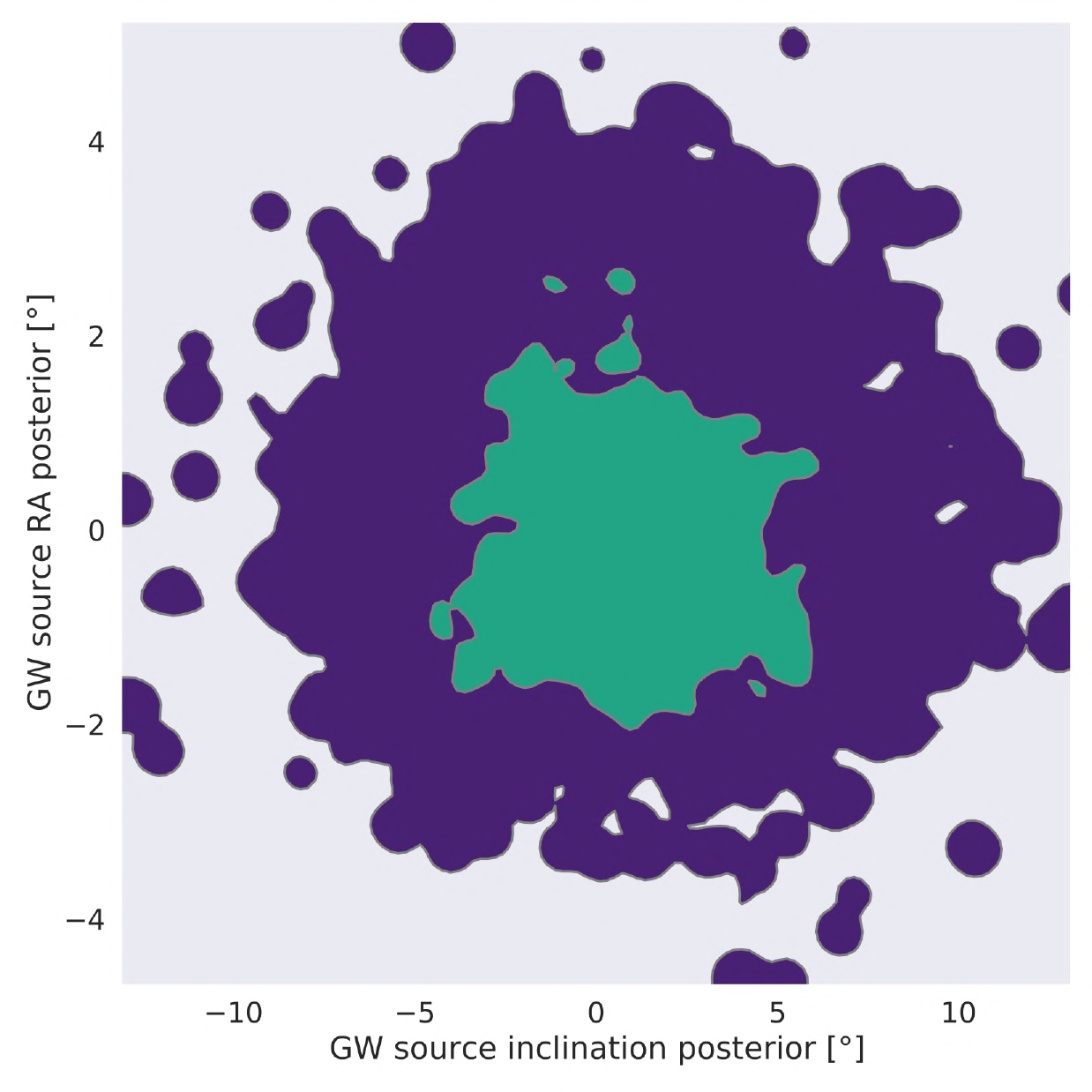}
         \caption{$i$ and $\alpha_{src}$}
         \label{ssubfig24}
     \end{subfigure}
     \begin{subfigure}{0.245\linewidth}
         \includegraphics[width=\linewidth, keepaspectratio]{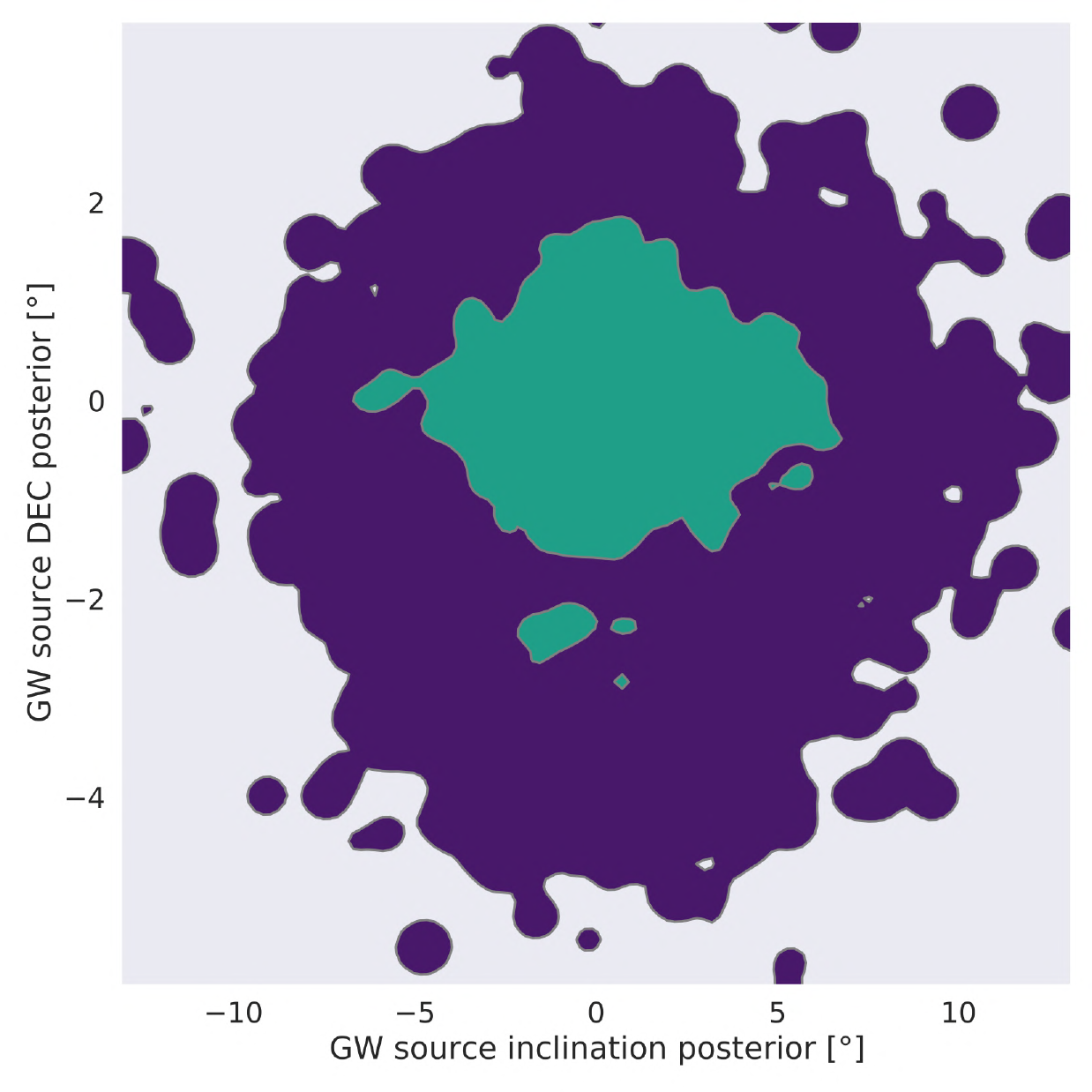}
         \caption{$i$ and $\delta_{src}$}
         \label{ssubfig25}
     \end{subfigure}
     \begin{subfigure}{0.245\linewidth}
         \includegraphics[width=\linewidth, keepaspectratio]{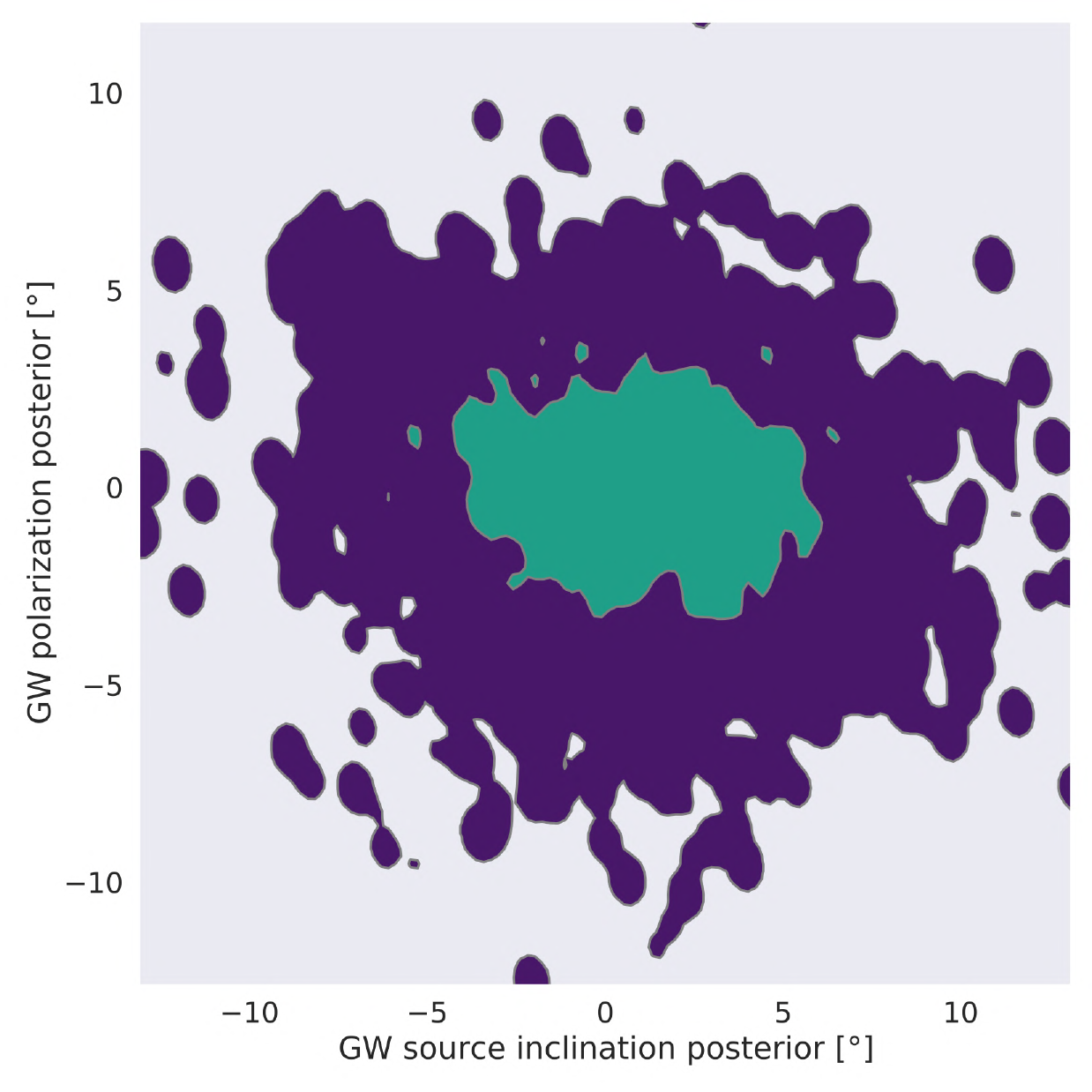}
         \caption{$i$ and $\psi_{eq}$}
         \label{ssubfig26}
     \end{subfigure}
     \begin{subfigure}{0.245\linewidth}
         \includegraphics[width=\linewidth, keepaspectratio]{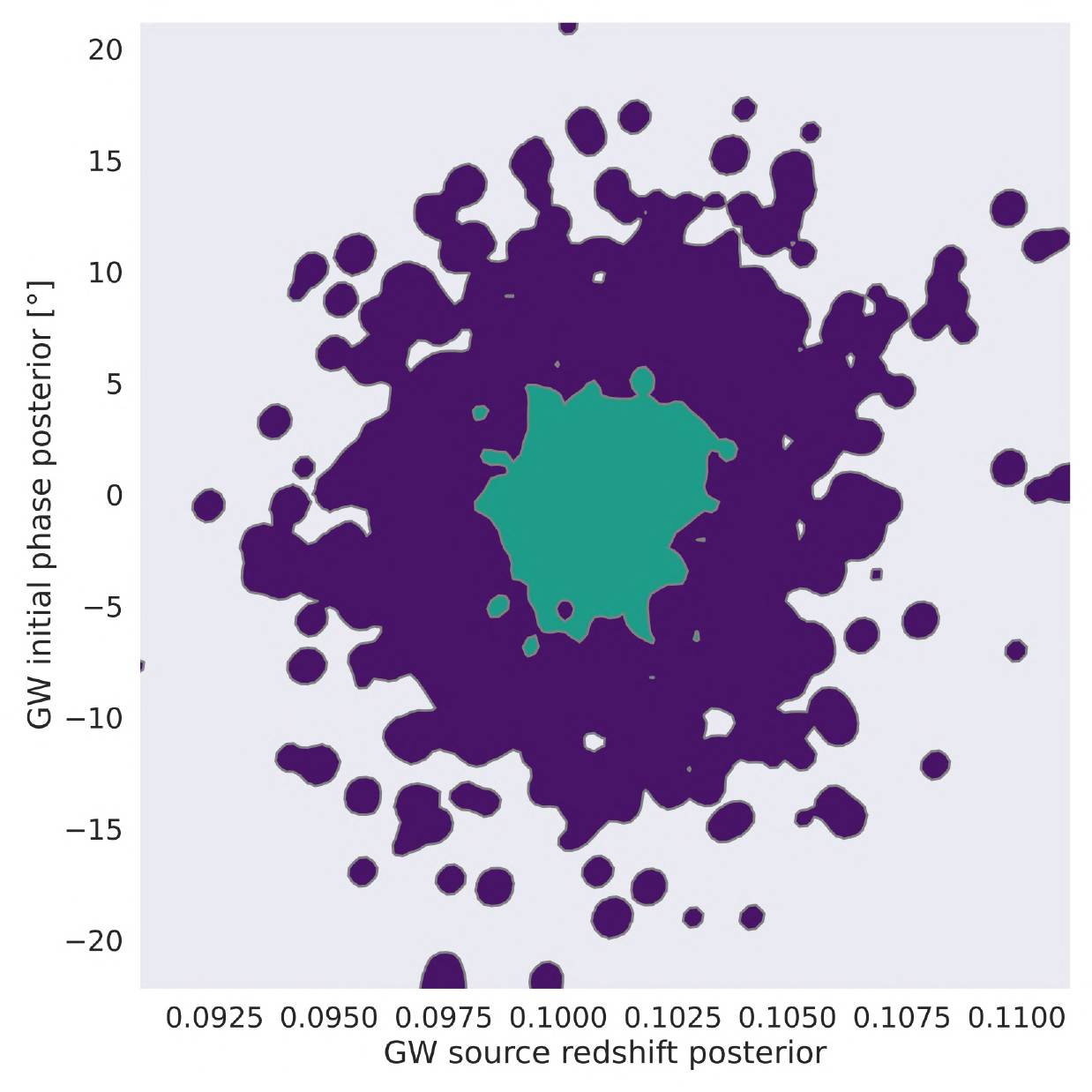}
         \caption{$z$ and $\delta_0$}
         \label{ssubfig27}
     \end{subfigure}
     \begin{subfigure}{0.245\linewidth}
         \includegraphics[width=\linewidth, keepaspectratio]{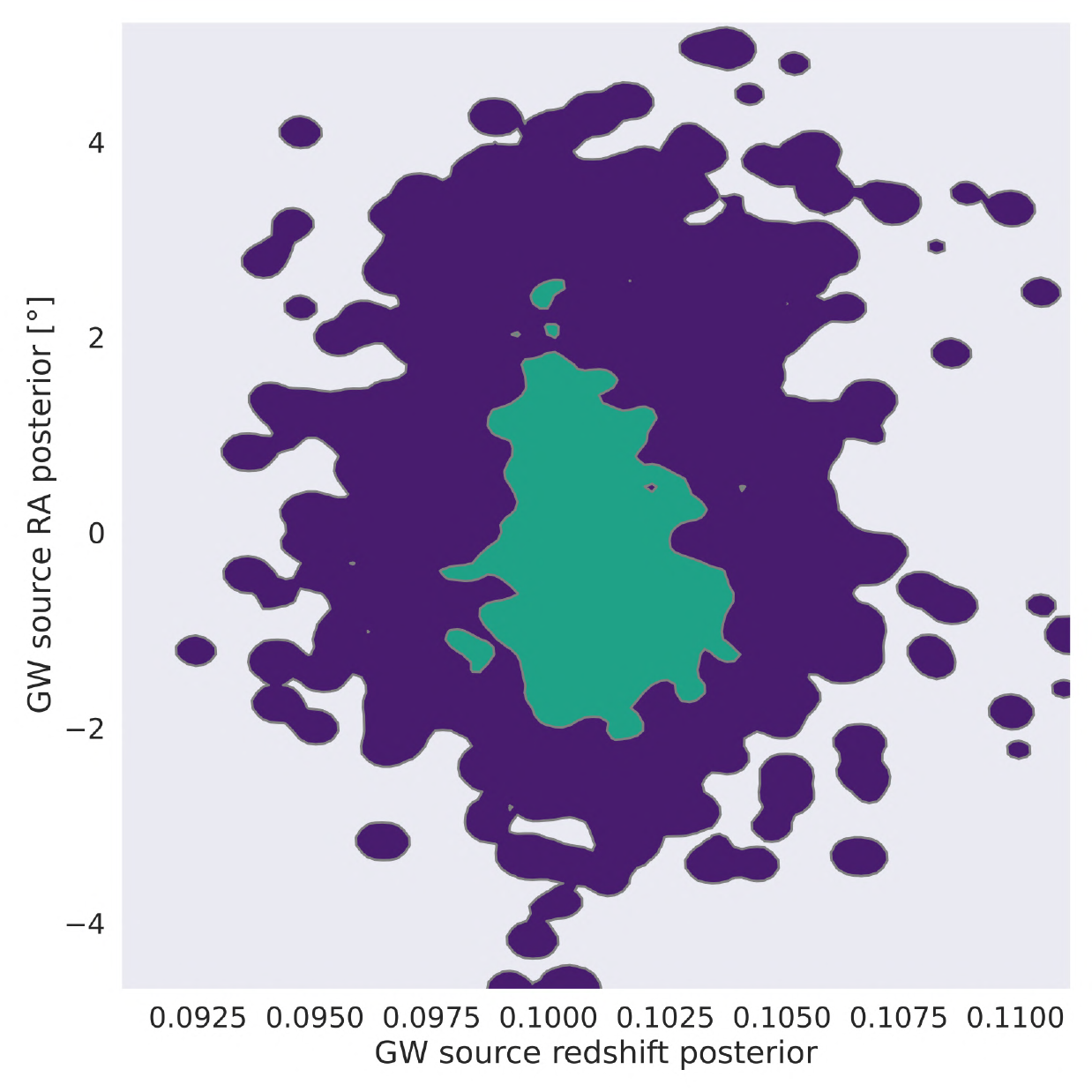}
         \caption{$z$ and $\alpha_{src}$}
         \label{ssubfig28}
     \end{subfigure}
     \begin{subfigure}{0.245\linewidth}
         \includegraphics[width=\linewidth, keepaspectratio]{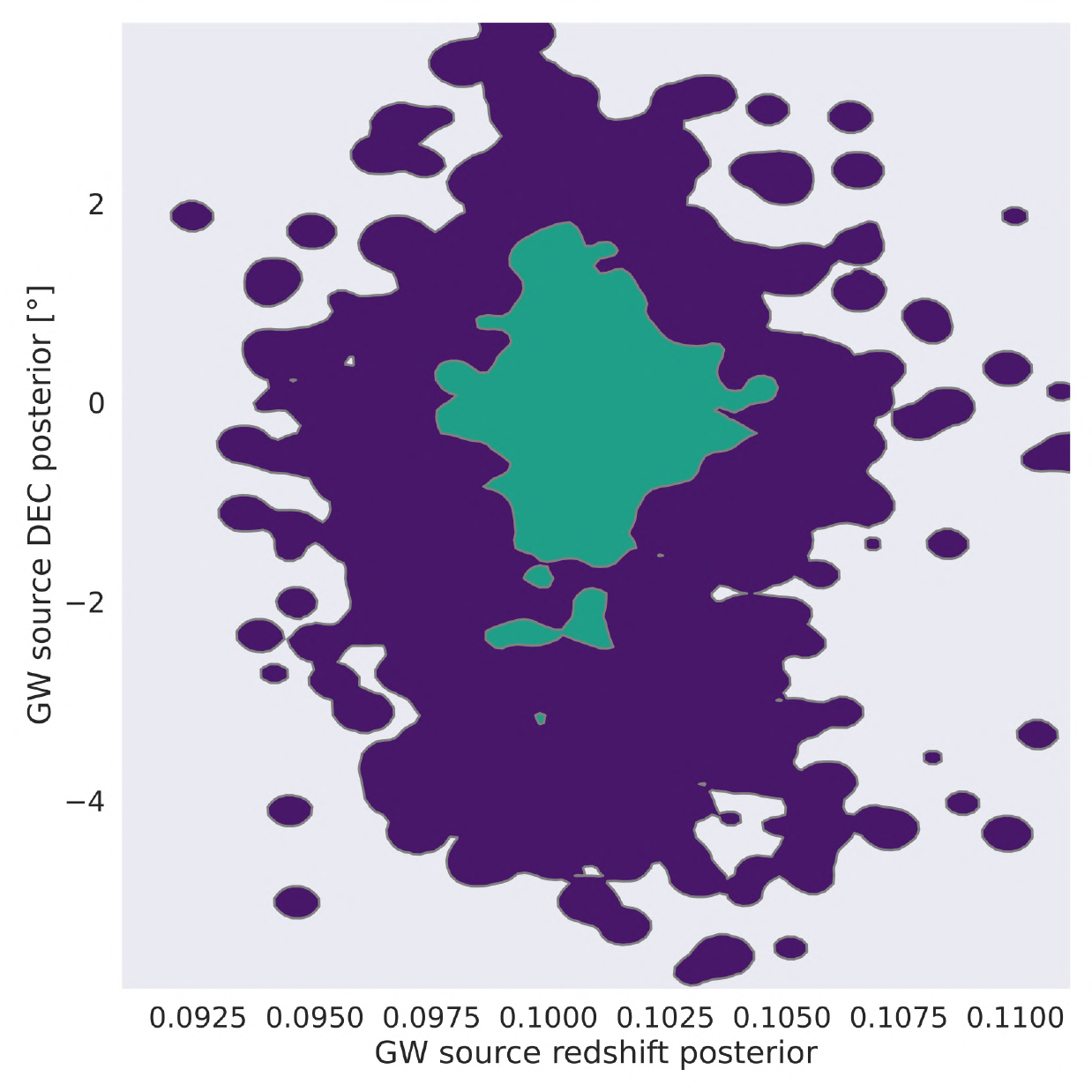}
         \caption{$z$ and $\delta_{src}$}
         \label{ssubfig29}
     \end{subfigure}
     \begin{subfigure}{0.245\linewidth}
         \includegraphics[width=\linewidth, keepaspectratio]{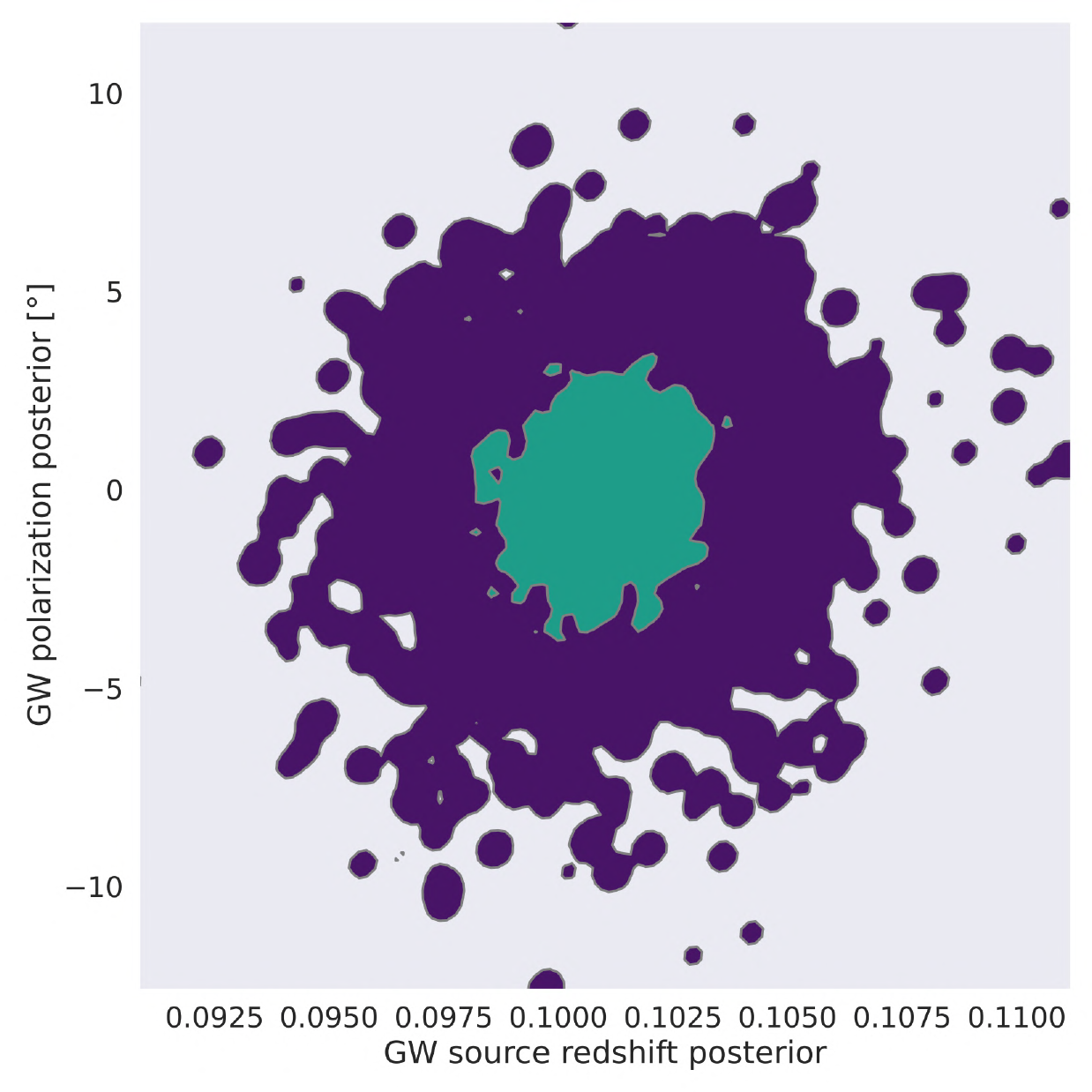}
         \caption{$z$ and $\psi_{eq}$}
         \label{ssubfig30}
     \end{subfigure}
     \begin{subfigure}{0.245\linewidth}
         \includegraphics[width=\linewidth, keepaspectratio]{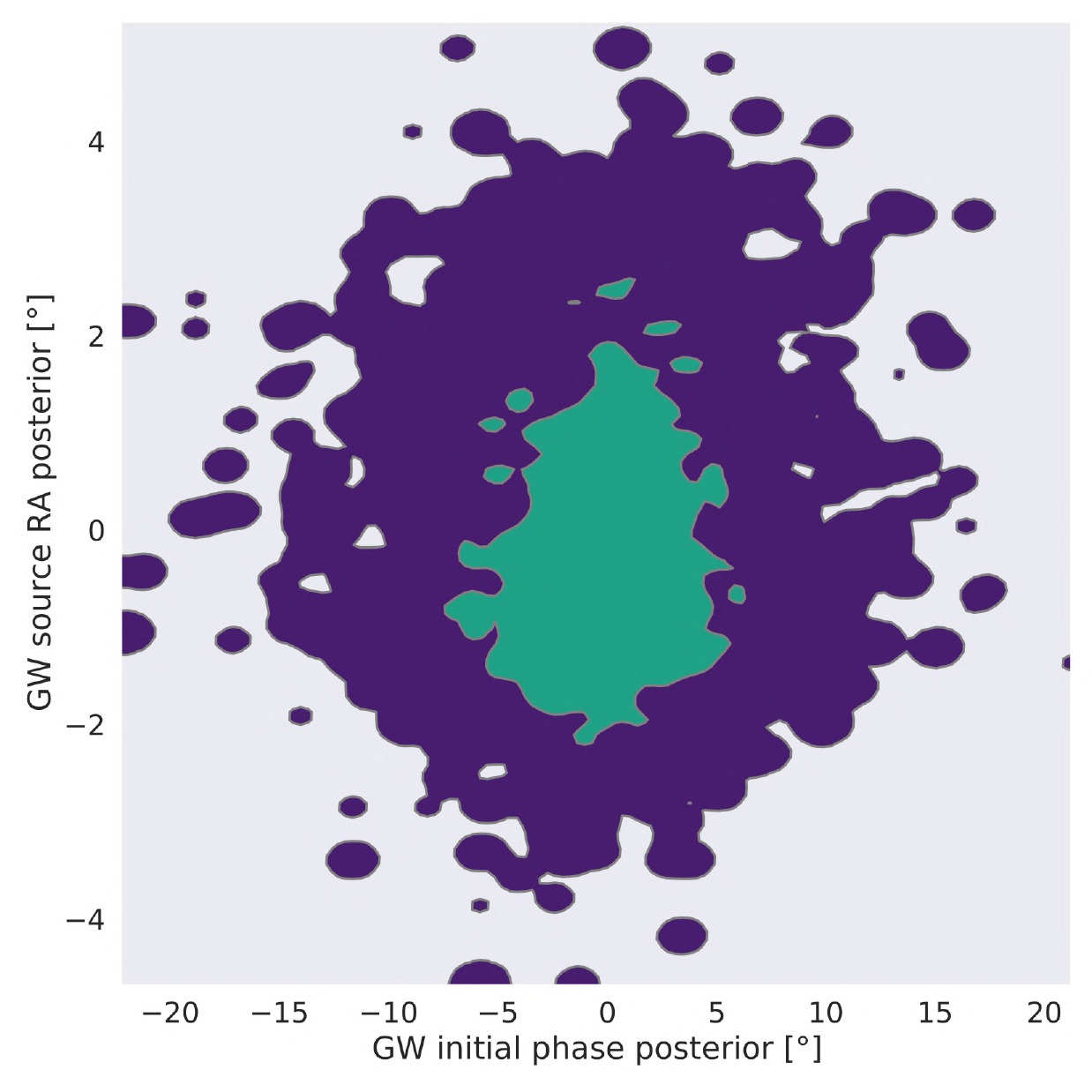}
         \caption{$\delta_0$ and $\alpha_{src}$}
         \label{ssubfig31}
     \end{subfigure}
     \begin{subfigure}{0.245\linewidth}
         \includegraphics[width=\linewidth, keepaspectratio]{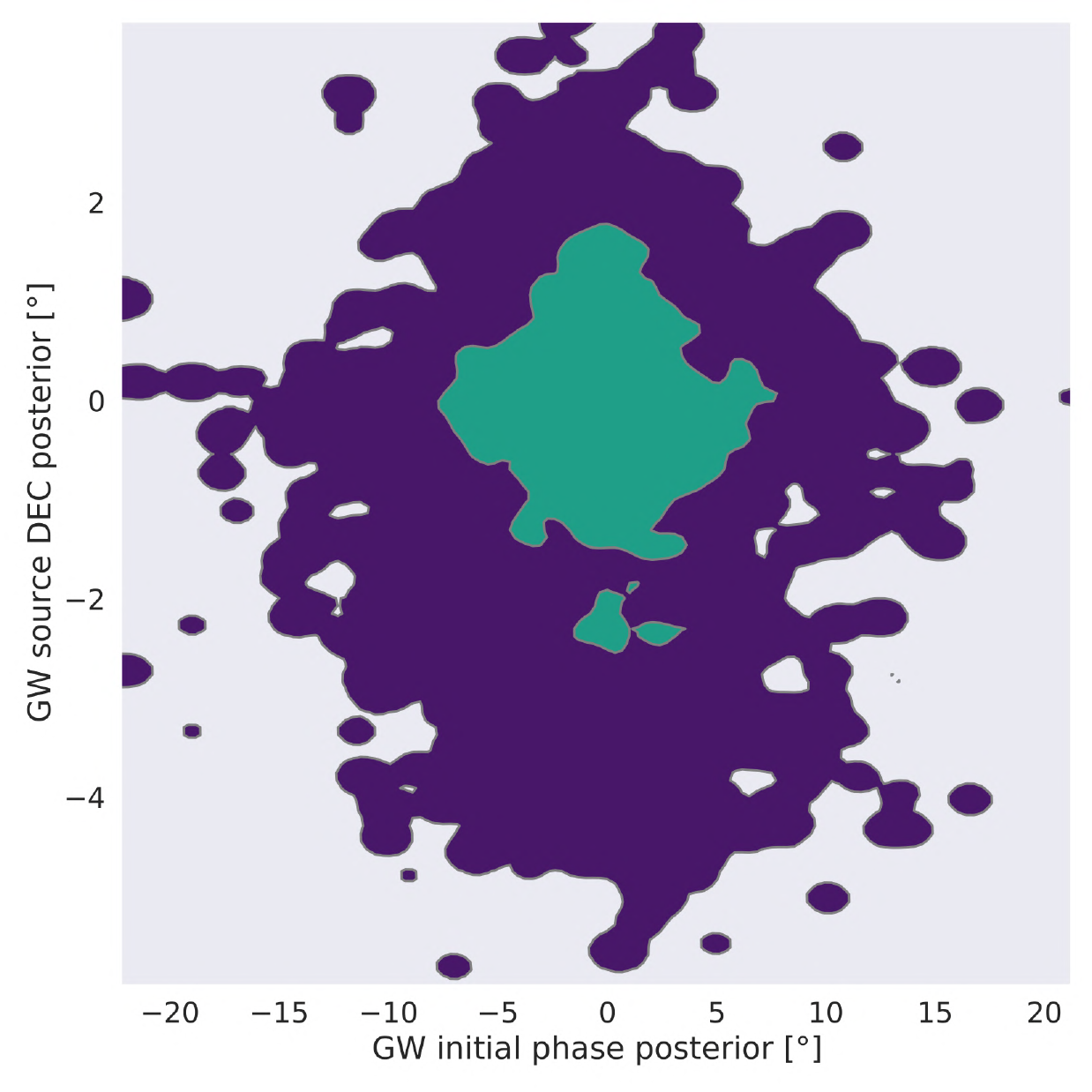}
         \caption{$\delta_0$ and $\delta_{src}$}
         \label{ssubfig32}
     \end{subfigure}
     \begin{subfigure}{0.245\linewidth}
         \includegraphics[width=\linewidth, keepaspectratio]{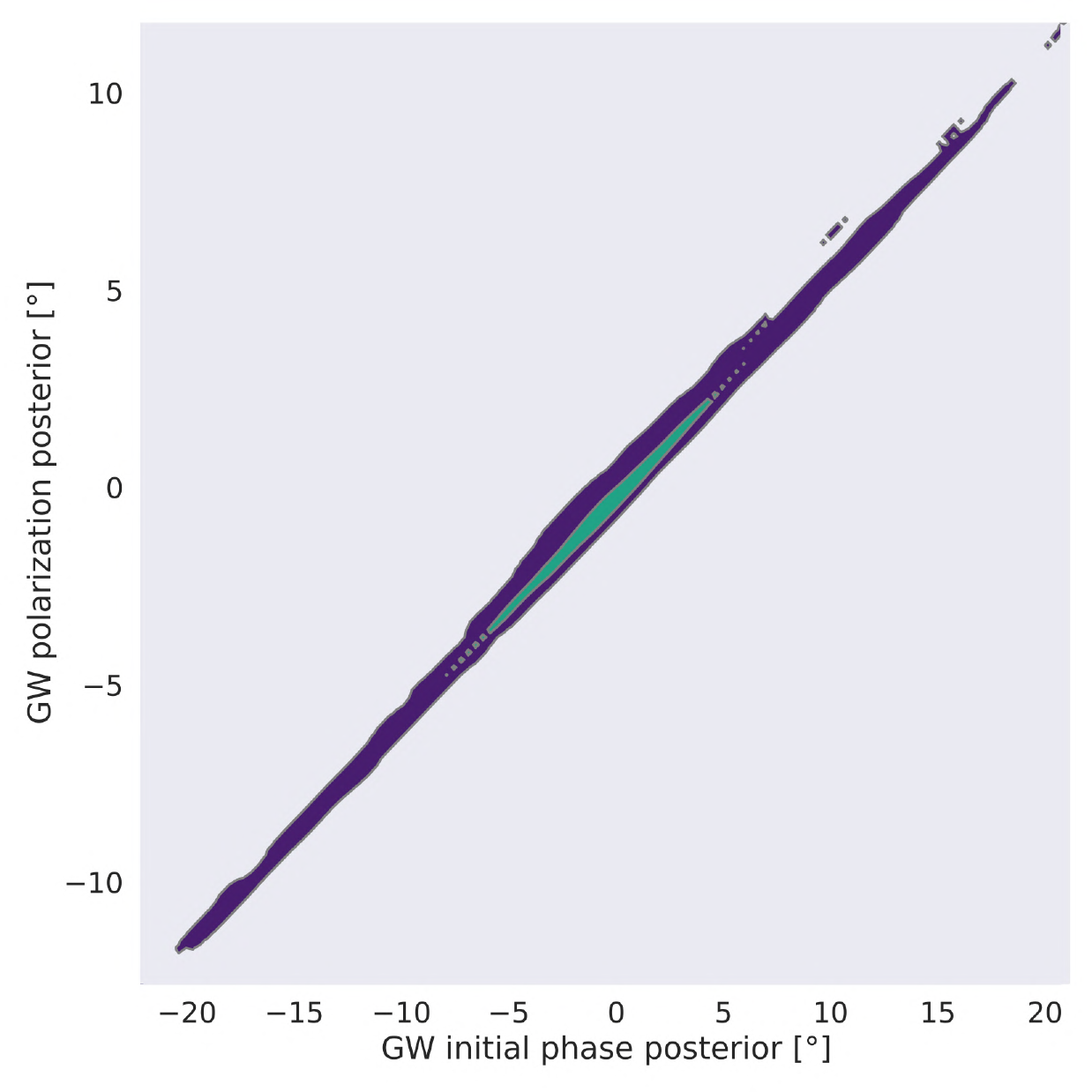}
         \caption{$\delta_0$ and $\psi_{eq}$}
         \label{ssubfig33}
     \end{subfigure}
     \begin{subfigure}{0.245\linewidth}
         \includegraphics[width=\linewidth, keepaspectratio]{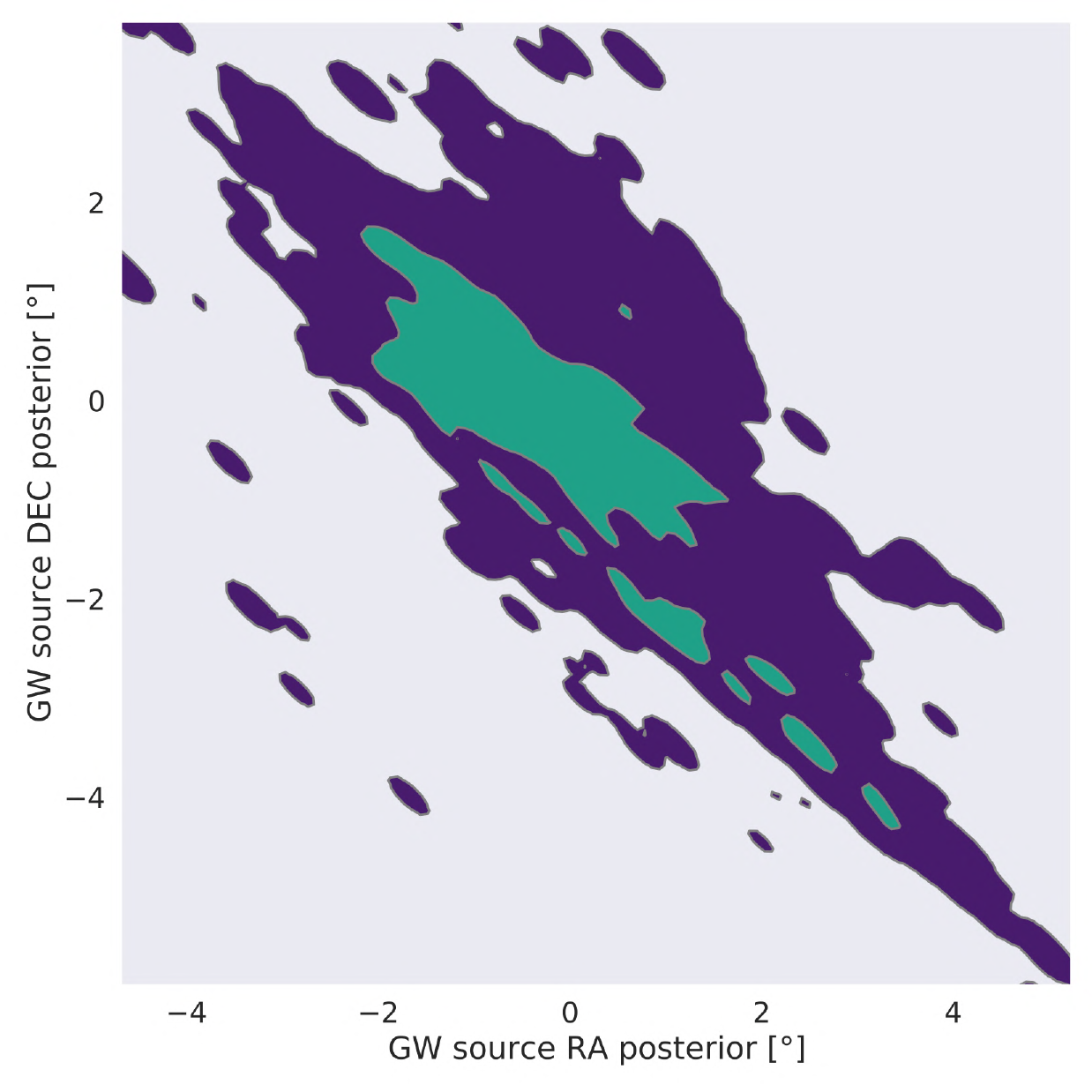}
         \caption{$\alpha_{src}$ and $\delta_{src}$}
         \label{ssubfig34}
     \end{subfigure}
     \begin{subfigure}{0.245\linewidth}
         \includegraphics[width=\linewidth, keepaspectratio]{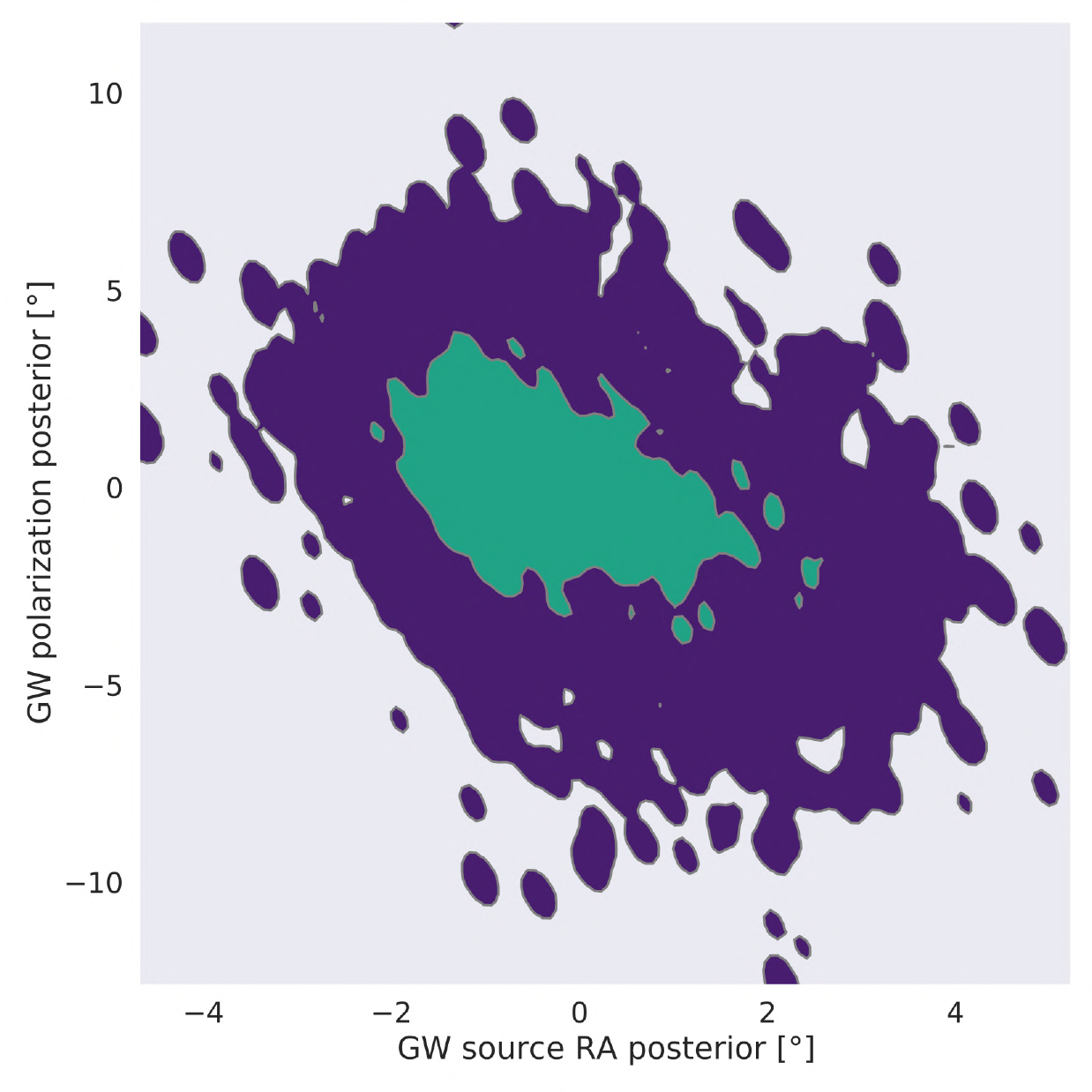}
         \caption{$\alpha_{src}$ and $\psi_{eq}$}
         \label{ssubfig35}
     \end{subfigure}
     \begin{subfigure}{0.245\linewidth}
         \includegraphics[width=\linewidth, keepaspectratio]{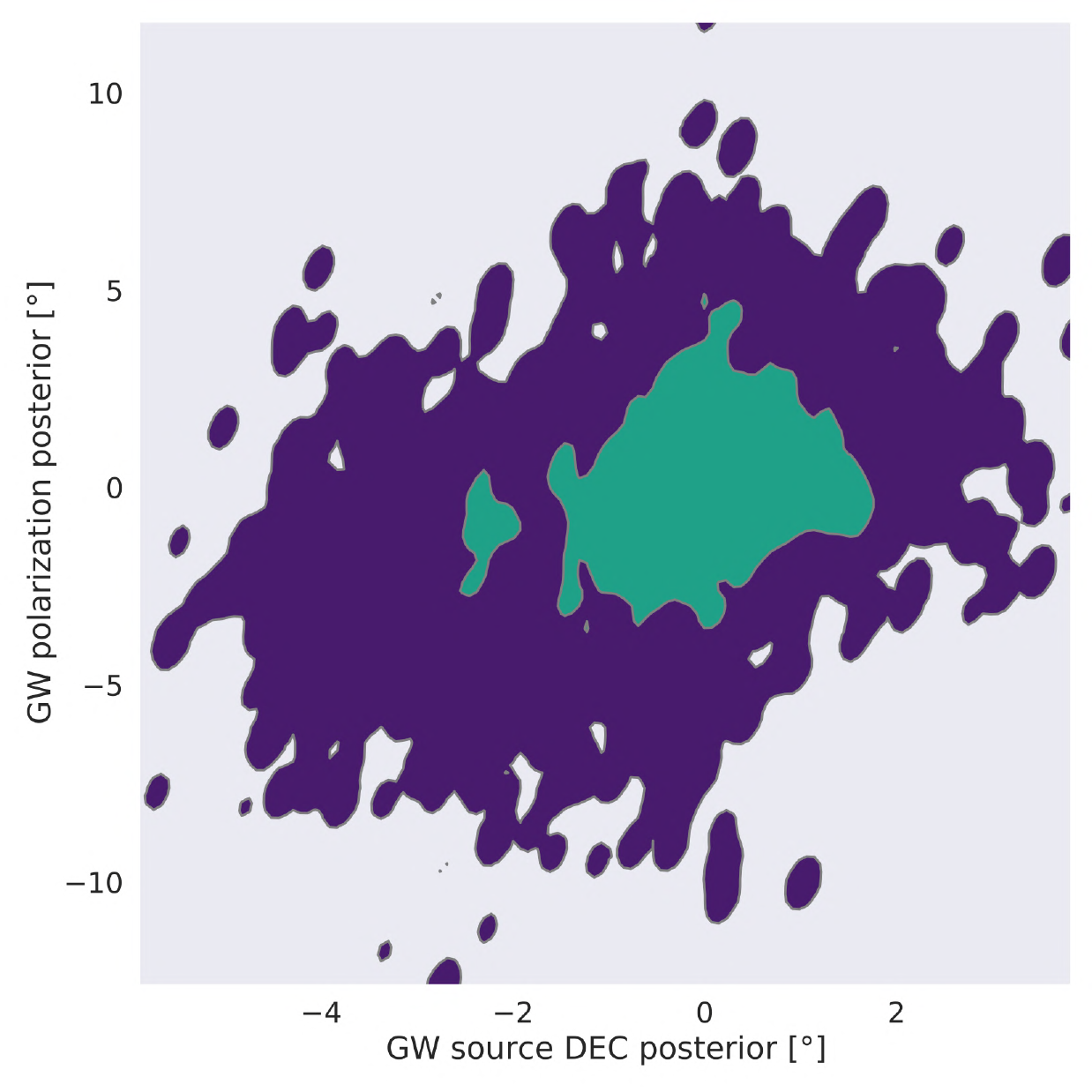}
         \caption{$\delta_{src}$ and $\psi_{eq}$}
         \label{ssubfig36}
     \end{subfigure} 
\caption{Individually shown joint posteriors of the $9 \times 9$ corner plot obtained after MCMC parameter estimation, for true parameter values described in Appendix \ref{app_B}.}
\label{fig_last}
\end{figure*}

\clearpage

\begin{thebibliography}{81}%
\makeatletter
\providecommand \@ifxundefined [1]{%
 \@ifx{#1\undefined}
}%
\providecommand \@ifnum [1]{%
 \ifnum #1\expandafter \@firstoftwo
 \else \expandafter \@secondoftwo
 \fi
}%
\providecommand \@ifx [1]{%
 \ifx #1\expandafter \@firstoftwo
 \else \expandafter \@secondoftwo
 \fi
}%
\providecommand \natexlab [1]{#1}%
\providecommand \enquote  [1]{``#1''}%
\providecommand \bibnamefont  [1]{#1}%
\providecommand \bibfnamefont [1]{#1}%
\providecommand \citenamefont [1]{#1}%
\providecommand \href@noop [0]{\@secondoftwo}%
\providecommand \href [0]{\begingroup \@sanitize@url \@href}%
\providecommand \@href[1]{\@@startlink{#1}\@@href}%
\providecommand \@@href[1]{\endgroup#1\@@endlink}%
\providecommand \@sanitize@url [0]{\catcode `\\12\catcode `\$12\catcode
  `\&12\catcode `\#12\catcode `\^12\catcode `\_12\catcode `\%12\relax}%
\providecommand \@@startlink[1]{}%
\providecommand \@@endlink[0]{}%
\providecommand \url  [0]{\begingroup\@sanitize@url \@url }%
\providecommand \@url [1]{\endgroup\@href {#1}{\urlprefix }}%
\providecommand \urlprefix  [0]{URL }%
\providecommand \Eprint [0]{\href }%
\providecommand \doibase [0]{https://doi.org/}%
\providecommand \selectlanguage [0]{\@gobble}%
\providecommand \bibinfo  [0]{\@secondoftwo}%
\providecommand \bibfield  [0]{\@secondoftwo}%
\providecommand \translation [1]{[#1]}%
\providecommand \BibitemOpen [0]{}%
\providecommand \bibitemStop [0]{}%
\providecommand \bibitemNoStop [0]{.\EOS\space}%
\providecommand \EOS [0]{\spacefactor3000\relax}%
\providecommand \BibitemShut  [1]{\csname bibitem#1\endcsname}%
\let\auto@bib@innerbib\@empty
\bibitem [{\citenamefont {Zer-Zion}(1998)}]{Zer-Zion:355151}%
  \BibitemOpen
  \bibfield  {author} {\bibinfo {author} {\bibfnamefont {D.}~\bibnamefont
  {Zer-Zion}},\ }\href {https://cds.cern.ch/record/355151} {\emph {\bibinfo
  {title} {{On the detection of gravitational waves through their interaction
  with particles in storage rings}}}},\ \bibinfo {type} {Tech. Rep.}\ (\bibinfo
   {institution} {CERN},\ \bibinfo {address} {Geneva},\ \bibinfo {year}
  {1998})\BibitemShut {NoStop}%
\bibitem [{\citenamefont {van Holten}(1999)}]{gr-qc/9906117}%
  \BibitemOpen
  \bibfield  {author} {\bibinfo {author} {\bibfnamefont {J.~W.}\ \bibnamefont
  {van Holten}},\ }\href@noop {} {\bibinfo {title} {Cyclotron motion in a
  gravitational-wave background}} (\bibinfo {year} {1999}),\ \Eprint
  {https://arxiv.org/abs/arXiv:gr-qc/9906117} {arXiv:gr-qc/9906117}
  \BibitemShut {NoStop}%
\bibitem [{\citenamefont {Ivanov}\ \emph {et~al.}(2002)\citenamefont {Ivanov},
  \citenamefont {Kobushkin},\ and\ \citenamefont {Wellenzohn}}]{gr-qc/0210091}%
  \BibitemOpen
  \bibfield  {author} {\bibinfo {author} {\bibfnamefont {A.~N.}\ \bibnamefont
  {Ivanov}}, \bibinfo {author} {\bibfnamefont {A.~P.}\ \bibnamefont
  {Kobushkin}},\ and\ \bibinfo {author} {\bibfnamefont {M.}~\bibnamefont
  {Wellenzohn}},\ }\href@noop {} {\bibinfo {title} {Storage rings as detectors
  for relic gravitational-wave background ?}} (\bibinfo {year} {2002}),\
  \Eprint {https://arxiv.org/abs/arXiv:gr-qc/0210091} {arXiv:gr-qc/0210091}
  \BibitemShut {NoStop}%
\bibitem [{\citenamefont {Dong}(2002)}]{Dong:2002sr}%
  \BibitemOpen
  \bibfield  {author} {\bibinfo {author} {\bibfnamefont {D.}~\bibnamefont
  {Dong}},\ }\bibfield  {title} {\bibinfo {title} {{Study of Charged Particle
  Storage Ring as Detector of Gravitational Waves}},\ }\href@noop {} {\bibfield
   {journal} {\bibinfo  {journal} {eConf}\ }\textbf {\bibinfo {volume}
  {C0211115}},\ \bibinfo {pages} {030} (\bibinfo {year} {2002})}\BibitemShut
  {NoStop}%
\bibitem [{\citenamefont {Faber}\ and\ \citenamefont {Suda}(2018)}]{Faber2018}%
  \BibitemOpen
  \bibfield  {author} {\bibinfo {author} {\bibfnamefont {M.}~\bibnamefont
  {Faber}}\ and\ \bibinfo {author} {\bibfnamefont {M.}~\bibnamefont {Suda}},\
  }\bibfield  {title} {\bibinfo {title} {Influence of gravitational waves on
  circular moving particles},\ }\href {https://doi.org/10.4236/jmp.2018.94045}
  {\bibfield  {journal} {\bibinfo  {journal} {Journal of Modern Physics}\
  }\textbf {\bibinfo {volume} {09}},\ \bibinfo {pages} {651} (\bibinfo {year}
  {2018})}\BibitemShut {NoStop}%
\bibitem [{\citenamefont {Rao}\ \emph {et~al.}(2020)\citenamefont {Rao},
  \citenamefont {Br\"uggen},\ and\ \citenamefont
  {Liske}}]{PhysRevD.102.122006}%
  \BibitemOpen
  \bibfield  {author} {\bibinfo {author} {\bibfnamefont {S.}~\bibnamefont
  {Rao}}, \bibinfo {author} {\bibfnamefont {M.}~\bibnamefont {Br\"uggen}},\
  and\ \bibinfo {author} {\bibfnamefont {J.}~\bibnamefont {Liske}},\ }\bibfield
   {title} {\bibinfo {title} {Detection of gravitational waves in circular
  particle accelerators},\ }\href {https://doi.org/10.1103/PhysRevD.102.122006}
  {\bibfield  {journal} {\bibinfo  {journal} {Phys. Rev. D}\ }\textbf {\bibinfo
  {volume} {102}},\ \bibinfo {pages} {122006} (\bibinfo {year}
  {2020})}\BibitemShut {NoStop}%
\bibitem [{\citenamefont {{Amaro-Seoane \textit{et al.,}} The
  LISA~Consortium}(2017)}]{2017arXiv170200786A}%
  \BibitemOpen
  \bibfield  {author} {\bibinfo {author} {\bibfnamefont {P.}~\bibnamefont
  {{Amaro-Seoane \textit{et al.,}} The LISA~Consortium}},\ }\bibfield  {title}
  {\bibinfo {title} {{Laser Interferometer Space Antenna}},\ }\href@noop {}
  {\bibfield  {journal} {\bibinfo  {journal} {arXiv e-prints}\ ,\ \bibinfo
  {eid} {arXiv:1702.00786}} (\bibinfo {year} {2017})},\ \Eprint
  {https://arxiv.org/abs/1702.00786} {arXiv:1702.00786 [astro-ph.IM]}
  \BibitemShut {NoStop}%
\bibitem [{\citenamefont {{Baker \textit{et al.,}} The
  LISA~Consortium}(2019)}]{2019arXiv190706482B}%
  \BibitemOpen
  \bibfield  {author} {\bibinfo {author} {\bibfnamefont {J.}~\bibnamefont
  {{Baker \textit{et al.,}} The LISA~Consortium}},\ }\bibfield  {title}
  {\bibinfo {title} {{The Laser Interferometer Space Antenna: Unveiling the
  Millihertz Gravitational Wave Sky}},\ }\href@noop {} {\bibfield  {journal}
  {\bibinfo  {journal} {arXiv e-prints}\ ,\ \bibinfo {eid} {arXiv:1907.06482}}
  (\bibinfo {year} {2019})},\ \Eprint {https://arxiv.org/abs/1907.06482}
  {arXiv:1907.06482 [astro-ph.IM]} \BibitemShut {NoStop}%
\bibitem [{\citenamefont {Dahal}(2020)}]{Dahal2020}%
  \BibitemOpen
  \bibfield  {author} {\bibinfo {author} {\bibfnamefont {P.~K.}\ \bibnamefont
  {Dahal}},\ }\bibfield  {title} {\bibinfo {title} {Review of pulsar timing
  array for gravitational wave research},\ }\bibfield  {journal} {\bibinfo
  {journal} {Journal of Astrophysics and Astronomy}\ }\textbf {\bibinfo
  {volume} {41}},\ \href {https://doi.org/10.1007/s12036-020-9625-y}
  {10.1007/s12036-020-9625-y} (\bibinfo {year} {2020})\BibitemShut {NoStop}%
\bibitem [{\citenamefont {Geiger}(2017)}]{Geiger2017}%
  \BibitemOpen
  \bibfield  {author} {\bibinfo {author} {\bibfnamefont {R.}~\bibnamefont
  {Geiger}},\ }\bibfield  {title} {\bibinfo {title} {Future gravitational wave
  detectors based on atom interferometry},\ }in\ \href
  {https://doi.org/10.1142/9789813141766_0008} {\emph {\bibinfo {booktitle} {An
  Overview of Gravitational Waves}}}\ (\bibinfo  {publisher} {{WORLD}
  {SCIENTIFIC}},\ \bibinfo {year} {2017})\ pp.\ \bibinfo {pages}
  {285--313}\BibitemShut {NoStop}%
\bibitem [{\citenamefont {Badurina}\ \emph {et~al.}(2020)\citenamefont
  {Badurina}, \citenamefont {Bentine}, \citenamefont {Blas}, \citenamefont
  {Bongs}, \citenamefont {Bortoletto}, \citenamefont {Bowcock}, \citenamefont
  {Bridges}, \citenamefont {Bowden}, \citenamefont {Buchmueller},\ and\
  \citenamefont {Burrage~\textit{et al.}}}]{Badurina_2020}%
  \BibitemOpen
  \bibfield  {author} {\bibinfo {author} {\bibfnamefont {L.}~\bibnamefont
  {Badurina}}, \bibinfo {author} {\bibfnamefont {E.}~\bibnamefont {Bentine}},
  \bibinfo {author} {\bibfnamefont {D.}~\bibnamefont {Blas}}, \bibinfo {author}
  {\bibfnamefont {K.}~\bibnamefont {Bongs}}, \bibinfo {author} {\bibfnamefont
  {D.}~\bibnamefont {Bortoletto}}, \bibinfo {author} {\bibfnamefont
  {T.}~\bibnamefont {Bowcock}}, \bibinfo {author} {\bibfnamefont
  {K.}~\bibnamefont {Bridges}}, \bibinfo {author} {\bibfnamefont
  {W.}~\bibnamefont {Bowden}}, \bibinfo {author} {\bibfnamefont
  {O.}~\bibnamefont {Buchmueller}},\ and\ \bibinfo {author} {\bibfnamefont
  {C.}~\bibnamefont {Burrage~\textit{et al.}}},\ }\bibfield  {title} {\bibinfo
  {title} {{AION}: an atom interferometer observatory and network},\ }\href
  {https://doi.org/10.1088/1475-7516/2020/05/011} {\bibfield  {journal}
  {\bibinfo  {journal} {Journal of Cosmology and Astroparticle Physics}\
  }\textbf {\bibinfo {volume} {2020}}\bibinfo  {number} { (05)},\ \bibinfo
  {pages} {011}}\BibitemShut {NoStop}%
\bibitem [{\citenamefont {Canuel}\ \emph {et~al.}(2018)\citenamefont {Canuel},
  \citenamefont {Bertoldi}, \citenamefont {Amand}, \citenamefont {Pozzo~di
  Borgo}, \citenamefont {Chantrait}, \citenamefont {Danquigny}, \citenamefont
  {Dovale~{\'A}lvarez}, \citenamefont {Fang}, \citenamefont {Freise},\ and\
  \citenamefont {Geiger~\textit{et al.}}}]{Canuel2018}%
  \BibitemOpen
\bibfield  {number} {  }\bibfield  {author} {\bibinfo {author} {\bibfnamefont
  {B.}~\bibnamefont {Canuel}}, \bibinfo {author} {\bibfnamefont
  {A.}~\bibnamefont {Bertoldi}}, \bibinfo {author} {\bibfnamefont
  {L.}~\bibnamefont {Amand}}, \bibinfo {author} {\bibfnamefont
  {E.}~\bibnamefont {Pozzo~di Borgo}}, \bibinfo {author} {\bibfnamefont
  {T.}~\bibnamefont {Chantrait}}, \bibinfo {author} {\bibfnamefont
  {C.}~\bibnamefont {Danquigny}}, \bibinfo {author} {\bibfnamefont
  {M.}~\bibnamefont {Dovale~{\'A}lvarez}}, \bibinfo {author} {\bibfnamefont
  {B.}~\bibnamefont {Fang}}, \bibinfo {author} {\bibfnamefont {A.}~\bibnamefont
  {Freise}},\ and\ \bibinfo {author} {\bibfnamefont {R.}~\bibnamefont
  {Geiger~\textit{et al.}}},\ }\bibfield  {title} {\bibinfo {title} {Exploring
  gravity with the {MIGA} large scale atom interferometer},\ }\href
  {https://doi.org/10.1038/s41598-018-32165-z} {\bibfield  {journal} {\bibinfo
  {journal} {Scientific Reports}\ }\textbf {\bibinfo {volume} {8}},\ \bibinfo
  {pages} {14064} (\bibinfo {year} {2018})}\BibitemShut {NoStop}%
\bibitem [{\citenamefont {Canuel~\textit{et al.}}(2020)}]{Canuel_2020}%
  \BibitemOpen
  \bibfield  {author} {\bibinfo {author} {\bibfnamefont {B.}~\bibnamefont
  {Canuel~\textit{et al.}}},\ }\bibfield  {title} {\bibinfo {title}
  {{ELGAR}{\textemdash}a european laboratory for gravitation and
  atom-interferometric research},\ }\href
  {https://doi.org/10.1088/1361-6382/aba80e} {\bibfield  {journal} {\bibinfo
  {journal} {Classical and Quantum Gravity}\ }\textbf {\bibinfo {volume}
  {37}},\ \bibinfo {pages} {225017} (\bibinfo {year} {2020})}\BibitemShut
  {NoStop}%
\bibitem [{\citenamefont {{Mahiro \textit{et al.}}}(2021)}]{Abe2021}%
  \BibitemOpen
  \bibfield  {author} {\bibinfo {author} {\bibfnamefont {A.}~\bibnamefont
  {{Mahiro \textit{et al.}}}},\ }\bibfield  {title} {\bibinfo {title}
  {Matter-wave atomic gradiometer interferometric sensor ({MAGIS}-100)},\
  }\href {https://doi.org/10.1088/2058-9565/abf719} {\bibfield  {journal}
  {\bibinfo  {journal} {Quantum Science and Technology}\ }\textbf {\bibinfo
  {volume} {6}},\ \bibinfo {pages} {044003} (\bibinfo {year}
  {2021})}\BibitemShut {NoStop}%
\bibitem [{\citenamefont {Berlin}\ \emph {et~al.}(2021)\citenamefont {Berlin},
  \citenamefont {Brüggen}, \citenamefont {Buchmueller}, \citenamefont {Chen},
  \citenamefont {D'Agnolo}, \citenamefont {Deng}, \citenamefont {Ellis},
  \citenamefont {Ellis}, \citenamefont {Franchetti},\ and\ \citenamefont
  {Ivanov~\textit{et al.}}}]{2105.00992}%
  \BibitemOpen
  \bibfield  {author} {\bibinfo {author} {\bibfnamefont {A.}~\bibnamefont
  {Berlin}}, \bibinfo {author} {\bibfnamefont {M.}~\bibnamefont {Brüggen}},
  \bibinfo {author} {\bibfnamefont {O.}~\bibnamefont {Buchmueller}}, \bibinfo
  {author} {\bibfnamefont {P.}~\bibnamefont {Chen}}, \bibinfo {author}
  {\bibfnamefont {R.~T.}\ \bibnamefont {D'Agnolo}}, \bibinfo {author}
  {\bibfnamefont {R.}~\bibnamefont {Deng}}, \bibinfo {author} {\bibfnamefont
  {J.~R.}\ \bibnamefont {Ellis}}, \bibinfo {author} {\bibfnamefont
  {S.}~\bibnamefont {Ellis}}, \bibinfo {author} {\bibfnamefont
  {G.}~\bibnamefont {Franchetti}},\ and\ \bibinfo {author} {\bibfnamefont
  {A.}~\bibnamefont {Ivanov~\textit{et al.}}},\ }\href@noop {} {\bibinfo
  {title} {Storage rings and gravitational waves: Summary and outlook}}
  (\bibinfo {year} {2021}),\ \Eprint {https://arxiv.org/abs/arXiv:2105.00992}
  {arXiv:2105.00992} \BibitemShut {NoStop}%
\bibitem [{\citenamefont {Habs}\ and\ \citenamefont {Grimm}(1995)}]{Habs1995}%
  \BibitemOpen
  \bibfield  {author} {\bibinfo {author} {\bibfnamefont {D.}~\bibnamefont
  {Habs}}\ and\ \bibinfo {author} {\bibfnamefont {R.}~\bibnamefont {Grimm}},\
  }\bibfield  {title} {\bibinfo {title} {Crystalline ion beams},\ }\href
  {https://doi.org/10.1146/annurev.ns.45.120195.002135} {\bibfield  {journal}
  {\bibinfo  {journal} {Annual Review of Nuclear and Particle Science}\
  }\textbf {\bibinfo {volume} {45}},\ \bibinfo {pages} {391} (\bibinfo {year}
  {1995})}\BibitemShut {NoStop}%
\bibitem [{\citenamefont {Sch\"{a}tz}\ \emph {et~al.}(2001)\citenamefont
  {Sch\"{a}tz}, \citenamefont {Schramm},\ and\ \citenamefont
  {Habs}}]{Schtz2001}%
  \BibitemOpen
  \bibfield  {author} {\bibinfo {author} {\bibfnamefont {T.}~\bibnamefont
  {Sch\"{a}tz}}, \bibinfo {author} {\bibfnamefont {U.}~\bibnamefont
  {Schramm}},\ and\ \bibinfo {author} {\bibfnamefont {D.}~\bibnamefont
  {Habs}},\ }\bibfield  {title} {\bibinfo {title} {Crystalline ion beams},\
  }\href {https://doi.org/10.1038/35089045} {\bibfield  {journal} {\bibinfo
  {journal} {Nature}\ }\textbf {\bibinfo {volume} {412}},\ \bibinfo {pages}
  {717} (\bibinfo {year} {2001})}\BibitemShut {NoStop}%
\bibitem [{\citenamefont {Schramm}\ \emph {et~al.}(2001)\citenamefont
  {Schramm}, \citenamefont {Sch\"atz},\ and\ \citenamefont
  {Habs}}]{PhysRevLett.87.184801}%
  \BibitemOpen
  \bibfield  {author} {\bibinfo {author} {\bibfnamefont {U.}~\bibnamefont
  {Schramm}}, \bibinfo {author} {\bibfnamefont {T.}~\bibnamefont {Sch\"atz}},\
  and\ \bibinfo {author} {\bibfnamefont {D.}~\bibnamefont {Habs}},\ }\bibfield
  {title} {\bibinfo {title} {Bunched crystalline ion beams},\ }\href
  {https://doi.org/10.1103/PhysRevLett.87.184801} {\bibfield  {journal}
  {\bibinfo  {journal} {Phys. Rev. Lett.}\ }\textbf {\bibinfo {volume} {87}},\
  \bibinfo {pages} {184801} (\bibinfo {year} {2001})}\BibitemShut {NoStop}%
\bibitem [{\citenamefont {Nagaitsev}\ \emph {et~al.}(2019)\citenamefont
  {Nagaitsev}, \citenamefont {Arodzero}, \citenamefont {Lobach}, \citenamefont
  {Murokh}, \citenamefont {Romanov}, \citenamefont {Ruelas}, \citenamefont
  {Shaftan},\ and\ \citenamefont {Stancari}}]{Nagaitsev:IPAC2019-MOPRB089}%
  \BibitemOpen
  \bibfield  {author} {\bibinfo {author} {\bibfnamefont {S.}~\bibnamefont
  {Nagaitsev}}, \bibinfo {author} {\bibfnamefont {A.}~\bibnamefont {Arodzero}},
  \bibinfo {author} {\bibfnamefont {I.}~\bibnamefont {Lobach}}, \bibinfo
  {author} {\bibfnamefont {A.}~\bibnamefont {Murokh}}, \bibinfo {author}
  {\bibfnamefont {A.}~\bibnamefont {Romanov}}, \bibinfo {author} {\bibfnamefont
  {M.}~\bibnamefont {Ruelas}}, \bibinfo {author} {\bibfnamefont
  {T.}~\bibnamefont {Shaftan}},\ and\ \bibinfo {author} {\bibfnamefont
  {G.}~\bibnamefont {Stancari}},\ }\bibfield  {title} {{
  \bibinfo {title} {Experimental study of a single electron in a
  storage ring via undulator radiation}},\ }in\ \href
  {https://doi.org/doi:10.18429/JACoW-IPAC2019-MOPRB089} {{
  \emph {\bibinfo {booktitle} {Proc. 10th International Particle
  Accelerator Conference (IPAC'19), Melbourne, Australia, 19-24 May 2019}}}},\
  \bibinfo {series and number} {\bibinfo {series} {International Particle
  Accelerator Conference}\ No.~\bibinfo {number} {10}}\ (\bibinfo  {publisher}
  {JACoW Publishing},\ \bibinfo {address} {Geneva, Switzerland},\ \bibinfo
  {year} {2019})\ pp.\ \bibinfo {pages} {781--784},\ \bibinfo {note}
  {https://doi.org/10.18429/JACoW-IPAC2019-MOPRB089}\BibitemShut {NoStop}%
\bibitem [{\citenamefont {Frisch}(2016)}]{Frisch:2016hju}%
  \BibitemOpen
  \bibfield  {author} {\bibinfo {author} {\bibfnamefont {J.}~\bibnamefont
  {Frisch}},\ }\bibfield  {title} {\bibinfo {title} {{Beam Arrival Time
  Monitors}},\ }in\ \href {https://doi.org/10.18429/JACoW-IBIC2015-TUALA01}
  {\emph {\bibinfo {booktitle} {{4th International Beam Instrumentation
  Conference}}}}\ (\bibinfo {year} {2016})\ p.\ \bibinfo {pages}
  {TUALA01}\BibitemShut {NoStop}%
\bibitem [{\citenamefont {Abbott~\textit{et al.}}\ \emph
  {et~al.}(2009)\citenamefont {Abbott~\textit{et al.}}, \citenamefont {{LIGO
  Scientific Collaboration}},\ and\ \citenamefont {{Virgo
  Collaboration}}}]{Abbott2009}%
  \BibitemOpen
  \bibfield  {author} {\bibinfo {author} {\bibfnamefont {B.~P.}\ \bibnamefont
  {Abbott~\textit{et al.}}}, \bibinfo {author} {\bibnamefont {{LIGO Scientific
  Collaboration}}},\ and\ \bibinfo {author} {\bibnamefont {{Virgo
  Collaboration}}},\ }\bibfield  {title} {\bibinfo {title} {{LIGO}: the {L}aser
  {I}nterferometer {G}ravitational-wave {O}bservatory},\ }\href
  {https://doi.org/10.1088/0034-4885/72/7/076901} {\bibfield  {journal}
  {\bibinfo  {journal} {Reports on Progress in Physics}\ }\textbf {\bibinfo
  {volume} {72}},\ \bibinfo {pages} {076901} (\bibinfo {year}
  {2009})}\BibitemShut {NoStop}%
\bibitem [{\citenamefont {Weber}(1966)}]{PhysRevLett.17.1228}%
  \BibitemOpen
  \bibfield  {author} {\bibinfo {author} {\bibfnamefont {J.}~\bibnamefont
  {Weber}},\ }\bibfield  {title} {\bibinfo {title} {Observation of the thermal
  fluctuations of a gravitational-wave detector},\ }\href
  {https://doi.org/10.1103/PhysRevLett.17.1228} {\bibfield  {journal} {\bibinfo
   {journal} {Phys. Rev. Lett.}\ }\textbf {\bibinfo {volume} {17}},\ \bibinfo
  {pages} {1228} (\bibinfo {year} {1966})}\BibitemShut {NoStop}%
\bibitem [{\citenamefont {Weber}(1967)}]{PhysRevLett.18.498}%
  \BibitemOpen
  \bibfield  {author} {\bibinfo {author} {\bibfnamefont {J.}~\bibnamefont
  {Weber}},\ }\bibfield  {title} {\bibinfo {title} {Gravitational radiation},\
  }\href {https://doi.org/10.1103/PhysRevLett.18.498} {\bibfield  {journal}
  {\bibinfo  {journal} {Phys. Rev. Lett.}\ }\textbf {\bibinfo {volume} {18}},\
  \bibinfo {pages} {498} (\bibinfo {year} {1967})}\BibitemShut {NoStop}%
\bibitem [{\citenamefont {Astone}\ \emph {et~al.}(2013)\citenamefont {Astone},
  \citenamefont {Bassan}, \citenamefont {Coccia}, \citenamefont {D'Antonio},
  \citenamefont {Fafone}, \citenamefont {Giordano}, \citenamefont {Marini},
  \citenamefont {Minenkov}, \citenamefont {Modena},\ and\ \citenamefont
  {Moleti~\textit{et al.}}}]{PhysRevD.87.082002}%
  \BibitemOpen
  \bibfield  {author} {\bibinfo {author} {\bibfnamefont {P.}~\bibnamefont
  {Astone}}, \bibinfo {author} {\bibfnamefont {M.}~\bibnamefont {Bassan}},
  \bibinfo {author} {\bibfnamefont {E.}~\bibnamefont {Coccia}}, \bibinfo
  {author} {\bibfnamefont {S.}~\bibnamefont {D'Antonio}}, \bibinfo {author}
  {\bibfnamefont {V.}~\bibnamefont {Fafone}}, \bibinfo {author} {\bibfnamefont
  {G.}~\bibnamefont {Giordano}}, \bibinfo {author} {\bibfnamefont
  {A.}~\bibnamefont {Marini}}, \bibinfo {author} {\bibfnamefont
  {Y.}~\bibnamefont {Minenkov}}, \bibinfo {author} {\bibfnamefont
  {I.}~\bibnamefont {Modena}},\ and\ \bibinfo {author} {\bibfnamefont
  {A.}~\bibnamefont {Moleti~\textit{et al.}}},\ }\bibfield  {title} {\bibinfo
  {title} {Analysis of 3 years of data from the gravitational wave detectors
  {EXPLORER} and {NAUTILUS}},\ }\href
  {https://doi.org/10.1103/PhysRevD.87.082002} {\bibfield  {journal} {\bibinfo
  {journal} {Phys. Rev. D}\ }\textbf {\bibinfo {volume} {87}},\ \bibinfo
  {pages} {082002} (\bibinfo {year} {2013})}\BibitemShut {NoStop}%
\bibitem [{\citenamefont {Mauceli}\ \emph {et~al.}(1996)\citenamefont
  {Mauceli}, \citenamefont {Geng}, \citenamefont {Hamilton}, \citenamefont
  {Johnson}, \citenamefont {Merkowitz}, \citenamefont {Morse}, \citenamefont
  {Price},\ and\ \citenamefont {Solomonson}}]{PhysRevD.54.1264}%
  \BibitemOpen
  \bibfield  {author} {\bibinfo {author} {\bibfnamefont {E.}~\bibnamefont
  {Mauceli}}, \bibinfo {author} {\bibfnamefont {Z.~K.}\ \bibnamefont {Geng}},
  \bibinfo {author} {\bibfnamefont {W.~O.}\ \bibnamefont {Hamilton}}, \bibinfo
  {author} {\bibfnamefont {W.~W.}\ \bibnamefont {Johnson}}, \bibinfo {author}
  {\bibfnamefont {S.}~\bibnamefont {Merkowitz}}, \bibinfo {author}
  {\bibfnamefont {A.}~\bibnamefont {Morse}}, \bibinfo {author} {\bibfnamefont
  {B.}~\bibnamefont {Price}},\ and\ \bibinfo {author} {\bibfnamefont
  {N.}~\bibnamefont {Solomonson}},\ }\bibfield  {title} {\bibinfo {title} {The
  {A}llegro gravitational wave detector: Data acquisition and analysis},\
  }\href {https://doi.org/10.1103/PhysRevD.54.1264} {\bibfield  {journal}
  {\bibinfo  {journal} {Phys. Rev. D}\ }\textbf {\bibinfo {volume} {54}},\
  \bibinfo {pages} {1264} (\bibinfo {year} {1996})}\BibitemShut {NoStop}%
\bibitem [{\citenamefont {Sathyaprakash}\ and\ \citenamefont
  {Schutz}(2009)}]{Sathyaprakash2009}%
  \BibitemOpen
  \bibfield  {author} {\bibinfo {author} {\bibfnamefont {B.~S.}\ \bibnamefont
  {Sathyaprakash}}\ and\ \bibinfo {author} {\bibfnamefont {B.~F.}\ \bibnamefont
  {Schutz}},\ }\bibfield  {title} {\bibinfo {title} {Physics, astrophysics and
  cosmology with gravitational waves},\ }\bibfield  {journal} {\bibinfo
  {journal} {Living Reviews in Relativity}\ }\textbf {\bibinfo {volume} {12}},\
  \href {https://doi.org/10.12942/lrr-2009-2} {10.12942/lrr-2009-2} (\bibinfo
  {year} {2009})\BibitemShut {NoStop}%
\bibitem [{Note1()}]{Note1}%
  \BibitemOpen
  \bibinfo {note} {In paper~I, Eq.~(B1), we used the right ascension and
  declination of the GW propagation direction instead, which points in the
  opposite direction to the GW source}\BibitemShut {NoStop}%
\bibitem [{Note2()}]{Note2}%
  \BibitemOpen
  \bibinfo {note} {In paper~I, Eq.~(B1) contains a minor error, which is fixed
  by replacing all $\protect \qopname \relax o{sin}{\psi _0}$ with $ - \protect
  \qopname \relax o{sin}{\psi _0}$. We have corrected this in the erratum for
  paper~I, and here, we have also changed the notation from $\psi _0$ to $\phi
  _0$ for semantic reasons.}\BibitemShut {Stop}%
\bibitem [{\citenamefont
  {Rao}(2022)}]{Rao_Storage_Ring_Gravitational-wave_2022}%
  \BibitemOpen
  \bibfield  {author} {\bibinfo {author} {\bibfnamefont {S.}~\bibnamefont
  {Rao}},\ }\href {https://doi.org/10.5281/zenodo.7485016} {\bibinfo {title}
  {{Storage Ring Gravitational-wave Observatory (SRGO) simulation code,
  doi:10.5281/zenodo.7485016}}} (\bibinfo {year} {2022})\BibitemShut {NoStop}%
\bibitem [{\citenamefont {Blanchet}\ \emph {et~al.}(1996)\citenamefont
  {Blanchet}, \citenamefont {Iyer}, \citenamefont {Will},\ and\ \citenamefont
  {Wiseman}}]{Blanchet_1996}%
  \BibitemOpen
  \bibfield  {author} {\bibinfo {author} {\bibfnamefont {L.}~\bibnamefont
  {Blanchet}}, \bibinfo {author} {\bibfnamefont {B.~R.}\ \bibnamefont {Iyer}},
  \bibinfo {author} {\bibfnamefont {C.~M.}\ \bibnamefont {Will}},\ and\
  \bibinfo {author} {\bibfnamefont {A.~G.}\ \bibnamefont {Wiseman}},\
  }\bibfield  {title} {\bibinfo {title} {Gravitational waveforms from
  inspiralling compact binaries to second-post-newtonian order},\ }\href
  {https://doi.org/10.1088/0264-9381/13/4/002} {\bibfield  {journal} {\bibinfo
  {journal} {Classical and Quantum Gravity}\ }\textbf {\bibinfo {volume}
  {13}},\ \bibinfo {pages} {575} (\bibinfo {year} {1996})}\BibitemShut
  {NoStop}%
\bibitem [{\citenamefont {Blanchet}(2014)}]{Blanchet2014}%
  \BibitemOpen
  \bibfield  {author} {\bibinfo {author} {\bibfnamefont {L.}~\bibnamefont
  {Blanchet}},\ }\bibfield  {title} {\bibinfo {title} {Gravitational radiation
  from post-newtonian sources and inspiralling compact binaries},\ }\bibfield
  {journal} {\bibinfo  {journal} {Living Reviews in Relativity}\ }\textbf
  {\bibinfo {volume} {17}},\ \href {https://doi.org/10.12942/lrr-2014-2}
  {10.12942/lrr-2014-2} (\bibinfo {year} {2014})\BibitemShut {NoStop}%
\bibitem [{\citenamefont {Pen}(1999)}]{Pen_1999}%
  \BibitemOpen
  \bibfield  {author} {\bibinfo {author} {\bibfnamefont {U.-L.}\ \bibnamefont
  {Pen}},\ }\bibfield  {title} {\bibinfo {title} {Analytical fit to the
  luminosity distance for flat cosmologies with a cosmological constant},\
  }\href {https://doi.org/10.1086/313167} {\bibfield  {journal} {\bibinfo
  {journal} {The Astrophysical Journal Supplement Series}\ }\textbf {\bibinfo
  {volume} {120}},\ \bibinfo {pages} {49} (\bibinfo {year} {1999})}\BibitemShut
  {NoStop}%
\bibitem [{\citenamefont {Thrane}\ and\ \citenamefont
  {Talbot}(2019)}]{thrane_talbot_2019}%
  \BibitemOpen
  \bibfield  {author} {\bibinfo {author} {\bibfnamefont {E.}~\bibnamefont
  {Thrane}}\ and\ \bibinfo {author} {\bibfnamefont {C.}~\bibnamefont
  {Talbot}},\ }\bibfield  {title} {\bibinfo {title} {An introduction to
  bayesian inference in gravitational-wave astronomy: Parameter estimation,
  model selection, and hierarchical models},\ }\href
  {https://doi.org/10.1017/pasa.2019.2} {\bibfield  {journal} {\bibinfo
  {journal} {Publications of the Astronomical Society of Australia}\ }\textbf
  {\bibinfo {volume} {36}},\ \bibinfo {pages} {e010} (\bibinfo {year}
  {2019})}\BibitemShut {NoStop}%
\bibitem [{\citenamefont {Ubale}(2012)}]{Ubale2012}%
  \BibitemOpen
  \bibfield  {author} {\bibinfo {author} {\bibfnamefont {P.}~\bibnamefont
  {Ubale}},\ }\bibfield  {title} {\bibinfo {title} {Numerical solution of
  boole's rule in numerical integration by using general quadrature formula},\
  }\href {https://doi.org/10.18052/www.scipress.com/bsmass.2.1} {\bibfield
  {journal} {\bibinfo  {journal} {The Bulletin of Society for Mathematical
  Services and Standards}\ }\textbf {\bibinfo {volume} {2}},\ \bibinfo {pages}
  {1} (\bibinfo {year} {2012})}\BibitemShut {NoStop}%
\bibitem [{\citenamefont {van Ravenzwaaij}\ \emph {et~al.}(2016)\citenamefont
  {van Ravenzwaaij}, \citenamefont {Cassey},\ and\ \citenamefont
  {Brown}}]{vanRavenzwaaij2016}%
  \BibitemOpen
  \bibfield  {author} {\bibinfo {author} {\bibfnamefont {D.}~\bibnamefont {van
  Ravenzwaaij}}, \bibinfo {author} {\bibfnamefont {P.}~\bibnamefont {Cassey}},\
  and\ \bibinfo {author} {\bibfnamefont {S.~D.}\ \bibnamefont {Brown}},\
  }\bibfield  {title} {\bibinfo {title} {A simple introduction to markov chain
  monte{\textendash}carlo sampling},\ }\href
  {https://doi.org/10.3758/s13423-016-1015-8} {\bibfield  {journal} {\bibinfo
  {journal} {Psychonomic Bulletin {\&} Review}\ }\textbf {\bibinfo {volume}
  {25}},\ \bibinfo {pages} {143} (\bibinfo {year} {2016})}\BibitemShut
  {NoStop}%
\bibitem [{\citenamefont {Speagle}(2019)}]{1909.12313}%
  \BibitemOpen
  \bibfield  {author} {\bibinfo {author} {\bibfnamefont {J.~S.}\ \bibnamefont
  {Speagle}},\ }\href@noop {} {\bibinfo {title} {A conceptual introduction to
  markov chain monte carlo methods}} (\bibinfo {year} {2019}),\ \Eprint
  {https://arxiv.org/abs/arXiv:1909.12313} {arXiv:1909.12313} \BibitemShut
  {NoStop}%
\bibitem [{\citenamefont {Luengo}\ \emph {et~al.}(2020)\citenamefont {Luengo},
  \citenamefont {Martino}, \citenamefont {Bugallo}, \citenamefont {Elvira},\
  and\ \citenamefont {S\"{a}rkk\"{a}}}]{Luengo2020}%
  \BibitemOpen
  \bibfield  {author} {\bibinfo {author} {\bibfnamefont {D.}~\bibnamefont
  {Luengo}}, \bibinfo {author} {\bibfnamefont {L.}~\bibnamefont {Martino}},
  \bibinfo {author} {\bibfnamefont {M.}~\bibnamefont {Bugallo}}, \bibinfo
  {author} {\bibfnamefont {V.}~\bibnamefont {Elvira}},\ and\ \bibinfo {author}
  {\bibfnamefont {S.}~\bibnamefont {S\"{a}rkk\"{a}}},\ }\bibfield  {title}
  {\bibinfo {title} {A survey of monte carlo methods for parameter
  estimation},\ }\bibfield  {journal} {\bibinfo  {journal} {{EURASIP} Journal
  on Advances in Signal Processing}\ }\textbf {\bibinfo {volume} {2020}},\
  \href {https://doi.org/10.1186/s13634-020-00675-6}
  {10.1186/s13634-020-00675-6} (\bibinfo {year} {2020})\BibitemShut {NoStop}%
\bibitem [{\citenamefont {Braak}(2006)}]{Braak2006}%
  \BibitemOpen
  \bibfield  {author} {\bibinfo {author} {\bibfnamefont {C.~J. F.~T.}\
  \bibnamefont {Braak}},\ }\bibfield  {title} {\bibinfo {title} {A markov chain
  monte carlo version of the genetic algorithm differential evolution: easy
  bayesian computing for real parameter spaces},\ }\href
  {https://doi.org/10.1007/s11222-006-8769-1} {\bibfield  {journal} {\bibinfo
  {journal} {Statistics and Computing}\ }\textbf {\bibinfo {volume} {16}},\
  \bibinfo {pages} {239} (\bibinfo {year} {2006})}\BibitemShut {NoStop}%
\bibitem [{\citenamefont {Salvatier}\ \emph {et~al.}(2016)\citenamefont
  {Salvatier}, \citenamefont {Wiecki},\ and\ \citenamefont
  {Fonnesbeck}}]{Salvatier2016}%
  \BibitemOpen
  \bibfield  {author} {\bibinfo {author} {\bibfnamefont {J.}~\bibnamefont
  {Salvatier}}, \bibinfo {author} {\bibfnamefont {T.~V.}\ \bibnamefont
  {Wiecki}},\ and\ \bibinfo {author} {\bibfnamefont {C.}~\bibnamefont
  {Fonnesbeck}},\ }\bibfield  {title} {\bibinfo {title} {Probabilistic
  programming in python using {PyMC}3},\ }\href
  {https://doi.org/10.7717/peerj-cs.55} {\bibfield  {journal} {\bibinfo
  {journal} {{PeerJ} Computer Science}\ }\textbf {\bibinfo {volume} {2}},\
  \bibinfo {pages} {e55} (\bibinfo {year} {2016})}\BibitemShut {NoStop}%
\bibitem [{\citenamefont {Moore}\ \emph {et~al.}(2014)\citenamefont {Moore},
  \citenamefont {Cole},\ and\ \citenamefont {Berry}}]{Moore_2015}%
  \BibitemOpen
  \bibfield  {author} {\bibinfo {author} {\bibfnamefont {C.~J.}\ \bibnamefont
  {Moore}}, \bibinfo {author} {\bibfnamefont {R.~H.}\ \bibnamefont {Cole}},\
  and\ \bibinfo {author} {\bibfnamefont {C.~P.~L.}\ \bibnamefont {Berry}},\
  }\bibfield  {title} {\bibinfo {title} {Gravitational-wave sensitivity
  curves},\ }\href {https://doi.org/10.1088/0264-9381/32/1/015014} {\bibfield
  {journal} {\bibinfo  {journal} {Classical and Quantum Gravity}\ }\textbf
  {\bibinfo {volume} {32}},\ \bibinfo {pages} {015014} (\bibinfo {year}
  {2014})}\BibitemShut {NoStop}%
\bibitem [{Note3()}]{Note3}%
  \BibitemOpen
  \bibinfo {note} {The `LHC-GW' sensitivity curve shown in Fig.~1 of paper~I
  was not a conventional one, since it was not normalized for a fixed
  observation time. It could not be directly compared with the sensitivity
  curves of other GW detectors, and was therefore shown in a separate plot. In
  this study, since we plot the strain amplitude spectral density, this issue
  is resolved.}\BibitemShut {Stop}%
\bibitem [{\citenamefont {{Pourbaix, D.}}\ \emph {et~al.}(2004)\citenamefont
  {{Pourbaix, D.}}, \citenamefont {{Tokovinin, A. A.}}, \citenamefont {{Batten,
  A. H.}}, \citenamefont {{Fekel, F. C.}}, \citenamefont {{Hartkopf, W. I.}},
  \citenamefont {{Levato, H.}}, \citenamefont {{Morrell, N. I.}}, \citenamefont
  {{Torres, G.}},\ and\ \citenamefont {{Udry, S.}}}]{sb9}%
  \BibitemOpen
  \bibfield  {author} {\bibinfo {author} {\bibnamefont {{Pourbaix, D.}}},
  \bibinfo {author} {\bibnamefont {{Tokovinin, A. A.}}}, \bibinfo {author}
  {\bibnamefont {{Batten, A. H.}}}, \bibinfo {author} {\bibnamefont {{Fekel, F.
  C.}}}, \bibinfo {author} {\bibnamefont {{Hartkopf, W. I.}}}, \bibinfo
  {author} {\bibnamefont {{Levato, H.}}}, \bibinfo {author} {\bibnamefont
  {{Morrell, N. I.}}}, \bibinfo {author} {\bibnamefont {{Torres, G.}}},\ and\
  \bibinfo {author} {\bibnamefont {{Udry, S.}}},\ }\bibfield  {title} {\bibinfo
  {title} {Sb9: The ninth catalogue of spectroscopic binary orbits},\ }\href
  {https://doi.org/10.1051/0004-6361:20041213} {\bibfield  {journal} {\bibinfo
  {journal} {A\&A}\ }\textbf {\bibinfo {volume} {424}},\ \bibinfo {pages} {727}
  (\bibinfo {year} {2004})}\BibitemShut {NoStop}%
\bibitem [{\citenamefont {Prusti~\textit{et al.,}
  Gaia~Collaboration}(2016)}]{gaia1}%
  \BibitemOpen
  \bibfield  {author} {\bibinfo {author} {\bibfnamefont {T.}~\bibnamefont
  {Prusti~\textit{et al.,} Gaia~Collaboration}},\ }\bibfield  {title} {\bibinfo
  {title} {The gaia mission},\ }\href
  {https://doi.org/10.1051/0004-6361/201629272} {\bibfield  {journal} {\bibinfo
   {journal} {A\&A}\ }\textbf {\bibinfo {volume} {595}},\ \bibinfo {pages} {A1}
  (\bibinfo {year} {2016})}\BibitemShut {NoStop}%
\bibitem [{\citenamefont {Babusiaux}\ \emph {et~al.}(2022)\citenamefont
  {Babusiaux}, \citenamefont {Fabricius}, \citenamefont {Khanna}, \citenamefont
  {Muraveva}, \citenamefont {Reyl{\'{e}}}, \citenamefont {Spoto},\ and\
  \citenamefont {Vallenari}}]{gaia2}%
  \BibitemOpen
  \bibfield  {author} {\bibinfo {author} {\bibfnamefont {C.}~\bibnamefont
  {Babusiaux}}, \bibinfo {author} {\bibfnamefont {C.}~\bibnamefont
  {Fabricius}}, \bibinfo {author} {\bibfnamefont {S.}~\bibnamefont {Khanna}},
  \bibinfo {author} {\bibfnamefont {T.}~\bibnamefont {Muraveva}}, \bibinfo
  {author} {\bibfnamefont {C.}~\bibnamefont {Reyl{\'{e}}}}, \bibinfo {author}
  {\bibfnamefont {F.}~\bibnamefont {Spoto}},\ and\ \bibinfo {author}
  {\bibfnamefont {A.}~\bibnamefont {Vallenari}},\ }\bibfield  {title} {\bibinfo
  {title} {Gaia data release 3. catalogue validation},\ }\bibfield  {journal}
  {\bibinfo  {journal} {A\&A}\ }\href
  {https://doi.org/10.1051/0004-6361/202243790} {10.1051/0004-6361/202243790}
  (\bibinfo {year} {2022})\BibitemShut {NoStop}%
\bibitem [{\citenamefont {Vallenari}\ \emph {et~al.}(2022)\citenamefont
  {Vallenari}, \citenamefont {Brown},\ and\ \citenamefont {Prusti}}]{gaia3}%
  \BibitemOpen
  \bibfield  {author} {\bibinfo {author} {\bibfnamefont {A.}~\bibnamefont
  {Vallenari}}, \bibinfo {author} {\bibfnamefont {A.}~\bibnamefont {Brown}},\
  and\ \bibinfo {author} {\bibfnamefont {T.}~\bibnamefont {Prusti}},\
  }\bibfield  {title} {\bibinfo {title} {Gaia data release 3. {S}ummary of the
  content and survey properties},\ }\bibfield  {journal} {\bibinfo  {journal}
  {A\&A}\ }\href {https://doi.org/10.1051/0004-6361/202243940}
  {10.1051/0004-6361/202243940} (\bibinfo {year} {2022})\BibitemShut {NoStop}%
\bibitem [{\citenamefont {Farrow}\ \emph {et~al.}(2019)\citenamefont {Farrow},
  \citenamefont {Zhu},\ and\ \citenamefont {Thrane}}]{Farrow_2019}%
  \BibitemOpen
  \bibfield  {author} {\bibinfo {author} {\bibfnamefont {N.}~\bibnamefont
  {Farrow}}, \bibinfo {author} {\bibfnamefont {X.-J.}\ \bibnamefont {Zhu}},\
  and\ \bibinfo {author} {\bibfnamefont {E.}~\bibnamefont {Thrane}},\
  }\bibfield  {title} {\bibinfo {title} {The mass distribution of galactic
  double neutron stars},\ }\href {https://doi.org/10.3847/1538-4357/ab12e3}
  {\bibfield  {journal} {\bibinfo  {journal} {The Astrophysical Journal}\
  }\textbf {\bibinfo {volume} {876}},\ \bibinfo {pages} {18} (\bibinfo {year}
  {2019})}\BibitemShut {NoStop}%
\bibitem [{\citenamefont {Tauris}\ \emph {et~al.}(2017)\citenamefont {Tauris},
  \citenamefont {Kramer}, \citenamefont {Freire}, \citenamefont {Wex},
  \citenamefont {Janka}, \citenamefont {Langer}, \citenamefont {Podsiadlowski},
  \citenamefont {Bozzo}, \citenamefont {Chaty},\ and\ \citenamefont
  {Kruckow~\textit{et al.}}}]{Tauris_2017}%
  \BibitemOpen
  \bibfield  {author} {\bibinfo {author} {\bibfnamefont {T.~M.}\ \bibnamefont
  {Tauris}}, \bibinfo {author} {\bibfnamefont {M.}~\bibnamefont {Kramer}},
  \bibinfo {author} {\bibfnamefont {P.~C.~C.}\ \bibnamefont {Freire}}, \bibinfo
  {author} {\bibfnamefont {N.}~\bibnamefont {Wex}}, \bibinfo {author}
  {\bibfnamefont {H.-T.}\ \bibnamefont {Janka}}, \bibinfo {author}
  {\bibfnamefont {N.}~\bibnamefont {Langer}}, \bibinfo {author} {\bibfnamefont
  {P.}~\bibnamefont {Podsiadlowski}}, \bibinfo {author} {\bibfnamefont
  {E.}~\bibnamefont {Bozzo}}, \bibinfo {author} {\bibfnamefont
  {S.}~\bibnamefont {Chaty}},\ and\ \bibinfo {author} {\bibfnamefont {M.~U.}\
  \bibnamefont {Kruckow~\textit{et al.}}},\ }\bibfield  {title} {\bibinfo
  {title} {Formation of double neutron star systems},\ }\href
  {https://doi.org/10.3847/1538-4357/aa7e89} {\bibfield  {journal} {\bibinfo
  {journal} {The Astrophysical Journal}\ }\textbf {\bibinfo {volume} {846}},\
  \bibinfo {pages} {170} (\bibinfo {year} {2017})}\BibitemShut {NoStop}%
\bibitem [{\citenamefont {Cameron}\ \emph {et~al.}(2018)\citenamefont
  {Cameron}, \citenamefont {Champion}, \citenamefont {Kramer}, \citenamefont
  {Bailes}, \citenamefont {Barr}, \citenamefont {Bassa}, \citenamefont
  {Bhandari}, \citenamefont {Bhat}, \citenamefont {Burgay},\ and\ \citenamefont
  {Burke-Spolaor~\textit{et al.}}}]{cameron10.1093/mnrasl/sly003}%
  \BibitemOpen
  \bibfield  {author} {\bibinfo {author} {\bibfnamefont {A.~D.}\ \bibnamefont
  {Cameron}}, \bibinfo {author} {\bibfnamefont {D.~J.}\ \bibnamefont
  {Champion}}, \bibinfo {author} {\bibfnamefont {M.}~\bibnamefont {Kramer}},
  \bibinfo {author} {\bibfnamefont {M.}~\bibnamefont {Bailes}}, \bibinfo
  {author} {\bibfnamefont {E.~D.}\ \bibnamefont {Barr}}, \bibinfo {author}
  {\bibfnamefont {C.~G.}\ \bibnamefont {Bassa}}, \bibinfo {author}
  {\bibfnamefont {S.}~\bibnamefont {Bhandari}}, \bibinfo {author}
  {\bibfnamefont {N.~D.~R.}\ \bibnamefont {Bhat}}, \bibinfo {author}
  {\bibfnamefont {M.}~\bibnamefont {Burgay}},\ and\ \bibinfo {author}
  {\bibfnamefont {S.}~\bibnamefont {Burke-Spolaor~\textit{et al.}}},\
  }\bibfield  {title} {\bibinfo {title} {{The High Time Resolution Universe
  Pulsar Survey -- XIII. PSR J1757--1854, the most accelerated binary
  pulsar}},\ }\href {https://doi.org/10.1093/mnrasl/sly003} {\bibfield
  {journal} {\bibinfo  {journal} {Monthly Notices of the Royal Astronomical
  Society: Letters}\ }\textbf {\bibinfo {volume} {475}},\ \bibinfo {pages}
  {L57} (\bibinfo {year} {2018})},\ \Eprint
  {https://arxiv.org/abs/https://academic.oup.com/mnrasl/article-pdf/475/1/L57/24841638/sly003.pdf}
  {https://academic.oup.com/mnrasl/article-pdf/475/1/L57/24841638/sly003.pdf}
  \BibitemShut {NoStop}%
\bibitem [{\citenamefont {Stovall}\ \emph {et~al.}(2018)\citenamefont
  {Stovall}, \citenamefont {Freire}, \citenamefont {Chatterjee}, \citenamefont
  {Demorest}, \citenamefont {Lorimer}, \citenamefont {McLaughlin},
  \citenamefont {Pol}, \citenamefont {van Leeuwen}, \citenamefont {Wharton},\
  and\ \citenamefont {Allen~\textit{et al.}}}]{Stovall_2018}%
  \BibitemOpen
  \bibfield  {author} {\bibinfo {author} {\bibfnamefont {K.}~\bibnamefont
  {Stovall}}, \bibinfo {author} {\bibfnamefont {P.~C.~C.}\ \bibnamefont
  {Freire}}, \bibinfo {author} {\bibfnamefont {S.}~\bibnamefont {Chatterjee}},
  \bibinfo {author} {\bibfnamefont {P.~B.}\ \bibnamefont {Demorest}}, \bibinfo
  {author} {\bibfnamefont {D.~R.}\ \bibnamefont {Lorimer}}, \bibinfo {author}
  {\bibfnamefont {M.~A.}\ \bibnamefont {McLaughlin}}, \bibinfo {author}
  {\bibfnamefont {N.}~\bibnamefont {Pol}}, \bibinfo {author} {\bibfnamefont
  {J.}~\bibnamefont {van Leeuwen}}, \bibinfo {author} {\bibfnamefont {R.~S.}\
  \bibnamefont {Wharton}},\ and\ \bibinfo {author} {\bibfnamefont
  {B.}~\bibnamefont {Allen~\textit{et al.}}},\ }\bibfield  {title} {\bibinfo
  {title} {{PALFA} discovery of a highly relativistic double neutron star
  binary},\ }\href {https://doi.org/10.3847/2041-8213/aaad06} {\bibfield
  {journal} {\bibinfo  {journal} {The Astrophysical Journal}\ }\textbf
  {\bibinfo {volume} {854}},\ \bibinfo {pages} {L22} (\bibinfo {year}
  {2018})}\BibitemShut {NoStop}%
\bibitem [{\citenamefont {Lynch}\ \emph {et~al.}(2018)\citenamefont {Lynch},
  \citenamefont {Swiggum}, \citenamefont {Kondratiev}, \citenamefont {Kaplan},
  \citenamefont {Stovall}, \citenamefont {Fonseca}, \citenamefont {Roberts},
  \citenamefont {Levin}, \citenamefont {DeCesar},\ and\ \citenamefont
  {Cui~\textit{et al.}}}]{Lynch_2018}%
  \BibitemOpen
  \bibfield  {author} {\bibinfo {author} {\bibfnamefont {R.~S.}\ \bibnamefont
  {Lynch}}, \bibinfo {author} {\bibfnamefont {J.~K.}\ \bibnamefont {Swiggum}},
  \bibinfo {author} {\bibfnamefont {V.~I.}\ \bibnamefont {Kondratiev}},
  \bibinfo {author} {\bibfnamefont {D.~L.}\ \bibnamefont {Kaplan}}, \bibinfo
  {author} {\bibfnamefont {K.}~\bibnamefont {Stovall}}, \bibinfo {author}
  {\bibfnamefont {E.}~\bibnamefont {Fonseca}}, \bibinfo {author} {\bibfnamefont
  {M.~S.~E.}\ \bibnamefont {Roberts}}, \bibinfo {author} {\bibfnamefont
  {L.}~\bibnamefont {Levin}}, \bibinfo {author} {\bibfnamefont {M.~E.}\
  \bibnamefont {DeCesar}},\ and\ \bibinfo {author} {\bibfnamefont
  {B.}~\bibnamefont {Cui~\textit{et al.}}},\ }\bibfield  {title} {\bibinfo
  {title} {The {G}reen {B}ank {N}orth {C}elestial {C}ap pulsar survey. {III}.
  45 new pulsar timing solutions},\ }\href
  {https://doi.org/10.3847/1538-4357/aabf8a} {\bibfield  {journal} {\bibinfo
  {journal} {The Astrophysical Journal}\ }\textbf {\bibinfo {volume} {859}},\
  \bibinfo {pages} {93} (\bibinfo {year} {2018})}\BibitemShut {NoStop}%
\bibitem [{\citenamefont {Kaplan}\ \emph {et~al.}(2014)\citenamefont {Kaplan},
  \citenamefont {Boyles}, \citenamefont {Dunlap}, \citenamefont {Tendulkar},
  \citenamefont {Deller}, \citenamefont {Ransom}, \citenamefont {McLaughlin},
  \citenamefont {Lorimer},\ and\ \citenamefont {Stairs}}]{Kaplan_2014}%
  \BibitemOpen
  \bibfield  {author} {\bibinfo {author} {\bibfnamefont {D.~L.}\ \bibnamefont
  {Kaplan}}, \bibinfo {author} {\bibfnamefont {J.}~\bibnamefont {Boyles}},
  \bibinfo {author} {\bibfnamefont {B.~H.}\ \bibnamefont {Dunlap}}, \bibinfo
  {author} {\bibfnamefont {S.~P.}\ \bibnamefont {Tendulkar}}, \bibinfo {author}
  {\bibfnamefont {A.~T.}\ \bibnamefont {Deller}}, \bibinfo {author}
  {\bibfnamefont {S.~M.}\ \bibnamefont {Ransom}}, \bibinfo {author}
  {\bibfnamefont {M.~A.}\ \bibnamefont {McLaughlin}}, \bibinfo {author}
  {\bibfnamefont {D.~R.}\ \bibnamefont {Lorimer}},\ and\ \bibinfo {author}
  {\bibfnamefont {I.~H.}\ \bibnamefont {Stairs}},\ }\bibfield  {title}
  {\bibinfo {title} {A 1.05$m_{\odot}$ companion to {PSR}
  {J}2222\textemdash0137: The coolest known white dwarf?},\ }\href
  {https://doi.org/10.1088/0004-637x/789/2/119} {\bibfield  {journal} {\bibinfo
   {journal} {The Astrophysical Journal}\ }\textbf {\bibinfo {volume} {789}},\
  \bibinfo {pages} {119} (\bibinfo {year} {2014})}\BibitemShut {NoStop}%
\bibitem [{\citenamefont {Lyne}\ \emph {et~al.}(2004)\citenamefont {Lyne},
  \citenamefont {Burgay}, \citenamefont {Kramer}, \citenamefont {Possenti},
  \citenamefont {Manchester}, \citenamefont {Camilo}, \citenamefont
  {McLaughlin}, \citenamefont {Lorimer}, \citenamefont {D'Amico},\ and\
  \citenamefont {Joshi~\textit{et al.}}}]{Lyne2004}%
  \BibitemOpen
  \bibfield  {author} {\bibinfo {author} {\bibfnamefont {A.~G.}\ \bibnamefont
  {Lyne}}, \bibinfo {author} {\bibfnamefont {M.}~\bibnamefont {Burgay}},
  \bibinfo {author} {\bibfnamefont {M.}~\bibnamefont {Kramer}}, \bibinfo
  {author} {\bibfnamefont {A.}~\bibnamefont {Possenti}}, \bibinfo {author}
  {\bibfnamefont {R.}~\bibnamefont {Manchester}}, \bibinfo {author}
  {\bibfnamefont {F.}~\bibnamefont {Camilo}}, \bibinfo {author} {\bibfnamefont
  {M.~A.}\ \bibnamefont {McLaughlin}}, \bibinfo {author} {\bibfnamefont
  {D.~R.}\ \bibnamefont {Lorimer}}, \bibinfo {author} {\bibfnamefont
  {N.}~\bibnamefont {D'Amico}},\ and\ \bibinfo {author} {\bibfnamefont {B.~C.}\
  \bibnamefont {Joshi~\textit{et al.}}},\ }\bibfield  {title} {\bibinfo {title}
  {A double-pulsar system: A rare laboratory for relativistic gravity and
  plasma physics},\ }\href {https://doi.org/10.1126/science.1094645} {\bibfield
   {journal} {\bibinfo  {journal} {Science}\ }\textbf {\bibinfo {volume}
  {303}},\ \bibinfo {pages} {1153} (\bibinfo {year} {2004})},\ \Eprint
  {https://arxiv.org/abs/https://www.science.org/doi/pdf/10.1126/science.1094645}
  {https://www.science.org/doi/pdf/10.1126/science.1094645} \BibitemShut
  {NoStop}%
\bibitem [{\citenamefont {Kramer}\ \emph {et~al.}(2021)\citenamefont {Kramer},
  \citenamefont {Stairs}, \citenamefont {Manchester}, \citenamefont {Wex},
  \citenamefont {Deller}, \citenamefont {Coles}, \citenamefont {Ali},
  \citenamefont {Burgay}, \citenamefont {Camilo},\ and\ \citenamefont
  {Cognard~\textit{et al.}}}]{PhysRevX.11.041050}%
  \BibitemOpen
  \bibfield  {author} {\bibinfo {author} {\bibfnamefont {M.}~\bibnamefont
  {Kramer}}, \bibinfo {author} {\bibfnamefont {I.~H.}\ \bibnamefont {Stairs}},
  \bibinfo {author} {\bibfnamefont {R.~N.}\ \bibnamefont {Manchester}},
  \bibinfo {author} {\bibfnamefont {N.}~\bibnamefont {Wex}}, \bibinfo {author}
  {\bibfnamefont {A.~T.}\ \bibnamefont {Deller}}, \bibinfo {author}
  {\bibfnamefont {W.~A.}\ \bibnamefont {Coles}}, \bibinfo {author}
  {\bibfnamefont {M.}~\bibnamefont {Ali}}, \bibinfo {author} {\bibfnamefont
  {M.}~\bibnamefont {Burgay}}, \bibinfo {author} {\bibfnamefont
  {F.}~\bibnamefont {Camilo}},\ and\ \bibinfo {author} {\bibfnamefont
  {I.}~\bibnamefont {Cognard~\textit{et al.}}},\ }\bibfield  {title} {\bibinfo
  {title} {Strong-field gravity tests with the double pulsar},\ }\href
  {https://doi.org/10.1103/PhysRevX.11.041050} {\bibfield  {journal} {\bibinfo
  {journal} {Phys. Rev. X}\ }\textbf {\bibinfo {volume} {11}},\ \bibinfo
  {pages} {041050} (\bibinfo {year} {2021})}\BibitemShut {NoStop}%
\bibitem [{\citenamefont {Lamberts}\ \emph {et~al.}(2018)\citenamefont
  {Lamberts}, \citenamefont {Garrison-Kimmel}, \citenamefont {Hopkins},
  \citenamefont {Quataert}, \citenamefont {Bullock}, \citenamefont
  {Faucher-Giguère}, \citenamefont {Wetzel}, \citenamefont {Kereš},
  \citenamefont {Drango},\ and\ \citenamefont {Sanderson}}]{mwbhs}%
  \BibitemOpen
  \bibfield  {author} {\bibinfo {author} {\bibfnamefont {A.}~\bibnamefont
  {Lamberts}}, \bibinfo {author} {\bibfnamefont {S.}~\bibnamefont
  {Garrison-Kimmel}}, \bibinfo {author} {\bibfnamefont {P.~F.}\ \bibnamefont
  {Hopkins}}, \bibinfo {author} {\bibfnamefont {E.}~\bibnamefont {Quataert}},
  \bibinfo {author} {\bibfnamefont {J.~S.}\ \bibnamefont {Bullock}}, \bibinfo
  {author} {\bibfnamefont {C.-A.}\ \bibnamefont {Faucher-Giguère}}, \bibinfo
  {author} {\bibfnamefont {A.}~\bibnamefont {Wetzel}}, \bibinfo {author}
  {\bibfnamefont {D.}~\bibnamefont {Kereš}}, \bibinfo {author} {\bibfnamefont
  {K.}~\bibnamefont {Drango}},\ and\ \bibinfo {author} {\bibfnamefont {R.~E.}\
  \bibnamefont {Sanderson}},\ }\bibfield  {title} {\bibinfo {title}
  {{Predicting the binary black hole population of the Milky Way with
  cosmological simulations}},\ }\href {https://doi.org/10.1093/mnras/sty2035}
  {\bibfield  {journal} {\bibinfo  {journal} {Monthly Notices of the Royal
  Astronomical Society}\ }\textbf {\bibinfo {volume} {480}},\ \bibinfo {pages}
  {2704} (\bibinfo {year} {2018})},\ \Eprint
  {https://arxiv.org/abs/https://academic.oup.com/mnras/article-pdf/480/2/2704/25495274/sty2035.pdf}
  {https://academic.oup.com/mnras/article-pdf/480/2/2704/25495274/sty2035.pdf}
  \BibitemShut {NoStop}%
\bibitem [{\citenamefont {Sesana}\ \emph {et~al.}(2020)\citenamefont {Sesana},
  \citenamefont {Lamberts},\ and\ \citenamefont {Petiteau}}]{mwbbhs}%
  \BibitemOpen
  \bibfield  {author} {\bibinfo {author} {\bibfnamefont {A.}~\bibnamefont
  {Sesana}}, \bibinfo {author} {\bibfnamefont {A.}~\bibnamefont {Lamberts}},\
  and\ \bibinfo {author} {\bibfnamefont {A.}~\bibnamefont {Petiteau}},\
  }\bibfield  {title} {\bibinfo {title} {{Finding binary black holes in the
  Milky Way with LISA}},\ }\href {https://doi.org/10.1093/mnrasl/slaa039}
  {\bibfield  {journal} {\bibinfo  {journal} {Monthly Notices of the Royal
  Astronomical Society: Letters}\ }\textbf {\bibinfo {volume} {494}},\ \bibinfo
  {pages} {L75} (\bibinfo {year} {2020})},\ \Eprint
  {https://arxiv.org/abs/https://academic.oup.com/mnrasl/article-pdf/494/1/L75/32977056/slaa039.pdf}
  {https://academic.oup.com/mnrasl/article-pdf/494/1/L75/32977056/slaa039.pdf}
  \BibitemShut {NoStop}%
\bibitem [{\citenamefont {Sahu}\ \emph {et~al.}(2022)\citenamefont {Sahu},
  \citenamefont {Anderson}, \citenamefont {Casertano}, \citenamefont {Bond},
  \citenamefont {Udalski}, \citenamefont {Dominik}, \citenamefont {Calamida},
  \citenamefont {Bellini}, \citenamefont {Brown},\ and\ \citenamefont
  {Rejkuba~\textit{et al.}}}]{Sahu_2022}%
  \BibitemOpen
  \bibfield  {author} {\bibinfo {author} {\bibfnamefont {K.~C.}\ \bibnamefont
  {Sahu}}, \bibinfo {author} {\bibfnamefont {J.}~\bibnamefont {Anderson}},
  \bibinfo {author} {\bibfnamefont {S.}~\bibnamefont {Casertano}}, \bibinfo
  {author} {\bibfnamefont {H.~E.}\ \bibnamefont {Bond}}, \bibinfo {author}
  {\bibfnamefont {A.}~\bibnamefont {Udalski}}, \bibinfo {author} {\bibfnamefont
  {M.}~\bibnamefont {Dominik}}, \bibinfo {author} {\bibfnamefont
  {A.}~\bibnamefont {Calamida}}, \bibinfo {author} {\bibfnamefont
  {A.}~\bibnamefont {Bellini}}, \bibinfo {author} {\bibfnamefont {T.~M.}\
  \bibnamefont {Brown}},\ and\ \bibinfo {author} {\bibfnamefont
  {M.}~\bibnamefont {Rejkuba~\textit{et al.}}},\ }\bibfield  {title} {\bibinfo
  {title} {An isolated stellar-mass black hole detected through astrometric
  microlensing*},\ }\href {https://doi.org/10.3847/1538-4357/ac739e} {\bibfield
   {journal} {\bibinfo  {journal} {The Astrophysical Journal}\ }\textbf
  {\bibinfo {volume} {933}},\ \bibinfo {pages} {83} (\bibinfo {year}
  {2022})}\BibitemShut {NoStop}%
\bibitem [{\citenamefont {Tamfal}\ \emph {et~al.}(2018)\citenamefont {Tamfal},
  \citenamefont {Capelo}, \citenamefont {Kazantzidis}, \citenamefont {Mayer},
  \citenamefont {Potter}, \citenamefont {Stadel},\ and\ \citenamefont
  {Widrow}}]{imbh1}%
  \BibitemOpen
  \bibfield  {author} {\bibinfo {author} {\bibfnamefont {T.}~\bibnamefont
  {Tamfal}}, \bibinfo {author} {\bibfnamefont {P.~R.}\ \bibnamefont {Capelo}},
  \bibinfo {author} {\bibfnamefont {S.}~\bibnamefont {Kazantzidis}}, \bibinfo
  {author} {\bibfnamefont {L.}~\bibnamefont {Mayer}}, \bibinfo {author}
  {\bibfnamefont {D.}~\bibnamefont {Potter}}, \bibinfo {author} {\bibfnamefont
  {J.}~\bibnamefont {Stadel}},\ and\ \bibinfo {author} {\bibfnamefont {L.~M.}\
  \bibnamefont {Widrow}},\ }\bibfield  {title} {\bibinfo {title} {Formation of
  {LISA} black hole binaries in merging dwarf galaxies: The imprint of dark
  matter},\ }\href {https://doi.org/10.3847/2041-8213/aada4b} {\bibfield
  {journal} {\bibinfo  {journal} {The Astrophysical Journal}\ }\textbf
  {\bibinfo {volume} {864}},\ \bibinfo {pages} {L19} (\bibinfo {year}
  {2018})}\BibitemShut {NoStop}%
\bibitem [{\citenamefont {Weller}\ \emph {et~al.}(2022)\citenamefont {Weller},
  \citenamefont {Pacucci}, \citenamefont {Hernquist},\ and\ \citenamefont
  {Bose}}]{imbh2}%
  \BibitemOpen
  \bibfield  {author} {\bibinfo {author} {\bibfnamefont {E.~J.}\ \bibnamefont
  {Weller}}, \bibinfo {author} {\bibfnamefont {F.}~\bibnamefont {Pacucci}},
  \bibinfo {author} {\bibfnamefont {L.}~\bibnamefont {Hernquist}},\ and\
  \bibinfo {author} {\bibfnamefont {S.}~\bibnamefont {Bose}},\ }\bibfield
  {title} {\bibinfo {title} {{Dynamics of intermediate-mass black holes
  wandering in the milky way galaxy using the illustris TNG50 simulation}},\
  }\href {https://doi.org/10.1093/mnras/stac179} {\bibfield  {journal}
  {\bibinfo  {journal} {Monthly Notices of the Royal Astronomical Society}\
  }\textbf {\bibinfo {volume} {511}},\ \bibinfo {pages} {2229} (\bibinfo {year}
  {2022})},\ \Eprint
  {https://arxiv.org/abs/https://academic.oup.com/mnras/article-pdf/511/2/2229/42500268/stac179.pdf}
  {https://academic.oup.com/mnras/article-pdf/511/2/2229/42500268/stac179.pdf}
  \BibitemShut {NoStop}%
\bibitem [{\citenamefont {Fragione}\ \emph
  {et~al.}(2018{\natexlab{a}})\citenamefont {Fragione}, \citenamefont
  {Ginsburg},\ and\ \citenamefont {Kocsis}}]{imbh3}%
  \BibitemOpen
  \bibfield  {author} {\bibinfo {author} {\bibfnamefont {G.}~\bibnamefont
  {Fragione}}, \bibinfo {author} {\bibfnamefont {I.}~\bibnamefont {Ginsburg}},\
  and\ \bibinfo {author} {\bibfnamefont {B.}~\bibnamefont {Kocsis}},\
  }\bibfield  {title} {\bibinfo {title} {Gravitational waves and
  intermediate-mass black hole retention in globular clusters},\ }\href
  {https://doi.org/10.3847/1538-4357/aab368} {\bibfield  {journal} {\bibinfo
  {journal} {The Astrophysical Journal}\ }\textbf {\bibinfo {volume} {856}},\
  \bibinfo {pages} {92} (\bibinfo {year} {2018}{\natexlab{a}})}\BibitemShut
  {NoStop}%
\bibitem [{\citenamefont {Fragione}\ \emph
  {et~al.}(2018{\natexlab{b}})\citenamefont {Fragione}, \citenamefont {Leigh},
  \citenamefont {Ginsburg},\ and\ \citenamefont {Kocsis}}]{imbh4}%
  \BibitemOpen
  \bibfield  {author} {\bibinfo {author} {\bibfnamefont {G.}~\bibnamefont
  {Fragione}}, \bibinfo {author} {\bibfnamefont {N.~W.~C.}\ \bibnamefont
  {Leigh}}, \bibinfo {author} {\bibfnamefont {I.}~\bibnamefont {Ginsburg}},\
  and\ \bibinfo {author} {\bibfnamefont {B.}~\bibnamefont {Kocsis}},\
  }\bibfield  {title} {\bibinfo {title} {Tidal disruption events and
  gravitational waves from intermediate-mass black holes in evolving globular
  clusters across space and time},\ }\href
  {https://doi.org/10.3847/1538-4357/aae486} {\bibfield  {journal} {\bibinfo
  {journal} {The Astrophysical Journal}\ }\textbf {\bibinfo {volume} {867}},\
  \bibinfo {pages} {119} (\bibinfo {year} {2018}{\natexlab{b}})}\BibitemShut
  {NoStop}%
\bibitem [{\citenamefont {Ricarte}\ \emph {et~al.}(2021)\citenamefont
  {Ricarte}, \citenamefont {Tremmel}, \citenamefont {Natarajan},\ and\
  \citenamefont {Quinn}}]{Ricarte_2021}%
  \BibitemOpen
  \bibfield  {author} {\bibinfo {author} {\bibfnamefont {A.}~\bibnamefont
  {Ricarte}}, \bibinfo {author} {\bibfnamefont {M.}~\bibnamefont {Tremmel}},
  \bibinfo {author} {\bibfnamefont {P.}~\bibnamefont {Natarajan}},\ and\
  \bibinfo {author} {\bibfnamefont {T.}~\bibnamefont {Quinn}},\ }\bibfield
  {title} {\bibinfo {title} {Unveiling the population of wandering black holes
  via electromagnetic signatures},\ }\href
  {https://doi.org/10.3847/2041-8213/ac1170} {\bibfield  {journal} {\bibinfo
  {journal} {The Astrophysical Journal Letters}\ }\textbf {\bibinfo {volume}
  {916}},\ \bibinfo {pages} {L18} (\bibinfo {year} {2021})}\BibitemShut
  {NoStop}%
\bibitem [{\citenamefont {Collaboration}\ and\ \citenamefont
  {Akiyama~\textit{et al.}}(2022)}]{EHTC_2022}%
  \BibitemOpen
  \bibfield  {author} {\bibinfo {author} {\bibfnamefont {E.~H.~T.}\
  \bibnamefont {Collaboration}}\ and\ \bibinfo {author} {\bibfnamefont
  {K.}~\bibnamefont {Akiyama~\textit{et al.}}},\ }\bibfield  {title} {\bibinfo
  {title} {First sagittarius a* event horizon telescope results. i. the shadow
  of the supermassive black hole in the center of the milky way},\ }\href
  {https://doi.org/10.3847/2041-8213/ac6674} {\bibfield  {journal} {\bibinfo
  {journal} {The Astrophysical Journal Letters}\ }\textbf {\bibinfo {volume}
  {930}},\ \bibinfo {pages} {L12} (\bibinfo {year} {2022})}\BibitemShut
  {NoStop}%
\bibitem [{\citenamefont {Berti}(2006)}]{Berti_2006}%
  \BibitemOpen
  \bibfield  {author} {\bibinfo {author} {\bibfnamefont {E.}~\bibnamefont
  {Berti}},\ }\bibfield  {title} {\bibinfo {title} {{LISA} observations of
  massive black hole mergers: event rates and issues in waveform modelling},\
  }\href {https://doi.org/10.1088/0264-9381/23/19/s17} {\bibfield  {journal}
  {\bibinfo  {journal} {Classical and Quantum Gravity}\ }\textbf {\bibinfo
  {volume} {23}},\ \bibinfo {pages} {S785} (\bibinfo {year}
  {2006})}\BibitemShut {NoStop}%
\bibitem [{\citenamefont {Bender}\ \emph {et~al.}(2005)\citenamefont {Bender},
  \citenamefont {Kormendy}, \citenamefont {Bower}, \citenamefont {Green},
  \citenamefont {Thomas}, \citenamefont {Danks}, \citenamefont {Gull},
  \citenamefont {Hutchings}, \citenamefont {Joseph}, \citenamefont {Kaiser},
  \citenamefont {Lauer}, \citenamefont {Nelson}, \citenamefont {Richstone},
  \citenamefont {Weistrop},\ and\ \citenamefont {Woodgate}}]{Bender_2005}%
  \BibitemOpen
  \bibfield  {author} {\bibinfo {author} {\bibfnamefont {R.}~\bibnamefont
  {Bender}}, \bibinfo {author} {\bibfnamefont {J.}~\bibnamefont {Kormendy}},
  \bibinfo {author} {\bibfnamefont {G.}~\bibnamefont {Bower}}, \bibinfo
  {author} {\bibfnamefont {R.}~\bibnamefont {Green}}, \bibinfo {author}
  {\bibfnamefont {J.}~\bibnamefont {Thomas}}, \bibinfo {author} {\bibfnamefont
  {A.~C.}\ \bibnamefont {Danks}}, \bibinfo {author} {\bibfnamefont
  {T.}~\bibnamefont {Gull}}, \bibinfo {author} {\bibfnamefont {J.~B.}\
  \bibnamefont {Hutchings}}, \bibinfo {author} {\bibfnamefont {C.~L.}\
  \bibnamefont {Joseph}}, \bibinfo {author} {\bibfnamefont {M.~E.}\
  \bibnamefont {Kaiser}}, \bibinfo {author} {\bibfnamefont {T.~R.}\
  \bibnamefont {Lauer}}, \bibinfo {author} {\bibfnamefont {C.~H.}\ \bibnamefont
  {Nelson}}, \bibinfo {author} {\bibfnamefont {D.}~\bibnamefont {Richstone}},
  \bibinfo {author} {\bibfnamefont {D.}~\bibnamefont {Weistrop}},\ and\
  \bibinfo {author} {\bibfnamefont {B.}~\bibnamefont {Woodgate}},\ }\bibfield
  {title} {\bibinfo {title} {Hst stis spectroscopy of the triple nucleus of
  m31: Two nested disks in keplerian rotation around a supermassive black
  hole},\ }\href {https://doi.org/10.1086/432434} {\bibfield  {journal}
  {\bibinfo  {journal} {The Astrophysical Journal}\ }\textbf {\bibinfo {volume}
  {631}},\ \bibinfo {pages} {280} (\bibinfo {year} {2005})}\BibitemShut
  {NoStop}%
\bibitem [{\citenamefont {Merloni}\ and\ \citenamefont {Heinz}(2008)}]{smbh1}%
  \BibitemOpen
  \bibfield  {author} {\bibinfo {author} {\bibfnamefont {A.}~\bibnamefont
  {Merloni}}\ and\ \bibinfo {author} {\bibfnamefont {S.}~\bibnamefont
  {Heinz}},\ }\bibfield  {title} {\bibinfo {title} {{A synthesis model for AGN
  evolution: supermassive black holes growth and feedback modes}},\ }\href
  {https://doi.org/10.1111/j.1365-2966.2008.13472.x} {\bibfield  {journal}
  {\bibinfo  {journal} {Monthly Notices of the Royal Astronomical Society}\
  }\textbf {\bibinfo {volume} {388}},\ \bibinfo {pages} {1011} (\bibinfo {year}
  {2008})},\ \Eprint
  {https://arxiv.org/abs/https://academic.oup.com/mnras/article-pdf/388/3/1011/2802941/mnras0388-1011.pdf}
  {https://academic.oup.com/mnras/article-pdf/388/3/1011/2802941/mnras0388-1011.pdf}
  \BibitemShut {NoStop}%
\bibitem [{\citenamefont {Habouzit}\ \emph {et~al.}(2021)\citenamefont
  {Habouzit}, \citenamefont {Li}, \citenamefont {Somerville}, \citenamefont
  {Genel}, \citenamefont {Pillepich}, \citenamefont {Volonteri}, \citenamefont
  {Davé}, \citenamefont {Rosas-Guevara}, \citenamefont {McAlpine},\ and\
  \citenamefont {Sébastien~\textit{et al.}}}]{smbh2}%
  \BibitemOpen
  \bibfield  {author} {\bibinfo {author} {\bibfnamefont {M.}~\bibnamefont
  {Habouzit}}, \bibinfo {author} {\bibfnamefont {Y.}~\bibnamefont {Li}},
  \bibinfo {author} {\bibfnamefont {R.~S.}\ \bibnamefont {Somerville}},
  \bibinfo {author} {\bibfnamefont {S.}~\bibnamefont {Genel}}, \bibinfo
  {author} {\bibfnamefont {A.}~\bibnamefont {Pillepich}}, \bibinfo {author}
  {\bibfnamefont {M.}~\bibnamefont {Volonteri}}, \bibinfo {author}
  {\bibfnamefont {R.}~\bibnamefont {Davé}}, \bibinfo {author} {\bibfnamefont
  {Y.}~\bibnamefont {Rosas-Guevara}}, \bibinfo {author} {\bibfnamefont
  {S.}~\bibnamefont {McAlpine}},\ and\ \bibinfo {author} {\bibfnamefont
  {P.}~\bibnamefont {Sébastien~\textit{et al.}}},\ }\bibfield  {title}
  {\bibinfo {title} {{Supermassive black holes in cosmological simulations I:
  MBH -- M$_{\star}$ relation and black hole mass function}},\ }\href
  {https://doi.org/10.1093/mnras/stab496} {\bibfield  {journal} {\bibinfo
  {journal} {Monthly Notices of the Royal Astronomical Society}\ }\textbf
  {\bibinfo {volume} {503}},\ \bibinfo {pages} {1940} (\bibinfo {year}
  {2021})},\ \Eprint
  {https://arxiv.org/abs/https://academic.oup.com/mnras/article-pdf/503/2/1940/36678260/stab496.pdf}
  {https://academic.oup.com/mnras/article-pdf/503/2/1940/36678260/stab496.pdf}
  \BibitemShut {NoStop}%
\bibitem [{\citenamefont {Sicilia}\ \emph {et~al.}(2022)\citenamefont
  {Sicilia}, \citenamefont {Lapi}, \citenamefont {Boco}, \citenamefont
  {Shankar}, \citenamefont {Alexander}, \citenamefont {Allevato}, \citenamefont
  {Villforth}, \citenamefont {Massardi}, \citenamefont {Spera},\ and\
  \citenamefont {Bressan~\textit{et al.}}}]{smbh3}%
  \BibitemOpen
  \bibfield  {author} {\bibinfo {author} {\bibfnamefont {A.}~\bibnamefont
  {Sicilia}}, \bibinfo {author} {\bibfnamefont {A.}~\bibnamefont {Lapi}},
  \bibinfo {author} {\bibfnamefont {L.}~\bibnamefont {Boco}}, \bibinfo {author}
  {\bibfnamefont {F.}~\bibnamefont {Shankar}}, \bibinfo {author} {\bibfnamefont
  {D.~M.}\ \bibnamefont {Alexander}}, \bibinfo {author} {\bibfnamefont
  {V.}~\bibnamefont {Allevato}}, \bibinfo {author} {\bibfnamefont
  {C.}~\bibnamefont {Villforth}}, \bibinfo {author} {\bibfnamefont
  {M.}~\bibnamefont {Massardi}}, \bibinfo {author} {\bibfnamefont
  {M.}~\bibnamefont {Spera}},\ and\ \bibinfo {author} {\bibfnamefont
  {A.}~\bibnamefont {Bressan~\textit{et al.}}},\ }\bibfield  {title} {\bibinfo
  {title} {The black hole mass function across cosmic time. {II}. heavy seeds
  and (super)massive black holes},\ }\href
  {https://doi.org/10.3847/1538-4357/ac7873} {\bibfield  {journal} {\bibinfo
  {journal} {The Astrophysical Journal}\ }\textbf {\bibinfo {volume} {934}},\
  \bibinfo {pages} {66} (\bibinfo {year} {2022})}\BibitemShut {NoStop}%
\bibitem [{\citenamefont {Voggel}\ \emph {et~al.}(2022)\citenamefont {Voggel},
  \citenamefont {Seth}, \citenamefont {Baumgardt}, \citenamefont {Husemann},
  \citenamefont {Neumayer}, \citenamefont {Hilker}, \citenamefont {Pechetti},
  \citenamefont {Mieske}, \citenamefont {Dumont},\ and\ \citenamefont
  {Georgiev}}]{smbh4}%
  \BibitemOpen
  \bibfield  {author} {\bibinfo {author} {\bibfnamefont {K.~T.}\ \bibnamefont
  {Voggel}}, \bibinfo {author} {\bibfnamefont {A.~C.}\ \bibnamefont {Seth}},
  \bibinfo {author} {\bibfnamefont {H.}~\bibnamefont {Baumgardt}}, \bibinfo
  {author} {\bibfnamefont {B.}~\bibnamefont {Husemann}}, \bibinfo {author}
  {\bibfnamefont {N.}~\bibnamefont {Neumayer}}, \bibinfo {author}
  {\bibfnamefont {M.}~\bibnamefont {Hilker}}, \bibinfo {author} {\bibfnamefont
  {R.}~\bibnamefont {Pechetti}}, \bibinfo {author} {\bibfnamefont
  {S.}~\bibnamefont {Mieske}}, \bibinfo {author} {\bibfnamefont
  {A.}~\bibnamefont {Dumont}},\ and\ \bibinfo {author} {\bibfnamefont
  {I.}~\bibnamefont {Georgiev}},\ }\bibfield  {title} {\bibinfo {title} {First
  direct dynamical detection of a dual supermassive black hole system at
  sub-kiloparsec separation},\ }\href
  {https://doi.org/10.1051/0004-6361/202140827} {\bibfield  {journal} {\bibinfo
   {journal} {A\&A}\ }\textbf {\bibinfo {volume} {658}},\ \bibinfo {pages}
  {A152} (\bibinfo {year} {2022})}\BibitemShut {NoStop}%
\bibitem [{\citenamefont {Colpi}\ \emph {et~al.}(2019)\citenamefont {Colpi},
  \citenamefont {Holley-Bockelmann}, \citenamefont {Bogdanovic}, \citenamefont
  {Natarajan}, \citenamefont {Bellovary}, \citenamefont {Sesana}, \citenamefont
  {Tremmel}, \citenamefont {Schnittman}, \citenamefont {Comerford},\ and\
  \citenamefont {Barausse~\textit{et al.}}}]{Colpi2019}%
  \BibitemOpen
  \bibfield  {author} {\bibinfo {author} {\bibfnamefont {M.}~\bibnamefont
  {Colpi}}, \bibinfo {author} {\bibfnamefont {K.}~\bibnamefont
  {Holley-Bockelmann}}, \bibinfo {author} {\bibfnamefont {T.}~\bibnamefont
  {Bogdanovic}}, \bibinfo {author} {\bibfnamefont {P.}~\bibnamefont
  {Natarajan}}, \bibinfo {author} {\bibfnamefont {J.}~\bibnamefont
  {Bellovary}}, \bibinfo {author} {\bibfnamefont {A.}~\bibnamefont {Sesana}},
  \bibinfo {author} {\bibfnamefont {M.}~\bibnamefont {Tremmel}}, \bibinfo
  {author} {\bibfnamefont {J.}~\bibnamefont {Schnittman}}, \bibinfo {author}
  {\bibfnamefont {J.}~\bibnamefont {Comerford}},\ and\ \bibinfo {author}
  {\bibfnamefont {E.}~\bibnamefont {Barausse~\textit{et al.}}},\ }\href@noop {}
  {\bibinfo {title} {Astro2020 science white paper: The gravitational wave view
  of massive black holes}} (\bibinfo {year} {2019}),\ \Eprint
  {https://arxiv.org/abs/arXiv:1903.06867} {arXiv:1903.06867} \BibitemShut
  {NoStop}%
\bibitem [{\citenamefont {Saulson}(2013)}]{SAULSON2013288}%
  \BibitemOpen
  \bibfield  {author} {\bibinfo {author} {\bibfnamefont {P.~R.}\ \bibnamefont
  {Saulson}},\ }\bibfield  {title} {\bibinfo {title} {Gravitational wave
  detection: Principles and practice},\ }\href
  {https://doi.org/https://doi.org/10.1016/j.crhy.2013.01.007} {\bibfield
  {journal} {\bibinfo  {journal} {Comptes Rendus Physique}\ }\textbf {\bibinfo
  {volume} {14}},\ \bibinfo {pages} {288} (\bibinfo {year} {2013})},\ \bibinfo
  {note} {gravitational waves / Ondes gravitationnelles}\BibitemShut {NoStop}%
\bibitem [{\citenamefont {Wang}\ and\ \citenamefont {Hu}(2022)}]{2208.13351}%
  \BibitemOpen
  \bibfield  {author} {\bibinfo {author} {\bibfnamefont {R.}~\bibnamefont
  {Wang}}\ and\ \bibinfo {author} {\bibfnamefont {B.}~\bibnamefont {Hu}},\
  }\href@noop {} {\bibinfo {title} {Litepig: A lite parameter inference system
  for the gravitational wave in the millihertz band}} (\bibinfo {year}
  {2022}),\ \Eprint {https://arxiv.org/abs/arXiv:2208.13351} {arXiv:2208.13351}
  \BibitemShut {NoStop}%
\bibitem [{\citenamefont {Cutler}\ and\ \citenamefont
  {Flanagan}(1994)}]{PhysRevD.49.2658}%
  \BibitemOpen
  \bibfield  {author} {\bibinfo {author} {\bibfnamefont {C.}~\bibnamefont
  {Cutler}}\ and\ \bibinfo {author} {\bibfnamefont {E.~E.}\ \bibnamefont
  {Flanagan}},\ }\bibfield  {title} {\bibinfo {title} {Gravitational waves from
  merging compact binaries: How accurately can one extract the binary's
  parameters from the inspiral waveform?},\ }\href
  {https://doi.org/10.1103/PhysRevD.49.2658} {\bibfield  {journal} {\bibinfo
  {journal} {Phys. Rev. D}\ }\textbf {\bibinfo {volume} {49}},\ \bibinfo
  {pages} {2658} (\bibinfo {year} {1994})}\BibitemShut {NoStop}%
\bibitem [{\citenamefont {Cutler}\ \emph {et~al.}(1993)\citenamefont {Cutler},
  \citenamefont {Apostolatos}, \citenamefont {Bildsten}, \citenamefont {Finn},
  \citenamefont {Flanagan}, \citenamefont {Kennefick}, \citenamefont
  {Markovic}, \citenamefont {Ori}, \citenamefont {Poisson}, \citenamefont
  {Sussman},\ and\ \citenamefont {Thorne}}]{PhysRevLett.70.2984}%
  \BibitemOpen
  \bibfield  {author} {\bibinfo {author} {\bibfnamefont {C.}~\bibnamefont
  {Cutler}}, \bibinfo {author} {\bibfnamefont {T.~A.}\ \bibnamefont
  {Apostolatos}}, \bibinfo {author} {\bibfnamefont {L.}~\bibnamefont
  {Bildsten}}, \bibinfo {author} {\bibfnamefont {L.~S.}\ \bibnamefont {Finn}},
  \bibinfo {author} {\bibfnamefont {E.~E.}\ \bibnamefont {Flanagan}}, \bibinfo
  {author} {\bibfnamefont {D.}~\bibnamefont {Kennefick}}, \bibinfo {author}
  {\bibfnamefont {D.~M.}\ \bibnamefont {Markovic}}, \bibinfo {author}
  {\bibfnamefont {A.}~\bibnamefont {Ori}}, \bibinfo {author} {\bibfnamefont
  {E.}~\bibnamefont {Poisson}}, \bibinfo {author} {\bibfnamefont {G.~J.}\
  \bibnamefont {Sussman}},\ and\ \bibinfo {author} {\bibfnamefont {K.~S.}\
  \bibnamefont {Thorne}},\ }\bibfield  {title} {\bibinfo {title} {The last
  three minutes: Issues in gravitational-wave measurements of coalescing
  compact binaries},\ }\href {https://doi.org/10.1103/PhysRevLett.70.2984}
  {\bibfield  {journal} {\bibinfo  {journal} {Phys. Rev. Lett.}\ }\textbf
  {\bibinfo {volume} {70}},\ \bibinfo {pages} {2984} (\bibinfo {year}
  {1993})}\BibitemShut {NoStop}%
\bibitem [{\citenamefont {Varma}\ \emph {et~al.}(2013)\citenamefont {Varma},
  \citenamefont {Fujita}, \citenamefont {Choudhary},\ and\ \citenamefont
  {Iyer}}]{PhysRevD.88.024038}%
  \BibitemOpen
  \bibfield  {author} {\bibinfo {author} {\bibfnamefont {V.}~\bibnamefont
  {Varma}}, \bibinfo {author} {\bibfnamefont {R.}~\bibnamefont {Fujita}},
  \bibinfo {author} {\bibfnamefont {A.}~\bibnamefont {Choudhary}},\ and\
  \bibinfo {author} {\bibfnamefont {B.~R.}\ \bibnamefont {Iyer}},\ }\bibfield
  {title} {\bibinfo {title} {Comparison of post-newtonian templates for extreme
  mass ratio inspirals},\ }\href {https://doi.org/10.1103/PhysRevD.88.024038}
  {\bibfield  {journal} {\bibinfo  {journal} {Phys. Rev. D}\ }\textbf {\bibinfo
  {volume} {88}},\ \bibinfo {pages} {024038} (\bibinfo {year}
  {2013})}\BibitemShut {NoStop}%
\bibitem [{\citenamefont {Caprini}\ and\ \citenamefont
  {Figueroa}(2018)}]{Caprini_2018}%
  \BibitemOpen
  \bibfield  {author} {\bibinfo {author} {\bibfnamefont {C.}~\bibnamefont
  {Caprini}}\ and\ \bibinfo {author} {\bibfnamefont {D.~G.}\ \bibnamefont
  {Figueroa}},\ }\bibfield  {title} {\bibinfo {title} {Cosmological backgrounds
  of gravitational waves},\ }\href {https://doi.org/10.1088/1361-6382/aac608}
  {\bibfield  {journal} {\bibinfo  {journal} {Classical and Quantum Gravity}\
  }\textbf {\bibinfo {volume} {35}},\ \bibinfo {pages} {163001} (\bibinfo
  {year} {2018})}\BibitemShut {NoStop}%
\bibitem [{\citenamefont {Christensen}(2018)}]{Christensen_2019}%
  \BibitemOpen
  \bibfield  {author} {\bibinfo {author} {\bibfnamefont {N.}~\bibnamefont
  {Christensen}},\ }\bibfield  {title} {\bibinfo {title} {Stochastic
  gravitational wave backgrounds},\ }\href
  {https://doi.org/10.1088/1361-6633/aae6b5} {\bibfield  {journal} {\bibinfo
  {journal} {Reports on Progress in Physics}\ }\textbf {\bibinfo {volume}
  {82}},\ \bibinfo {pages} {016903} (\bibinfo {year} {2018})}\BibitemShut
  {NoStop}%
\bibitem [{\citenamefont {Renzini}\ \emph {et~al.}(2022)\citenamefont
  {Renzini}, \citenamefont {Goncharov}, \citenamefont {Jenkins},\ and\
  \citenamefont {Meyers}}]{galaxies10010034}%
  \BibitemOpen
  \bibfield  {author} {\bibinfo {author} {\bibfnamefont {A.~I.}\ \bibnamefont
  {Renzini}}, \bibinfo {author} {\bibfnamefont {B.}~\bibnamefont {Goncharov}},
  \bibinfo {author} {\bibfnamefont {A.~C.}\ \bibnamefont {Jenkins}},\ and\
  \bibinfo {author} {\bibfnamefont {P.~M.}\ \bibnamefont {Meyers}},\ }\bibfield
   {title} {\bibinfo {title} {Stochastic gravitational-wave backgrounds:
  Current detection efforts and future prospects},\ }\bibfield  {journal}
  {\bibinfo  {journal} {Galaxies}\ }\textbf {\bibinfo {volume} {10}},\ \href
  {https://doi.org/10.3390/galaxies10010034} {10.3390/galaxies10010034}
  (\bibinfo {year} {2022})\BibitemShut {NoStop}%
\bibitem [{\citenamefont {Rodriguez}\ \emph {et~al.}(2013)\citenamefont
  {Rodriguez}, \citenamefont {Farr}, \citenamefont {Farr},\ and\ \citenamefont
  {Mandel}}]{PhysRevD.88.084013}%
  \BibitemOpen
  \bibfield  {author} {\bibinfo {author} {\bibfnamefont {C.~L.}\ \bibnamefont
  {Rodriguez}}, \bibinfo {author} {\bibfnamefont {B.}~\bibnamefont {Farr}},
  \bibinfo {author} {\bibfnamefont {W.~M.}\ \bibnamefont {Farr}},\ and\
  \bibinfo {author} {\bibfnamefont {I.}~\bibnamefont {Mandel}},\ }\bibfield
  {title} {\bibinfo {title} {Inadequacies of the fisher information matrix in
  gravitational-wave parameter estimation},\ }\href
  {https://doi.org/10.1103/PhysRevD.88.084013} {\bibfield  {journal} {\bibinfo
  {journal} {Phys. Rev. D}\ }\textbf {\bibinfo {volume} {88}},\ \bibinfo
  {pages} {084013} (\bibinfo {year} {2013})}\BibitemShut {NoStop}%
\bibitem [{\citenamefont {Bernardi}\ \emph {et~al.}(2022)\citenamefont
  {Bernardi}, \citenamefont {Brost}, \citenamefont {Denisov}, \citenamefont
  {Landsberg}, \citenamefont {Aleksa}, \citenamefont {d'Enterria},
  \citenamefont {Janot}, \citenamefont {Mangano}, \citenamefont {Selvaggi},\
  and\ \citenamefont {Zimmermann~\textit{et al.}}}]{2203.06520}%
  \BibitemOpen
  \bibfield  {author} {\bibinfo {author} {\bibfnamefont {G.}~\bibnamefont
  {Bernardi}}, \bibinfo {author} {\bibfnamefont {E.}~\bibnamefont {Brost}},
  \bibinfo {author} {\bibfnamefont {D.}~\bibnamefont {Denisov}}, \bibinfo
  {author} {\bibfnamefont {G.}~\bibnamefont {Landsberg}}, \bibinfo {author}
  {\bibfnamefont {M.}~\bibnamefont {Aleksa}}, \bibinfo {author} {\bibfnamefont
  {D.}~\bibnamefont {d'Enterria}}, \bibinfo {author} {\bibfnamefont
  {P.}~\bibnamefont {Janot}}, \bibinfo {author} {\bibfnamefont {M.~L.}\
  \bibnamefont {Mangano}}, \bibinfo {author} {\bibfnamefont {M.}~\bibnamefont
  {Selvaggi}},\ and\ \bibinfo {author} {\bibfnamefont {F.}~\bibnamefont
  {Zimmermann~\textit{et al.}}},\ }\href@noop {} {\bibinfo {title} {The future
  circular collider: a summary for the us 2021 snowmass process}} (\bibinfo
  {year} {2022}),\ \Eprint {https://arxiv.org/abs/arXiv:2203.06520}
  {arXiv:2203.06520} \BibitemShut {NoStop}%
\bibitem [{\citenamefont {Piro}\ \emph {et~al.}(2022)\citenamefont {Piro},
  \citenamefont {Colpi}, \citenamefont {Aird}, \citenamefont {Mangiagli},
  \citenamefont {Fabian}, \citenamefont {Guainazzi}, \citenamefont {Marsat},
  \citenamefont {Sesana}, \citenamefont {McNamara},\ and\ \citenamefont
  {Bonetti~\textit{et al.}}}]{2211.13759}%
  \BibitemOpen
  \bibfield  {author} {\bibinfo {author} {\bibfnamefont {L.}~\bibnamefont
  {Piro}}, \bibinfo {author} {\bibfnamefont {M.}~\bibnamefont {Colpi}},
  \bibinfo {author} {\bibfnamefont {J.}~\bibnamefont {Aird}}, \bibinfo {author}
  {\bibfnamefont {A.}~\bibnamefont {Mangiagli}}, \bibinfo {author}
  {\bibfnamefont {A.~C.}\ \bibnamefont {Fabian}}, \bibinfo {author}
  {\bibfnamefont {M.}~\bibnamefont {Guainazzi}}, \bibinfo {author}
  {\bibfnamefont {S.}~\bibnamefont {Marsat}}, \bibinfo {author} {\bibfnamefont
  {A.}~\bibnamefont {Sesana}}, \bibinfo {author} {\bibfnamefont
  {P.}~\bibnamefont {McNamara}},\ and\ \bibinfo {author} {\bibfnamefont
  {M.}~\bibnamefont {Bonetti~\textit{et al.}}},\ }\href@noop {} {\bibinfo
  {title} {Chasing super-massive black hole merging events with $athena$ and
  lisa}} (\bibinfo {year} {2022}),\ \Eprint
  {https://arxiv.org/abs/arXiv:2211.13759} {arXiv:2211.13759} \BibitemShut
  {NoStop}%
\bibitem [{\citenamefont {Baker}\ \emph {et~al.}(2019)\citenamefont {Baker},
  \citenamefont {Haiman}, \citenamefont {Rossi}, \citenamefont {Berger},
  \citenamefont {Brandt}, \citenamefont {Breedt}, \citenamefont {Breivik},
  \citenamefont {Charisi}, \citenamefont {Derdzinski},\ and\ \citenamefont
  {D'Orazio~\textit{et al.}}}]{Baker2019Multimessenger}%
  \BibitemOpen
  \bibfield  {author} {\bibinfo {author} {\bibfnamefont {J.}~\bibnamefont
  {Baker}}, \bibinfo {author} {\bibfnamefont {Z.}~\bibnamefont {Haiman}},
  \bibinfo {author} {\bibfnamefont {E.~M.}\ \bibnamefont {Rossi}}, \bibinfo
  {author} {\bibfnamefont {E.}~\bibnamefont {Berger}}, \bibinfo {author}
  {\bibfnamefont {N.}~\bibnamefont {Brandt}}, \bibinfo {author} {\bibfnamefont
  {E.}~\bibnamefont {Breedt}}, \bibinfo {author} {\bibfnamefont
  {K.}~\bibnamefont {Breivik}}, \bibinfo {author} {\bibfnamefont
  {M.}~\bibnamefont {Charisi}}, \bibinfo {author} {\bibfnamefont
  {A.}~\bibnamefont {Derdzinski}},\ and\ \bibinfo {author} {\bibfnamefont
  {D.~J.}\ \bibnamefont {D'Orazio~\textit{et al.}}},\ }\bibfield  {title}
  {\bibinfo {title} {Multimessenger science opportunities with {mHz}
  gravitational waves},\ }\href@noop {} {\bibfield  {journal} {\bibinfo
  {journal} {Bulletin of the AAS}\ }\textbf {\bibinfo {volume} {51}} (\bibinfo
  {year} {2019})},\ \bibinfo {note}
  {https://baas.aas.org/pub/2020n3i123}\BibitemShut {NoStop}%
\bibitem [{\citenamefont {Eracleous}\ \emph {et~al.}(2019)\citenamefont
  {Eracleous}, \citenamefont {Gezari}, \citenamefont {Sesana}, \citenamefont
  {Bogdanovic}, \citenamefont {MacLeod}, \citenamefont {Roth},\ and\
  \citenamefont {Dai}}]{Eracleous2019Arena}%
  \BibitemOpen
  \bibfield  {author} {\bibinfo {author} {\bibfnamefont {M.}~\bibnamefont
  {Eracleous}}, \bibinfo {author} {\bibfnamefont {S.}~\bibnamefont {Gezari}},
  \bibinfo {author} {\bibfnamefont {A.}~\bibnamefont {Sesana}}, \bibinfo
  {author} {\bibfnamefont {T.}~\bibnamefont {Bogdanovic}}, \bibinfo {author}
  {\bibfnamefont {M.}~\bibnamefont {MacLeod}}, \bibinfo {author} {\bibfnamefont
  {N.}~\bibnamefont {Roth}},\ and\ \bibinfo {author} {\bibfnamefont
  {L.}~\bibnamefont {Dai}},\ }\bibfield  {title} {\bibinfo {title} {An {Arena}
  for {Multi}-{Messenger} {Astrophysics}: Inspiral and {Tidal} {Disruption} of
  {White} {Dwarfs} by {Massive} {Black} {Holes}},\ }\href@noop {} {\bibfield
  {journal} {\bibinfo  {journal} {Bulletin of the AAS}\ }\textbf {\bibinfo
  {volume} {51}} (\bibinfo {year} {2019})},\ \bibinfo {note}
  {https://baas.aas.org/pub/2020n3i010}\BibitemShut {NoStop}%
\bibitem [{\citenamefont {Ruan}\ \emph {et~al.}(2020)\citenamefont {Ruan},
  \citenamefont {Liu}, \citenamefont {Guo}, \citenamefont {Wu},\ and\
  \citenamefont {Cai}}]{Ruan2020}%
  \BibitemOpen
  \bibfield  {author} {\bibinfo {author} {\bibfnamefont {W.-H.}\ \bibnamefont
  {Ruan}}, \bibinfo {author} {\bibfnamefont {C.}~\bibnamefont {Liu}}, \bibinfo
  {author} {\bibfnamefont {Z.-K.}\ \bibnamefont {Guo}}, \bibinfo {author}
  {\bibfnamefont {Y.-L.}\ \bibnamefont {Wu}},\ and\ \bibinfo {author}
  {\bibfnamefont {R.-G.}\ \bibnamefont {Cai}},\ }\bibfield  {title} {\bibinfo
  {title} {The lisa--taiji network},\ }\href
  {https://doi.org/10.1038/s41550-019-1008-4} {\bibfield  {journal} {\bibinfo
  {journal} {Nature Astronomy}\ }\textbf {\bibinfo {volume} {4}},\ \bibinfo
  {pages} {108} (\bibinfo {year} {2020})}\BibitemShut {NoStop}%
\bibitem [{\citenamefont {Graham}\ and\ \citenamefont
  {Jung}(2018)}]{PhysRevD.97.024052}%
  \BibitemOpen
  \bibfield  {author} {\bibinfo {author} {\bibfnamefont {P.~W.}\ \bibnamefont
  {Graham}}\ and\ \bibinfo {author} {\bibfnamefont {S.}~\bibnamefont {Jung}},\
  }\bibfield  {title} {\bibinfo {title} {Localizing gravitational wave sources
  with single-baseline atom interferometers},\ }\href
  {https://doi.org/10.1103/PhysRevD.97.024052} {\bibfield  {journal} {\bibinfo
  {journal} {Phys. Rev. D}\ }\textbf {\bibinfo {volume} {97}},\ \bibinfo
  {pages} {024052} (\bibinfo {year} {2018})}\BibitemShut {NoStop}%
\bibitem [{\citenamefont {Villarino}(2005)}]{math/0506384}%
  \BibitemOpen
  \bibfield  {author} {\bibinfo {author} {\bibfnamefont {M.~B.}\ \bibnamefont
  {Villarino}},\ }\href@noop {} {\bibinfo {title} {Ramanujan's perimeter of an
  ellipse}} (\bibinfo {year} {2005}),\ \Eprint
  {https://arxiv.org/abs/arXiv:math/0506384} {arXiv:math/0506384} \BibitemShut
  {NoStop}%
\end{thebibliography}
%

\end{document}